\documentclass[a4paper,11pt]{article}
\pdfoutput=1 % if your are submitting a pdflatex (i.e. if you have
\usepackage{jheppub} % for details on the use of the package, please
\usepackage[T1]{fontenc} % if needed

\title{\boldmath Simulations of Domain Walls in Two Higgs Doublet Models}

\author{Richard A. Battye,}
\author{Apostolos Pilaftsis}
\author{and Dominic G. Viatic}
\affiliation{Department of Physics and Astronomy, University of Manchester, Manchester M13 9PL, United\\Kingdom}

\emailAdd{richard.battye@manchester.ac.uk}
\emailAdd{apostolos.pilaftsis@manchester.ac.uk}
\emailAdd{dominic.viatic@postgrad.manchester.ac.uk}
\abstract{The Two Higgs Doublet Model predicts the emergence of 3 distinct domain wall solutions arising from the breaking of 3 accidental global symmetries, $Z_2$, CP1 and CP2, at the electroweak scale for specific choices of the model parameters.
We present numerical kink solutions to the field equations in all three cases along with dynamical simulations of the models in (2+1) and (3+1) dimensions.
For each kink solution we define an associated topological current.
In all three cases simulations produce a network of domain walls which deviates from power law scaling in Minkowski and FRW simulations.
This deviation is attributed to a winding of the electroweak group parameters around the domain walls in our simulations.
We observe a local violation of the neutral vacuum condition on the domain walls in our simulations.
This violation is attributed to relative electroweak transformations across the domain walls which is a general feature emerging from random initial conditions.}

\begin{document} 
\maketitle
\flushbottom

% Introduction
\section{Introduction}
\label{sec:intro}

Although the Standard Model (SM) of particle physics has been conspicuously successful when confronted with experimental data, it is still not satisfactory as a complete or fundamental theory of nature.
Several observed phenomena in cosmology cannot be accounted for in the SM, including dark energy~\cite{Copeland:2006wr}, dark matter~\cite{Ellis:1983ew} and the observed baryon asymmetry of the Universe~\cite{Pilaftsis:1998pd,Buchmuller:2005eh}.
This necessitates some new physics beyond the Standard Model (BSM) to explain these phenomena.
There are also motivations from particle physics to extend the SM.
One interesting idea was that the electroweak and strong forces are unified into some grand unified theory (GUT) at high energy \cite{Lazarides:1980nt,Nath:2006ut}. Other BSM theories aim to address the so-called gauge hierarchy problem of the strength of forces in the SM and gravity~\cite{Djouadi:2005gj}, or provide mechanisms by which neutrinos can acquire mass~\cite{Minkowski:1977sc,Yanagida:1979as,Mohapatra:1979ia,Schechter:1980gr,Lazarides:1980nt} as first established by the measurement of neutrino oscillations~\cite{Fukuda1998,Ahmad2001}.
Therefore, it is important to consider what extensions of the SM can provide compelling explanations for these observations whilst remaining consistent with current experimental measurements.

A minimal and theoretically well-motivated extension of the SM is the so-called Two Higgs Doublet Model (2HDM) first suggested in 1973 \cite{Lee1973}, where the scalar sector of the SM is extended to include one additional complex scalar doublet (for a review, see~\cite{Branco2012}).
The 2HDM has been studied extensively in the contexts of CP violation~\cite{Pilaftsis1999,Grzadkowski2009,Grzadkowski2010,Keus2015}, dark matter and electroweak baryogenesis~\cite{Cohen:1993nk,Keus2014,Hindmarsh2015,Dorsch2017,Carena2015}.
The Higgs mechanism for electroweak symmetry breaking was verified at the LHC in 2012 by the observation of a Higgs boson \cite{Aad2012,Chatrchyan2012} with a mass of 125 GeV \cite{Aad2015}.
Properties of this scalar sector so far match those of the SM \cite{Djouadi:2005gi,Khachatryan2016}. However, current experimental measurements do not rule out the existence of additional scalar particles.
The 2HDM predicts five physical scalar particles: two neutral CP-even states, $h$ and $H$, one of which may be identified with the SM Higgs; a CP-odd neutral state, $A$, and two charged states, $H^\pm$.

Furthermore, it is already well known~\cite{Brawn2011} that the 2HDM predicts the emergence of a variety of topological defects, such as domain walls, vortices and global monopoles, from the breaking of accidental symmetries which the model can possess under certain parameter choices.
Symmetries which are currently broken, such as the electroweak symmetry, should be restored at sufficiently high energies \cite{Garagounis2003} when the system exceeds the relevant critical temperature \cite{Shellard1994}.
These broken symmetries are no longer directly observable. However, they should have been intact in the early Universe when temperatures were far higher than at present \cite{Garagounis2003}.
This suggests that the Universe has undergone a series of symmetry breaking phase transitions in its history.
These phase transitions can leave relic topological defects which could be used to probe high energy physics in the early Universe~\cite{Kibble:1982dd,Garagounis2003, Nakayama2017}.

Our focus in this paper will be on discrete symmetries of the 2HDM, whose spontaneous breaking predicts domain wall solutions. In particular, we are interested in numerical simulations of such topological defects by solving the relevant equations of motion~\cite{Garagounis2003,Moss2006,Pearson2009,Pearson2010}.
The results of these simulations can guide cosmological searches and reveal cosmological implications for new models providing phenomenological constraints on BSM theories complementary to those coming from particle physics experiments.

Domain walls are topological defects which emerge due to the breaking of discrete symmetries \cite{Garagounis2003}.
The breaking of a discrete symmetry results in a vacuum manifold containing topologically disconnected points corresponding to degenerate vacua.
Phase transitions producing domain walls occur at a finite rate, therefore, causally disconnected regions of space can select different vacua.
This divides the universe into so-called \textit{domains} the interfaces between which are \textit{domain walls} \cite{Shellard1994}.
Defects which emerge from the breaking of a global symmetry such as domain walls enter a regime of dynamical scaling such that the number of defects is constant per Hubble horizon \cite{Nakayama2017}.
Domain walls follow a power law scaling with an exponent close to -1 as has been shown in simulations of the Goldstone model \cite{Garagounis2003,Pearson2009}.
This is due to the fact that the walls have a surface tension
%,
%\begin{equation}
%	\sigma = \int {T^0}_0\;dx,
%\end{equation}
under which they collapse as quickly as causality permits.

Domain walls would present an undesirable fate for the Universe were they to exist in nature.
The energy density of matter and radiation in their respective epochs of domination decrease proportionally to $(\text{time})^{-2}$.
However, the energy density of domain walls decreases proportionally to $(\text{time})^{-1}$ \cite{Lazanu2015}.
This means that the energy density of domain walls grows relative to matter and radiation in their respective epochs and, therefore, dominates the energy density of the Universe at some late time \cite{Zeldovich1974,Larsson1996,Press1989}.
This is the so-called \textit{domain wall problem}.
If cosmic domain walls are to be realized in nature, constraints are placed on domain wall-forming models such that domain wall domination does not occur \cite{Lazanu2015} or, at least, occurs after present day.

The Two Higgs Doublet Model predicts three domain wall solutions which arise from the breaking of accidental $Z_2$, CP1 and CP2 symmetries.
We have obtained numerical kink solutions to the field equations in all three cases along with dynamical simulations of the models in (2+1) and (3+1) dimensions.
For all three symmetries, we find a stable network of domain walls which deviates from power law scaling in both Minkowski and Friedmann-Robertson-Walker (FRW) simulations. Most remarkably, we observe a local violation of the neutral vacuum condition on the domain walls in our simulations.

The remainder of this article exhibits the following structure. In Section~\ref{sec:2HDM}, we outline the scalar sector of the 2HDM, derive the relevant equations of motion and discuss the conserved electroweak currents. In addition, we discuss 
the different types of electroweak vacua, as well
as possible accidental discrete symmetries which lead to domain wall solutions. In Section~\ref{sec:sims}, we give a short overview of our simulation techniques used to analyze the evolution of domain-wall networks resulting from the breaking 
of the accidental symmetries, $Z_2$, CP1 and CP2.
Thus, Section~\ref{sec:Z2} presents results for 
$Z_2$ domain-wall simulations starting from random initial conditions. As explicitly demonstrated 
in Appendix~\ref{sec:Rescaling}, our analysis was aided by a suitable parametrization of the 2HDM potential in terms of the physical scalar masses and their mixings. Likewise, Sections~\ref{sec:CP1} and~\ref{sec:CP2} present kink solutions and results of domain wall simulations from the breaking of CP1 and CP2 accidental symmetries, respectively. For all three possible discrete symmetries, $Z_2$, CP1 and CP2, we discuss the dynamical features of the emerging domain walls, including their cosmological scaling. Finally, Section~\ref{sec:discussion} summarizes the main results of our study and presents an outlook for future work.

% The 2HDM
\section{The Two Higgs Doublet Model}
\label{sec:2HDM}

\subsection{Introduction}

The Higgs mechanism for electroweak symmetry breaking was confirmed at the LHC by the discovery of a Higgs boson of mass $\simeq$ 125 GeV \cite{Aad2015}.
Properties of this scalar so far match those of the SM~\cite{Djouadi:2005gi}, nonetheless, current data does not rule out the existence of more scalar particles.
Indeed the suggestion of a more complicated scalar sector far pre-dates the SM Higgs discovery \cite{Lee1973}.
There have been numerous studies, both theoretical and experimental, made into the possibility of extensions to the SM involving additional scalar particles.
One of the simplest models is the 2HDM where one additional complex scalar doublet is added to the SM.
The 2HDM has 5 physical scalar particles: 2 neutral CP-even states, $h$ and $H$, one CP-odd neutral state, $A$, and 2 charged states, $H^\pm$ \cite{Pilaftsis1999}.
The other three scalar degrees of freedom correspond to Goldstone bosons which are absorbed into the longitudinal components of the electroweak gauge bosons, $W^\pm$ and $Z^0$.
As stated earlier, there are currently several observed phenomena in cosmology which indicate the need for BSM physics including dark matter and the need for additional sources of CP violation.
The 2HDM has been used extensively in the literature to explore, for example, new particle physics phenomenology beyond the Standard Model which could explain these observations~\cite{Pilaftsis1999,Branco2012,Bian2016,Carena2015,Azevedo:2018llq, Basler:2019iuu}.

The Lagrangian of the electroweak-Higgs sector of the 2HDM is given by
\begin{equation}
    \mathcal{L} = -\frac{1}{4}W^{a,\,\mu\nu}W_{\mu\nu}^a - \frac{1}{4}B^{\mu\nu}B_{\mu\nu} + \left|D_\mu\Phi_1\right|^2 + \left|D_\mu\Phi_2\right|^2 - V(\Phi_1,\Phi_2)
\end{equation}
where $\Phi_1$ and $\Phi_2$ are the complex scalar Higgs doublets while
\begin{equation}
  \label{eq:Wmunu}
    W_{\mu\nu}^a = \partial_\mu W_\nu^a - \partial_\nu W_\mu^a - g \varepsilon^{abc} W_\mu^b W_\nu^c
\end{equation}
and
\begin{equation}
  \label{eq:Bmunu}
    B_{\mu\nu} = \partial_\mu B_\nu - \partial_\nu B_\mu
\end{equation}
are the $\text{SU}(2)_L$ and $\text{U}(1)_Y$ field strengths, respectively.
The gauge-covariant derivative is given by
\begin{equation}
    D_\mu = \partial_\mu + \frac{ig}{2} W_\mu^a \sigma^a + \frac{i g^\prime}{2}B_\mu.
\end{equation}
where $\sigma^a$ are the Pauli matrices.

The equations of motion for the Higgs and electroweak gauge fields can be derived to be
\begin{equation}
  \label{eq:ScalarEoM}
    D_\mu D^\mu \Phi_i + \frac{\partial V}{\partial \Phi_i^\dagger} = 0,
\end{equation}
\begin{equation}
  \label{eq:U1EoM}
    \partial_\mu B^{\mu\nu} = \frac{i g^\prime}{2}\left[\Phi_i^\dagger D^\nu \Phi_i - (D^\nu\Phi_i)^\dagger \Phi_i\right]
\end{equation}
and
\begin{equation}
   \label{eq:SU2EoM}
    \partial_\mu W^{a,\,\mu\nu} + g\varepsilon^{abc}W_\mu^b W^{c,\,\mu\nu} = \frac{ig}{2}\left[\Phi_i^\dagger\sigma^a D^\nu\Phi_i^\dagger - \left(D^\nu\Phi_i\right)^\dagger\sigma^a\Phi_i\right]\,,
\end{equation}
where summation over repeated indices is implied.
The right-hand sides~(RHS) of \eqref{eq:U1EoM} and \eqref{eq:SU2EoM} are $\text{U}(1)_{Y} $ and $\text{SU}(2)_{L}$ currents, respectively, 
which are defined as 
\begin{equation}\label{eq:U1current}
    J_B^\mu\ =\ \frac{ig^\prime}{2}\left[\Phi_i^\dagger (D^\mu \Phi_i)\: -\: (D^\mu\Phi_i)^\dagger\Phi_i\right]
\end{equation}
and
\begin{equation}\label{eq:SU2current}
    J_W^{a,\,\mu}\ =\ \frac{ig}{2}\left[\Phi_i^\dagger\sigma^a (D^\mu\Phi_i)\: -\: (D^\mu\Phi_i)^\dagger\sigma^a\Phi_i\right].
\end{equation}
The most general form of the tree-level 2HDM scalar potential is \cite{Lee1973,Branco2012}
\begin{align}
  \label{eq:2HDMPotential}
    V =& -\mu_1^2 (\Phi_1^\dagger \Phi_1) - \mu_2^2 (\Phi_2^\dagger \Phi_2) - \left[m_{12}^2(\Phi_1^\dagger\Phi_2) + \text{H.c.}\right] \nonumber\\
    &+ \lambda_1 (\Phi_1^\dagger \Phi_1)^2 + \lambda_2 (\Phi_2^\dagger \Phi_2)^2 + \lambda_3 (\Phi_1^\dagger \Phi_1) (\Phi_2^\dagger \Phi_2) + \lambda_4(\Phi_1^\dagger \Phi_2)(\Phi_2^\dagger \Phi_1)\nonumber\\
    &+ \left[\frac{\lambda_5}{2}(\Phi_1^\dagger\Phi_2)^2 + \lambda_6 (\Phi_1^\dagger \Phi_1)(\Phi_1^\dagger\Phi_2) + \lambda_7 (\Phi_2^\dagger \Phi_2)(\Phi_1^\dagger\Phi_2) + \text{H.c.}\right].
\end{align}
The general 2HDM potential has 14 free parameters where $\mu_1^2, \mu_2^2, \lambda_1, \lambda_2, \lambda_3$ and $\lambda_4$ are real and $m_{12}^2, \lambda_5, \lambda_6$ and $\lambda_7$ are complex.
It is common to impose some additional symmetry on the model, thereby reducing the number of free parameters.
In our context we note that the 2HDM can possess 3 accidental discrete symmetries which predict domain wall solutions under the parameter reductions shown in Table \ref{tab:2HDMConstraints} \cite{Brawn2011,Branco2012,Ivanov2007,Ferreira:2009wh,Dev:2014yca}.
\begin{table}
	\centering
	\begin{tabular}{| c || c | c | c | c | c | c | c | c | c | c |}
		\hline
		Symmetry & $ \mu_1^2 $ & $ \mu_2^2 $ & $ m_{12}^2 $ & $ \lambda_1 $ & $ \lambda_2 $ & $ \lambda_3 $ & $ \lambda_4 $ & $ \lambda_5 $ & $ \lambda_6 $ & $ \lambda_7 $ \\ \hline
		$ Z_2 $ & - & - & $ 0 $ & - & - & - & - & Real & $ 0 $ & $ 0 $ \\ \hline
		CP1 & - & - & Real & - & - & - & - & Real & Real & Real \\ \hline
		CP2 & - & $ \mu_1^2 $ & $ 0 $ & - & $ \lambda_1 $ & - & - & - & - & $ -\lambda_6 $ \\ \hline
	\end{tabular}
	\caption{Parameters in the 2HDM for which the model possesses an accidental discrete symmetry.
	Dashes signify the absence of a restriction on the corresponding parameter. See \cite{Brawn2011,Pilaftsis2011} for a complete classification of all possible symmetries.}
	\label{tab:2HDMConstraints}
\end{table}
It is useful to note that one can further reduce the 2HDM parameter space whilst keeping all 3 accidental symmetries of interest using the reduction $\lambda_6 = \lambda_7$ and $m_{12}^2, \lambda_5$ and $\lambda_6$ being real.
In this reduced basis the 3 discrete symmetries are obtained via the constraints given in Table~\ref{tab:Diag_Red_Constraints}.
However, in this basis, the restrictions for a CP2 symmetry are simply a subset of those for $Z_2$~\cite{Pilaftsis2011}. 
In fact, the CP2 model possesses an enhanced $S_2\times Z_2$ symmetry for $\mu_1^2 = \mu_2^2$,  $\lambda_1 = \lambda_2$ and $\lambda_6 = \lambda_7 = 0$~\cite{Darvishi:2020teg}.
\begin{table}
	\centering
	\begin{tabular}{| c || c | c | c | c | c | c | c | c | c | c |}
		\hline
		Symmetry & $ \mu_1^2 $ & $ \mu_2^2 $ & $ m_{12}^2 $ & $ \lambda_1 $ & $ \lambda_2 $ & $ \lambda_3 $ & $ \lambda_4 $ & $ \text{Re}\left(\lambda_5\right) $ & $ \lambda_6 = \lambda_7 $ \\ \hline
		$ Z_2 $ & - & - & $ 0 $ & - & - & - & - & - & $ 0 $ \\ \hline
		CP1 & - & - & - & - & - & - & - & - & - \\ \hline
		CP2 & - & $ \mu_1^2 $ & $ 0 $ & - & $ \lambda_1 $ & - & - & - & 0 \\ \hline
	\end{tabular}
	\caption{Parameters for the 2HDM in the reduced basis, $\text{Im}\left(m_{12}^2\right) = \text{Im}\left(\lambda_5\right) = \text{Im}\left(\lambda_6\right) = 0$ and $\lambda_6 = \lambda_7$, for which the model possesses one of the three accidental discrete symmetries.
	Dashes signify the absence of a restriction on the corresponding parameter.}
	\label{tab:Diag_Red_Constraints}
\end{table}

\subsection{Vacua}

The vacua in the 2HDM can be separated into 3 types \cite{Branco2012}: (i)~\textit{CP-preserving vacua} where the vacuum expectation values (VEVs) have no relative phase between them,
\begin{equation}\label{eq:NormalVac}
	\Phi_1^0 = \frac{1}{\sqrt{2}}\left(\begin{matrix}
	0 \\
	v_1
	\end{matrix}\right),\quad 
	\Phi_2^0 = \frac{1}{\sqrt{2}}\left(\begin{matrix}
	0 \\
	v_2
	\end{matrix}\right);
\end{equation}
(ii)~\textit{CP-breaking vacua} where the VEVs have a relative phase, $ \xi $,
\begin{equation}\label{eq:CPVac}
\Phi_1^0 = \frac{1}{\sqrt{2}}\left(\begin{matrix}
0 \\
v_1
\end{matrix}\right),\quad 
\Phi_2^0 = \frac{1}{\sqrt{2}}\left(\begin{matrix}
0 \\
v_2 e^{i\xi}
\end{matrix}\right);
\end{equation}
and (iii)~\textit{charge-breaking vacua} where there is a non-zero upper component in one of the doublets,
\begin{equation}\label{eq:ChargedVac}
\Phi_1^0 = \frac{1}{\sqrt{2}}\left(\begin{matrix}
0 \\
v_1
\end{matrix}\right),\quad 
\Phi_2^0 = \frac{1}{\sqrt{2}}\left(\begin{matrix}
v_+ \\
v_2 e^{i\xi}
\end{matrix}\right).
\end{equation}
The parameters $ v_1,\,v_2,\,v_+$ and $\xi$ are referred to as the \textit{vacuum manifold parameters}.
It can be demonstrated that charge-breaking vacua are the most general one would need by exploiting electroweak gauge freedom \cite{Branco2012}.
In other words one can produce a general parametrization of the Higgs doublets via an electroweak gauge transformation (EWGT) of the charge-breaking vacuum.

Let us therefore parametrize the Higgs doublets via an EWGT of the charge-breaking vacuum,
\begin{equation}\label{eq:gen_par}
    \Phi_i = U\Phi_i^0,
\end{equation}
with
\begin{equation}
  \label{eq:EWUMatrix}
    U = e^{i\theta} U_L = e^{i\theta}\exp\left(i \widehat{G}^a \frac{\sigma^a}{2}\right)\,,
\end{equation}
where $\widehat{G}^a \equiv {G^a}/{v_\text{SM}}$, and $\theta$ and $G^a$ are the would-be Goldstone bosons.
We can write the $U(1)_Y$ current as
\begin{equation}
  \label{eq:U(1)current}
    J_B^\mu = -\frac{1}{2}v_2^2\partial^\mu\xi + \frac{ig^\prime}{2}\Phi_i^{0\dagger}\left[U^\dagger D^\mu U - (D^\mu U)^\dagger U\right]\Phi_i^0.
\end{equation}
The U(1)$_Y$ current decomposes into a term involving the gradient of $\xi$ and terms involving covariant derivatives of the group parameters.
%As such a spatial variation of $\xi$ would yield a non-zero $J_B^\mu$.

Let us consider the gauge fields as pure gauges,
\begin{equation}
    \boldsymbol{W}_\mu\: \equiv\: W_\mu^a \frac{\sigma^a}{2}\ =\ \frac{1}{ig} U_L \partial _\mu U_L^\dagger\,, \qquad
    B_\mu\ =\ -\frac{2}{g^\prime}\partial_\mu\theta\,,
\end{equation}
obeying the gauge-fixing condition: $\boldsymbol{W}_0 = \boldsymbol{0}$ and $B_0 = 0$. As a consequence, the SU(2)$_L$ and U(1)$_Y$ field strengths given in~\eqref{eq:Wmunu} and~\eqref{eq:Bmunu} vanish identically, i.e.~$W^a_{\mu\nu} = B_{\mu\nu} = 0$. Moreover, in pure gauge, the fields $\boldsymbol{W}_\mu$ and $B_\mu$ can be removed from the electroweak currents, defined in~\eqref{eq:U1current} and~\eqref{eq:SU2current}, by an EWGT.
Upon an appropriate EWGT, the electroweak currents then take on the simpler 
form,
\begin{equation}
    J_B^\mu\ =\ \frac{ig^\prime}{2}\left[{\Phi}_i^\dagger (\partial^\mu {\Phi}_i)\: -\: (\partial ^\mu {\Phi}_i^\dagger){\Phi}_i\right]
\end{equation}
and
\begin{equation}
    J_W^{a,\,\mu}\ =\ \frac{ig^\prime}{2}\left[{\Phi}_i^\dagger \sigma^a (\partial^\mu {\Phi}_i)\: -\: (\partial ^\mu {\Phi}_i^\dagger) \sigma^a {\Phi}_i\right].
\end{equation}
Since the left-hand sides (LHS) of~\eqref{eq:U1EoM} and~\eqref{eq:SU2EoM} vanish, so should their RHSs.
Consequently, the unitary gauge transformation matrix $U$ in~\eqref{eq:EWUMatrix} and the spatial profile of the would-be Goldstone fields, $\theta$ and $G^a$, cannot be arbitrary. Nevertheless, employing the equation of motion of the scalar doublets~\eqref{eq:ScalarEoM}, one can show that the
electroweak currents are divergence\-less, i.e. $\partial_\mu J_B^\mu = 0$ and $\partial_\mu J_W^{a,\mu} = 0$.
Given that the electroweak currents $J_B^\mu$ and $J_W^{a,\mu}$ are zero in the boundaries, it is obvious that these currents must vanish every\-where, at least in the approximation with one spatial dimension.
Hence, the equations of motion for the electroweak gauge fields are automatically satisfied for these minimum energy kink solutions.
Moreover, the vanishing of the field strength tensors, $W_{\mu\nu}^a$ and $B_{\mu\nu}$, implies that their contribution to the vacuum energy is zero.
Hence, the equations of motion for the scalar sector~\eqref{eq:ScalarEoM}--\eqref{eq:SU2EoM}
enforce minimization of the energy density as well.

Taking into account the above considerations, the 
complete 2HDM Lagrangian of direct relevance to our study is
given by
\begin{equation}
  \label{eq:L2HDM}
    \mathcal{L}\ =\ \left|\partial_\mu\Phi_i\right|^2\: -\: V(\Phi_1,\Phi_2).
\end{equation}
It is the equations of motion derived from this Lagrangian which are solved numerically in our simulations of domain walls in the
next sections.

\subsection{Parametrizing the Global Field Theory}

An alternative representation of the scalar potential is the so-called bilinear field-space formalism \cite{Maniatis2007,Ivanov2007,Nishi:2006tg,Brawn2011},
\begin{equation}
	V = -\frac{1}{2}M_\mu R^\mu + \frac{1}{4}L_{\mu\nu}R^\mu R^\nu,
\end{equation}
where the first term contains the quadratic mass parameters and the second contains the quartic couplings.
The specific forms of the 4-vectors and tensor are
\begin{subequations}
	\begin{equation}\label{eq:RVector}
		R^\mu = \Phi^\dagger\sigma^\mu\Phi = \left(\begin{matrix}
		\Phi_1^\dagger\Phi_1 + \Phi_2^\dagger\Phi_2 \\
		\Phi_1^\dagger\Phi_2 + \Phi_2^\dagger\Phi_1 \\
		-i[\Phi_1^\dagger\Phi_2 - \Phi_2^\dagger\Phi_1] \\
		\Phi_1^\dagger\Phi_1 - \Phi_2^\dagger\Phi_2
		\end{matrix}\right),
	\end{equation}
	\begin{equation}
		M_\mu = \left(\mu_1^2 + \mu_2^2,\; 2\text{Re}\left(m_{12}^2\right),\; -2\text{Im}\left(m_{12}^2\right),\; \mu_1^2 - \mu_2^2\right),
	\end{equation}
	\begin{equation}\label{eq:Lmatrix}
		L_{\mu\nu} = \left(
		\begin{matrix}
		\lambda_{123} & \text{Re}\left(\lambda_{67}\right) & -\text{Im}\left(\lambda_{67}\right) & \bar{\lambda}_{12} \\
		\text{Re}\left(\lambda_{67}\right) & \lambda_4 + \text{Re}\left(\lambda_5\right) & -\text{Im}\left(\lambda_5\right) & \text{Re}\left(\bar{\lambda}_{67}\right) \\
		-\text{Im}\left(\lambda_{67}\right) & -\text{Im}\left(\lambda_5\right) & \lambda_4 - \text{Re}\left(\lambda_5\right) & -\text{Im}\left(\bar{\lambda}_{67}\right) \\
		\bar{\lambda}_{12} & \text{Re}\left(\bar{\lambda}_{67}\right) & -\text{Im}\left(\bar{\lambda}_{67}\right) & \bar{\lambda}_{123}
		\end{matrix}
		\right).
	\end{equation}
\end{subequations}
In \eqref{eq:RVector}, $ \Phi \equiv \left(\Phi_1,\Phi_2\right)^T $ and $ \sigma^\mu $ are the Pauli matrices including the identity, $\sigma^0 = I_2$.
Note that in \eqref{eq:Lmatrix} we have introduced the short-hand notations $\lambda_{ab} = \lambda_a + \lambda_b$, $\Bar{\lambda}_{ab} = \lambda_a - \lambda_b$, $\lambda_{abc} = \lambda_a + \lambda_b + \lambda_c $ and $\Bar{\lambda}_{abc} = \lambda_a + \lambda_b - \lambda_c$.
It should be noted that the 4-vector, $R^\mu$, is invariant under a unitary transformation and can, therefore, be equally written in terms of the vacuum manifold parameters:
\begin{equation}\label{eq:RVectorVac}
    R^\mu = \frac{1}{2}\left(\begin{matrix}
    v_1^2 + v_2^2 + v_+^2 \\
    2 v_1 v_2 \cos\xi \\
    2 v_1 v_2 \sin\xi \\
    v_1^2 - v_2^2 - v_+^2
    \end{matrix}\right).
\end{equation}
This allows us to relate the components of $R^\mu$ to the vacuum manifold parameters via the expressions
\begin{equation}
    v_1^2 = R^0 + R^3, \quad v_2^2 = \frac{(R^1)^2 + (R^2)^2}{R^0 + R^3},  \quad v_+^2 = \frac{R_\mu R^\mu}{R^0 + R^3}, \quad \tan\xi = \frac{R^2}{R^1}.
\end{equation}
%and $ \tan\xi = {R^2}/{R^1} $.
This general parametrization admits the possibility of charge-violating solutions. One can ensure that a solution respects charge symmetry by checking/demanding that $R^\mu$ have null norm, i.e. $R_\mu R^\mu = 0$.
This is the so-called \textit{neutral vacuum condition} \cite{Ivanov2007} and is achieved by having $v_+ = 0$.

It is apparent that there are several ways one can choose to represent the Higgs doublets in the 2HDM.
Possibly the most obvious representation is the so-called \textit{linear representation},
\begin{equation}\label{eq:LinRep}
    \Phi_1 = \left(\begin{matrix}
    \phi_1 + i\phi_2 \\
    \phi_3 + i\phi_4
    \end{matrix}\right),\quad
    \Phi_2 = \left(\begin{matrix}
    \phi_5 + i\phi_5 \\
    \phi_7 + i\phi_8
    \end{matrix}\right),
\end{equation}
where the components of the doublets are parametrized by 8 real scalar fields, $\phi_i$.
Generating a general parametrization of the doublets via a local electroweak transformation of the general vacuum \eqref{eq:ChargedVac} with the $\text{U}(1)_Y \times \text{SU}(2)_L$ matrix,
\begin{equation}\label{eq:EWGT}
    U = e^{i\theta}\left(\begin{matrix}
    \cos(\gamma_1) \exp({i\gamma_2}) & \sin({\gamma_1}) \exp({i\gamma_3}) \\
    -\sin({\gamma_1}) \exp({-i\gamma_3}) & \cos({\gamma_1}) \exp({-i\gamma_2})
    \end{matrix}\right),
\end{equation}
one produces an alternative 8-parameter representation of the doublets involving the vacuum manifold parameters, $v_1$, $v_2$, $\xi$ and $v_+$, and the electroweak group parameters $\gamma_1$, $\gamma_2$, $\gamma_3$ and $\theta$.
One could equally parametrize the doublets in terms of the components of $R^\mu$ and the 4 electroweak parameters $\gamma_1, \gamma_2, \gamma_3, \theta$.
We can relate the vacuum manifold parameters back to the linear representation via the expressions for $R^\mu$ in terms of the scalar fields $\phi_i$.
One can also introduce the SU(2)$_L$ invariant object $\Phi_1^T i \sigma^2 \Phi_2$ and promote $R^\mu$ to a 6-vector \cite{Brawn2011,Pilaftsis2011},
\begin{equation}\label{eq:RA}
    R^A = \left(\begin{matrix}
    \Phi_1^\dagger \Phi_1 + \Phi_2^\dagger \Phi_2 \\
    \Phi_1^\dagger \Phi_2 + \Phi_2^\dagger \Phi_1 \\
    -i\left[\Phi_1^\dagger \Phi_2 - \Phi_2^\dagger \Phi_1\right] \\
    \Phi_1^\dagger \Phi_1 - \Phi_2^\dagger \Phi_2 \\
    \Phi_1^T i \sigma^2 \Phi_2 - \Phi_2^\dagger i\sigma^2 \Phi_1^* \\
    -i\left[\Phi_1^T i \sigma^2 \Phi_2 + \Phi_2^\dagger i\sigma^2 \Phi_1^*\right]
    \end{matrix}\right) = 
    \frac{1}{2} \left(\begin{matrix}
    v_1^2 + v_2^2 + v_+^2 \\
    2 v_1 v_2 \cos\xi \\
    2 v_2 v_2 \sin\xi \\
    v_1^2 - v_2^2 - v_+^2 \\
    -2 v_1 v_+ \cos{2\theta}\\
    -2 v_1 v_+ \sin{2\theta}
    \end{matrix}\right),
\end{equation}
allowing us to exchange the group parameter $\theta$ in favour for $R^{4}$ and $R^5$,
\begin{equation}
    \tan(2\theta) = \frac{R^5}{R^4}.
\end{equation}
Hence, we can express the remaining electroweak parameters as
\begin{equation}
    \gamma_1 = \arctan\left(\sqrt{\frac{\phi_1^2 + \phi_2^2}{\phi_3^3 + \phi_4^2}}\right), \quad \gamma_2 = \theta - \arctan\left(\frac{\phi_4}{\phi_3}\right), \quad \gamma_3 = \arctan\left(\frac{\phi_2}{\phi_1}\right) - \theta.
\end{equation}
These parametrizations are summarized in Table~\ref{tab:DoubletParams}.
\begin{table}
    \centering
    \begin{tabular}{|c||c|}
        \hline
        Linear Representation & $\phi_1$, $\phi_2$, $\phi_3$, $\phi_4$, $\phi_5$, $\phi_6$, $\phi_7$, $\phi_8$ \\
        \hline
        Vacuum Manifold Parameters & $v_1$, $v_2$,  $\xi$, $v_+$, $\gamma_1$, $\gamma_2$,  $\gamma_3$, $\theta$ \\
        \hline
        $R$-space & $R^0$, $R^1$, $R^2$, $R^3$,  $\gamma_1$, $\gamma_2$, $\gamma_3$, $\theta$ \\
        \hline
        Alternative $R$-space & $R^0$, $R^1$, $R^2$, $R^3$, $R^4$, $\gamma_1$, $\gamma_2$,  $\gamma_3$ \\
        \hline
    \end{tabular}
    \caption{Alternative parametrizations of the vacuum manifold in the 2HDM both in the original scalar field space and  bilinear $R$-space. For definitions and relations among parameters, see text.}
    \label{tab:DoubletParams}
\end{table}

% Simulation Techniques
\section{Simulation Techniques}
\label{sec:sims}

In what follows we will want to perform two distinct tasks. The first is to calculate the minimum energy solutions to a set of equations (either the full 8 fields, or a restricted ansatz). This is done using the Gradient Flow technique. The other is to evolve the full equations of motion including the 2nd order time derivatives. 

\subsection{Gradient Flow}

Minimum energy field configurations for the full 8 field case, ie. $\phi_1,..,\phi_8$, can be obtained via variational techniques as the solution to the set of ordinary differential equations,
\begin{equation}
    \frac{d}{d x}\left(\frac{\partial\mathcal{E}}{\partial\left(d\phi_n/dx\right)}\right) - \frac{\partial\mathcal{E}}{\partial \phi_n} = 0,
\end{equation}
written here for one spatial dimension, appropriate for search for kink-like solutions.
This can be done by introducing an artificial \textit{simulation time}, $t$, and solve the \textit{gradient flow} equations (see, for example, \cite{Brawn2011}),
\begin{equation}\label{eq:GradientFlow}
\frac{\partial{\phi}_n}{\partial t} = -\frac{\delta E}{\delta \phi_n} = \frac{\partial}{\partial x}\left(\frac{\partial\mathcal{E}}{\partial\left(\partial\phi_n/\partial x\right)}\right) - \frac{\partial\mathcal{E}}{\partial \phi_n}.
\end{equation}
The solution to these equations gives the minimum energy solution in the long-time limit subject to appropriate boundary conditions where
\begin{equation}
    \lim_{t\rightarrow \infty}\frac{\partial \phi_n}{\partial t} = 0,
\end{equation}
which represent 8 coupled 1st order differential equations that can be discretized using standard techniques.
We will refer to solutions generated in this way as \textit{relaxed solutions}.
The gradient flow equations are evolved on a discrete grid where both spatial and temporal derivatives are approximated to second-order.

We will also do this for various ansatzes for the fields, $\phi_n$. The approach is exactly the same but typically this will reduce the dimensionality of the problem, and hence the number of coupled equations. 

\subsection{Full Dynamics in Two and Three Dimensions}

Although minimum energy kink solutions obtained via gradient flow are useful for investigating properties of domain walls such as their width and energy, more complex simulations of domain wall dynamics are required to determine features such as scaling behaviour and related domain wall interaction properties.
The equations of motion for the global scalar field theory are discretized and solved on a regular grid in (2+1) and (3+1) dimensions.
Simulations begin from normally distributed random initial conditions in each field of the linear representation, $\phi_1,..,\phi_8$.
These initial conditions are chosen to broadly represent the outcome of a 2nd order phase transition by randomly assigning the field configuration to one of the degenerate vacua at each grid point.
These initial conditions are not intended to model the detailed dynamics of realistic phase transition, rather, they are a minimal practical choice.
They initially produce large energy gradients between grid points. We smooth these discontinuities over adjacent grid points by introducing an artificial damping term ($\dot\phi)$ with coefficient, $\varepsilon$ \cite{Pearson2009,Pearson2010,Moss2006}.
This damping term is turned off after an appropriate, small number of time steps to restore the physical equations of motion.
In results presented later, we choose $ \varepsilon = 0.5 $ for the first 200 time steps before the damping effect is removed unless stated otherwise.
Spatial derivatives are approximated to fourth order and temporal derivatives to second order.
Our simulations use periodic boundary conditions.
Therefore, we must make sure to only consider results up to the so-called \textit{light-crossing time}, $ \tau = \frac{1}{2}P\Delta x $, where $ P $ is the number of spatial grid points and $ \Delta x $ is the grid spacing.
Beyond the light-crossing time the domain walls could have crossed the periodic boundaries and interacted with themselves making any further dynamics unphysical.

We also investigate the evolution of domain wall networks in an expanding universe.
However, the thickness of a domain wall is constant for a given set of model parameters, whereas the grid spacing will increase with time.
Therefore, in an expanding universe their comoving thickness will decrease in inverse proportion to the scale factor, $ a $, \cite{Sousa2010}.
This results in resolution issues as the lattice expands, with the width of the kink becoming smaller than the grid spacing. In order to ameliorate this effect, one modifies the equations of motion to give a constant comoving resolution,
\begin{equation}
	\ddot{\phi}_i + \alpha\mathcal{H}\dot{\phi}_i = \nabla^2\phi_i - a^\beta\frac{\partial V}{\partial\phi_i},
\end{equation}
where $ \alpha $ and $ \beta $ are constants\footnote{Note that a dot, here, signifies differentiation with respect to conformal time.}.
This is the so-called \textit{PRS algorithm} \cite{Press1989}.
Setting $ \beta = 0 $ produces a fixed comoving thickness for the domain walls and $ \alpha + \beta/2 = N $ yields unchanged dynamics for domain walls in $ (N + 1) $ dimensions \cite{Sousa2010,Press1989}.
In the results presented later, simulations start from an initial conformal time $ \eta_0 = 1 $ and use a power law, $ a = \eta^\gamma $, for the scale factor.

% Z2 Symmetry
\section{\boldmath $Z_2$ Symmetry}
\label{sec:Z2}

A $Z_2$ transformation of the Higgs doublets is given by
\begin{equation}
    \Phi_1 \rightarrow \Phi_1,\quad \Phi_2 \rightarrow -\Phi_2.
\end{equation}
The $Z_2$-symmetric 2HDM potential is obtained by applying the parameter restrictions of Table~\ref{tab:2HDMConstraints} on the general 2HDM potential, \eqref{eq:2HDMPotential}, and can be written as
\begin{align}
   \label{eq:Z2Potential}
    V &= -\mu_1^2 (\Phi_1^\dagger \Phi_1) - \mu_2^2 (\Phi_2^\dagger\Phi_2) + \lambda_1 (\Phi_1^\dagger\Phi_1)^2 + \lambda_2 (\Phi_2^\dagger \Phi_2)^2 + \lambda_3 (\Phi_1^\dagger \Phi_1)(\Phi_2^\dagger \Phi_2)\nonumber\\
    &\quad + (\lambda_4 - \left|\lambda_5\right|) \left[\text{Re}(\Phi_1^\dagger\Phi_2)\right]^2+ (\lambda_4 + \left|\lambda_5\right|) \left[\text{Im}(\Phi_1^\dagger\Phi_2)\right]^2.
\end{align}
For the CP-preserving scenario given in~\eqref{eq:NormalVac}, its VEVs can be calculated in terms of the potential parameters \cite{Brawn2011},
\begin{equation}\label{eq:Z2VEVs}
    v_1^2 = \frac{4\lambda_2\mu_1^2 - 2\tilde{\lambda}_{345}\mu_2^2}{4\lambda_1\lambda_2 - \tilde{\lambda}_{345}^2},\qquad
    v_2^2 = \frac{4\lambda_1\mu_2^2 - 2\tilde{\lambda}_{345}\mu_1^2}{4\lambda_1\lambda_2 - \tilde{\lambda}_{345}^2}
\end{equation}
where $\tilde{\lambda}_{345} = \lambda_3 + \lambda_4 - \left|\lambda_5\right|$.

In order to have a stable vacuum there must be an all-positive Higgs spectrum, i.e. we must have real, non-negative eigenvalues for all 5 physical Higgs states.
It is most convenient to compute the mass matrices using the representation,
\begin{equation}\label{eq:PWrep}
    \Phi_1 = \left(\begin{matrix}
    \varphi_1^+ \\
    \frac{1}{\sqrt{2}}v_1 + \varphi_1 + i a_1
    \end{matrix}\right),\quad
    \Phi_2 = e^{i\xi}\left(\begin{matrix}
    \varphi_2^+ \\
    \frac{1}{\sqrt{2}}v_2 + \varphi_2 + i a_2
    \end{matrix}\right),
\end{equation}
where $\varphi_i^+$ are complex scalar fields and the CP-even/odd fields can be expressed as
\begin{equation}\label{eq:massbasis}
    \left(\begin{matrix}
    \varphi_1 \\
    \varphi_2
    \end{matrix}\right) = \left(\begin{matrix}
    c_\alpha & -s_\alpha \\
    s_\alpha & c_\alpha
    \end{matrix}\right)\left(\begin{matrix}
    h \\
    H
    \end{matrix}\right), \qquad\left(\begin{matrix}
    a_1 \\
    a_2
    \end{matrix}\right) = \left(\begin{matrix}
    c_\beta & -s_\beta \\
    s_\beta & c_\beta
    \end{matrix}\right)\left(\begin{matrix}
    G^0 \\
    A
    \end{matrix}\right),
\end{equation}
respectively, where $c_\alpha=\cos\alpha$ and $s_\alpha=\sin\alpha$, and similarly for $c_\beta$ and $s_\beta$.
It should be obvious that the representation \eqref{eq:PWrep} reduces to \eqref{eq:CPVac} in the vacuum.
Since the $Z_2$-symmetric 2HDM is CP conserving, there is no mixing between the CP-even and CP-odd states and $\xi=0$.
Therefore, the mass matrix for the neutral sector becomes block diagonal.
The mass matrices in the $Z_2$-symmetric 2HDM are found to be
\begin{align}
   \label{eq:CPevenmassmatrix}
		\mathcal{M}_{h,H}^2 &= \frac{1}{2}\left[\frac{\partial^2 V}{\partial \varphi_i \partial\varphi_j}\right] = \left(\begin{matrix}
		2\lambda_1 v_1^2 & \tilde{\lambda}_{345} v_1 v_2 \\
		\tilde{\lambda}_{345} v_1 v_2 & 2\lambda_2 v_2^2
		\end{matrix}\right),\\
   \label{eq:Z2PseudoHessian}
		\mathcal{M}_A^2 &= \frac{1}{2}\left[\frac{\partial^2 V}{\partial a_i \partial a_j}\right] = \left|\lambda_5\right|\left(\begin{matrix}
		v_2^2 & -v_1 v_2 \\
		- v_1 v_2 & v_1^2
		\end{matrix}\right),\\
   \label{eq:Z2ChargedHessian}
		\mathcal{M}_{H^\pm}^2 &= \left[\frac{\partial^2 V}{\partial \varphi_i^+ \partial \varphi_j^-}\right] = -\frac{1}{2}\left(\lambda_4 - \left|\lambda_5\right|\right)\left(\begin{matrix}
		v_2^2 & -v_1 v_2 \\
		-v_1 v_2 & v_1^2
		\end{matrix}\right).
\end{align}
From the above matrix relations, it is not difficult
to calculate the physical squared masses for all scalar fields, i.e.
\begin{align}
  \label{eq:hmasses}
	M_{h}^2 &= \lambda_1 v_1^2 + \lambda_2 v_2^2 - \sqrt{\left(\lambda_1 v_1^2 - \lambda_2 v_2^2\right)^2 + \tilde{\lambda}_{345}^2 v_1^2 v_2^2},\\
  \label{eq:Hmasses}
	M_{H}^2 &= \lambda_1 v_1^2 + \lambda_2 v_2^2 + \sqrt{\left(\lambda_1 v_1^2 - \lambda_2 v_2^2\right)^2 + \tilde{\lambda}_{345}^2 v_1^2 v_2^2},\\
  \label{eq:pseudomassses}
	M_A^2 &= \left|\lambda_5\right| v_{\text{SM}}^2,\\
  \label{eq:chargedmasses}
    M_{H^\pm}^2 &= -\frac{1}{2}\left(\lambda_4 - \left|\lambda_5\right|\right)v_{\text{SM}}^2,
\end{align}
along with 2 zero eigenvalues corresponding to the would-be Goldstone bosons, $G^0$ and $G^\pm$.

One can re-express the scalar potential in terms of these masses and the CP-even/odd mixing angles $\alpha/\beta$,
\begin{align}
  \label{eq:Z2physpot}
    \Hat{V} &= - \frac{1}{2}\left[c_\alpha^2 + \Hat{M}_H^2 s_\alpha^2 + (1 - \Hat{M}_H^2) \tan\beta c_\alpha s_\alpha \right] (\hat{\Phi}_1^\dagger \hat{\Phi}_1)\nonumber\\
    &\quad - \frac{1}{2}\left[s_\alpha^2 + \Hat{M}_H^2 c_\alpha^2 + (1 - \Hat{M}_H^2) \cot\beta c_\alpha s_\alpha \right] (\hat{\Phi}_2^\dagger \hat{\Phi}_2) + \frac{c_\alpha^2 + \Hat{M}_H^2 s_\alpha^2}{2 c_\beta^2} (\hat{\Phi}_1^\dagger \hat{\Phi}_1)^2 \nonumber\\
    &\quad + \frac{s_\alpha^2 + \Hat{M}_H^2 c_\alpha^2}{2 s_\beta^2} (\hat{\Phi}_2^\dagger \hat{\Phi}_2)^2 + \frac{(1 - \Hat{M}_H^2) c_\alpha s_\alpha + 2 \Hat{M}_{H^\pm}^2 c_\beta s_\beta}{c_\beta s_\beta} (\hat{\Phi}_1^\dagger \hat{\Phi}_1) (\hat{\Phi}_2^\dagger \hat{\Phi}_2)\nonumber \\
   &\quad + (\Hat{M}_A^2 - 2\Hat{M}_{H^\pm}^2) (\hat{\Phi}_1^\dagger \hat{\Phi}_2) (\hat{\Phi}_2^\dagger \hat{\Phi}_1) - \frac{1}{2} \Hat{M}_A^2 \left[(\Hat{\Phi}_1^\dagger \Hat{\Phi}_2)^2 + \text{H.c.}\right],
\end{align}
where we have also rescaled for dimensionless energy per unit area\footnote{A circumflex denotes a dimensionless quantity.}.
This new parametrization contains the 5 dimensionless parameters, $\hat{M}_H, \hat{M}_A, \hat{M}_{H^\pm}, \alpha$ and $\beta$, with the measured values $M_h = 125\,{\rm GeV}$ and $v_{\text{SM}} = 246\,{\rm GeV}$ fixed. Typically, we will fix $\hat{M}_H=\hat{M}_A=\hat{M}_{H^\pm}$, so that there are just three parameters, $\hat{M}_H$, $\alpha$ and $\beta$ to vary. The details of the reparameterization and rescaling procedures are given in Appendix~\ref{sec:Rescaling}.

\subsection{Kinks Solutions} \label{sec:Z2kinks}

In \cite{Brawn2011} kink solutions were obtained as a function of the vacuum manifold parameters with  \eqref{eq:NormalVac} and \eqref{eq:CPVac} assumed as an ansatz reducing the number of gradient flow equations to two. Despite this ansatz being well-motivated, allowing one to obtain minimum energy solutions, it only allows exploration of a limited region of the field configuration space. Hence, the solutions available are restricted from the outset. In contrast, we make no assumption of the vacuum manifold parameterization in this work and obtain relaxed solutions from the general field configuration and the general vacuum ansatz \eqref{eq:ChargedVac}.

The general energy density of the $Z_2$-symmetric 2HDM in one dimension is given by
\begin{equation}
    \mathcal{E} = \frac{d\Phi_1^\dagger}{dx}\frac{d\Phi_1}{dx} + \frac{d\Phi_2^\dagger}{dx}\frac{d\Phi_2}{dx} + V(\Phi_1,\Phi_2).
\end{equation}
In the linear representation, the gradient flow equations which minimize the energy per unit area are given by
\begin{equation}\label{eq:Z2GenFlow}
    \frac{\partial \phi_i}{\partial t} = \frac{\partial^2 \phi_i}{\partial x^2} - \frac{\partial V}{\partial\phi_i}.
\end{equation}
The energy density in terms of the vacuum manifold parameters is,
\begin{equation}
    \begin{split}
        \mathcal{E} = & \frac{1}{2}\left(\frac{d v_1}{dx}\right)^2 + \frac{1}{2}\left(\frac{d v_2}{dx}\right)^2 + \frac{1}{2}v_2^2\left(\frac{d \xi}{dx}\right)^2 + \frac{1}{2}\left(\frac{d v_+}{dx}\right)^2 - \frac{1}{2}\mu_1^2 v_1^2 - \frac{1}{2}\mu_2^2\left(v_2^2 + v_+^2\right) \\
		& + \frac{1}{4}\lambda_1 v_1^4 + \frac{1}{4}\lambda_2\left(v_2^2 + v_+^2\right)^2 + \frac{1}{4}\lambda_3 v_1^2 v_+^2 + \frac{1}{4}\left(\lambda_{34} - \left|\lambda_5\right|c_{2\xi}\right) v_1^2 v_2^2.
    \end{split}
\end{equation}
The gradient flow equations for the vacuum manifold parameters are then given by
\begin{align}
  \label{eq:Z2VacFlow}
%	\begin{equation}
	\frac{\partial v_1}{\partial t} &= \frac{\partial^2 v_1}{\partial x^2} + \mu_1^2 v_1 - \lambda_1 v_1^3 - \frac{1}{2}\lambda_3 v_1 v_+^2 - \frac{1}{2}\left(\lambda_{34} - \left|\lambda_5\right|c_{2\xi}\right)v_1 v_2^2,\nonumber\\
%	\end{equation}
%	\begin{equation}
	\frac{\partial v_2}{\partial t} &= \frac{\partial^2 v_2}{\partial x^2} - v_2\left(\frac{\partial\xi}{\partial x}\right)^2 + \mu_2^2 v_2 - \lambda_2 v_2\left(v_2^2 + v_+^2\right) - \frac{1}{2}\left(\lambda_{34} - \left|\lambda_5\right|c_{2\xi}\right)v_1^2 v_2,\nonumber\\
%	\end{equation}
%	\begin{equation}
	\frac{\partial \xi}{\partial t} &= v_2^2\frac{\partial^2\xi}{\partial x^2} + 2v_2\left(\frac{\partial v_2}{\partial x}\right)\left(\frac{\partial \xi}{\partial x}\right) - \frac{1}{2}\left|\lambda_5\right|v_1^2 v_2^2 s_{2\xi},\\
%	\end{equation}
%	\begin{equation}
	\frac{\partial v_+}{\partial t} &= \frac{\partial^2 v_+}{\partial x^2} + \mu_2^2 v_+ - \lambda_2 v_+\left(v_2^2 + v_+^2\right) - \frac{1}{2}\lambda_3 v_1^2 v_+.\nonumber
%	\end{equation}
\end{align}
Note that the relations between the 2HDM potential parameters and physical parameters can be found in Appendix~\ref{sec:Rescaling}.

\begin{figure}
    \centering
    \includegraphics[width=\textwidth]{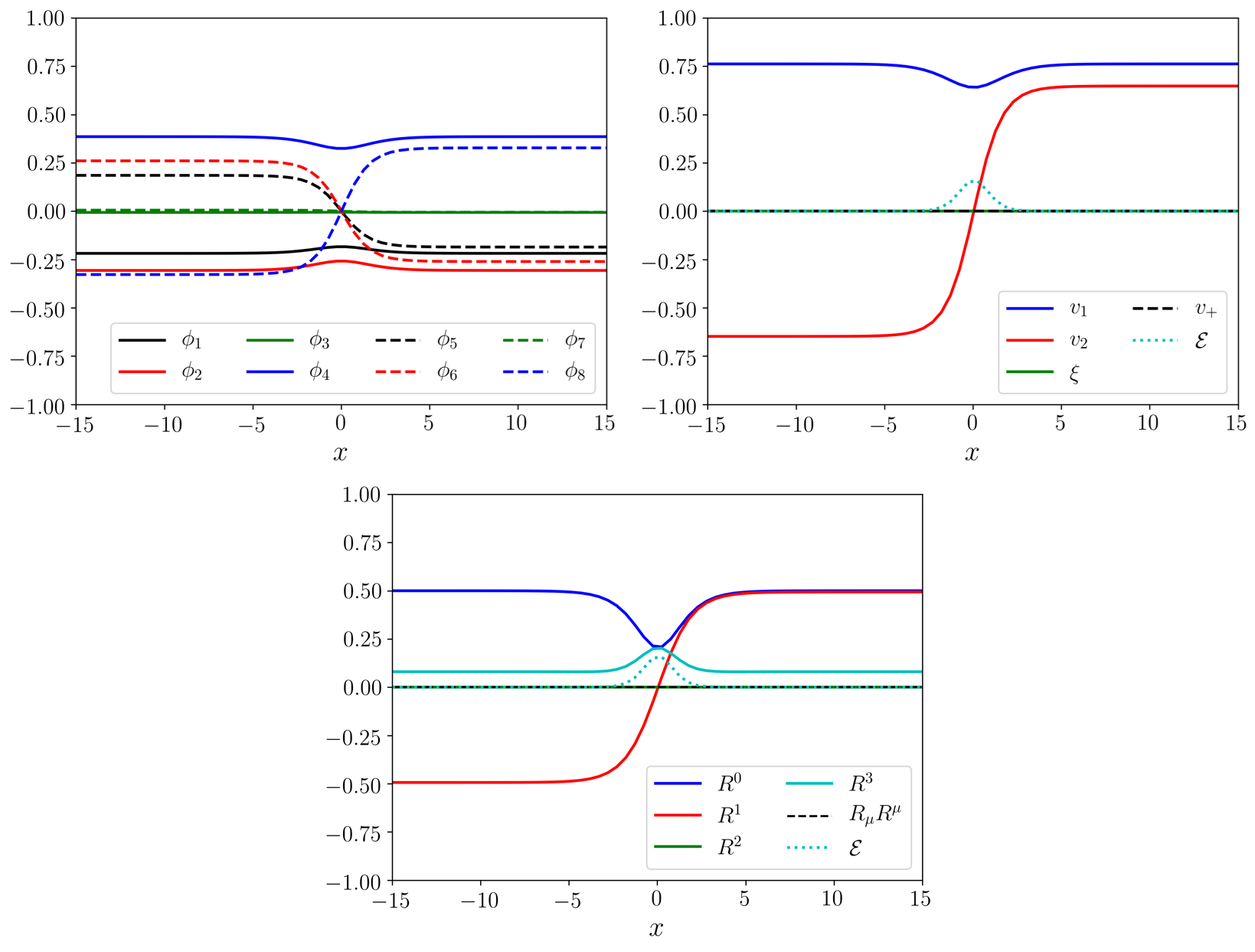}
    \caption{Kink solution of the $Z_2$-symmetric 2HDM in the linear representation (top, left) and vacuum manifold parameterization (top, right) along with corresponding $R$-space profiles (bottom). Parameters chosen were $M_H = M_A = M_{H^\pm} = 200$ GeV, $\tan\beta = 0.85$ and $\cos(\alpha - \beta) = 1.0$. Total energy of the solution is 0.697.}%437279593.}
    \label{fig:Z2FieldsFull2}
\end{figure}

The relaxed solutions to \eqref{eq:Z2GenFlow} and \eqref{eq:Z2VacFlow} are presented in Fig.~\ref{fig:Z2FieldsFull2}.
It can be seen that the solutions to these two systems of equations are equivalent, having the same energy and identical profiles for the components of $R^\mu$, for specific model parameters, and this is the case for all parameter choices which we have tried.
Moreover, the relaxed solution in the full field configuration of Fig.~\ref{fig:Z2FieldsFull2} can be brought to the same form as the solution for the vacuum manifold parameters via a global transformation of the fields multiplying by $U\in \text{SU}(2)_L\times \text{U}(1)_Y$ - a global electroweak transformation. In other words, there are a myriad of equivalent minimum energy solutions in the general parameterization which are related via global electroweak transformations. One of these is the solution found in \cite{Brawn2011} which used the restricted ansatz with $\xi\equiv 0$ and $v_+\equiv 0$, and indeed the minimum energy solutions of \eqref{eq:Z2VacFlow} have this property.

In both cases a kink forms in $R^1$ and both have the same vacuum values for the components of $R^\mu$ given by \eqref{eq:RVectorVac}.
The quantities $R^1$ and $R^2$ are odd under a $Z_2$ transformation whilst $R^0$ and $R^3$ are even.
We may then define a gauge-invariant topological current,
\begin{equation}
  \label{eq:Z2Jmu}
    J^\mu = \frac{1}{2\langle R^1 \rangle}\varepsilon^{\mu\nu}\partial_\nu R^1,
\end{equation}
and corresponding topological charge,
\begin{equation}
   \label{eq:Z2Q}
    Q = \int_{-\infty}^\infty dx J^0 = \frac{1}{2\langle R^1 \rangle}\left[R^1(\infty) - R^1(-\infty)\right].
\end{equation}
Note that $Q=1$ corresponds to a kink and $Q=-1$ to an anti-kink.

\subsection{Dynamical Simulations from Random Initial Conditions}

\begin{figure}
    \centering
    \hspace*{-0.71cm}
    \includegraphics[width=0.96\textwidth]{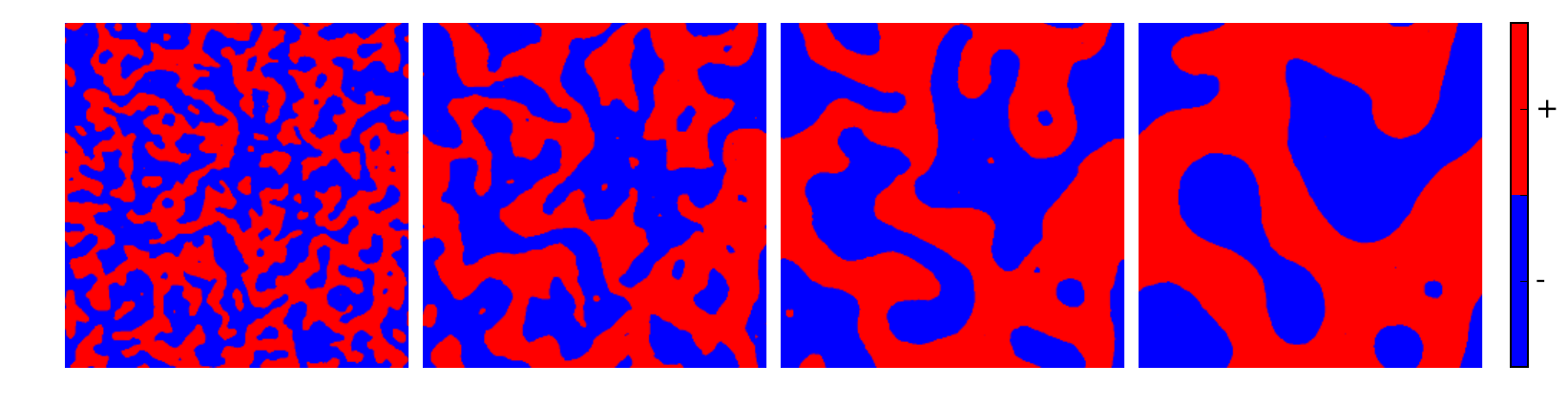} \\
    \includegraphics[width=\textwidth]{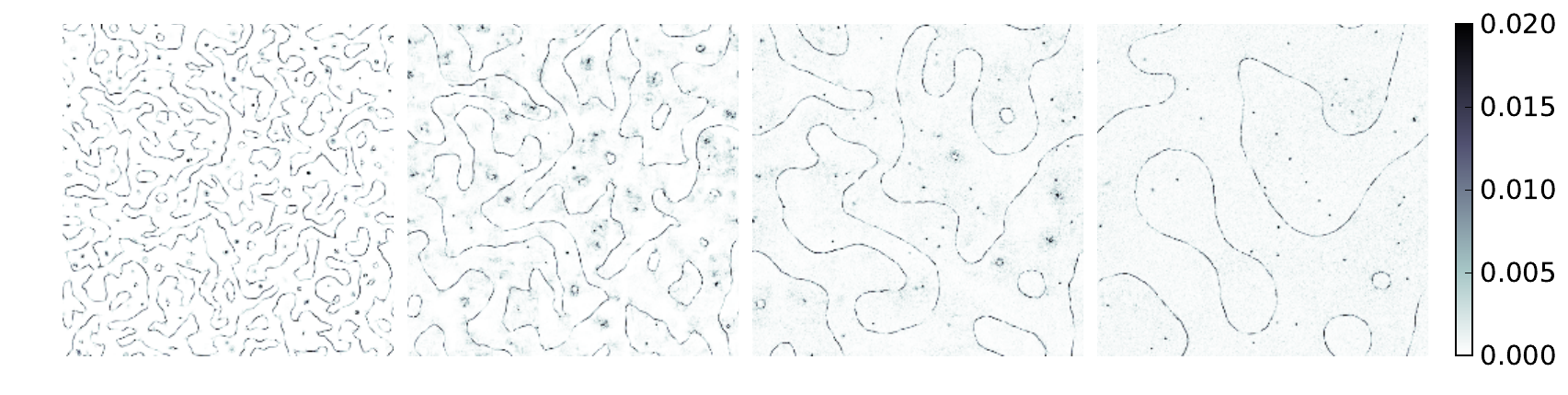}
    \caption{2D simulation of the evolution of domain walls in the $Z_2$-symmetric 2HDM with Minkowski metric. Parameters are $M_H = M_A = M_{H^\pm} = 200\;\text{GeV}$, $\tan\beta = 0.85$ and $\cos(\alpha - \beta) = 1.0$.
    Simulation was run for time, $t=480$ with temporal grid spacing, $\Delta t = 0.2$ and spatial grid size, $P = 1024$ with spacing, $\Delta x = 0.9$.
    Plots progress in time left-to-right and each plot is at double the timestep of the previous. In a simulation where the number of walls is reducing $\propto t^{-1}$ one would expect to see the number of walls visually reducing by a factor of two in subsequent plots from left to right. The upper figures present the sign of $R^1$ and the lower ones are $R^\mu R_\mu$ clearly showing that it is $\ne 0$ on the walls where the sign of $R^1$ changes.}
    \label{fig:Z2domains}
\end{figure}

\begin{figure}
    \centering
    \includegraphics[width=\textwidth]{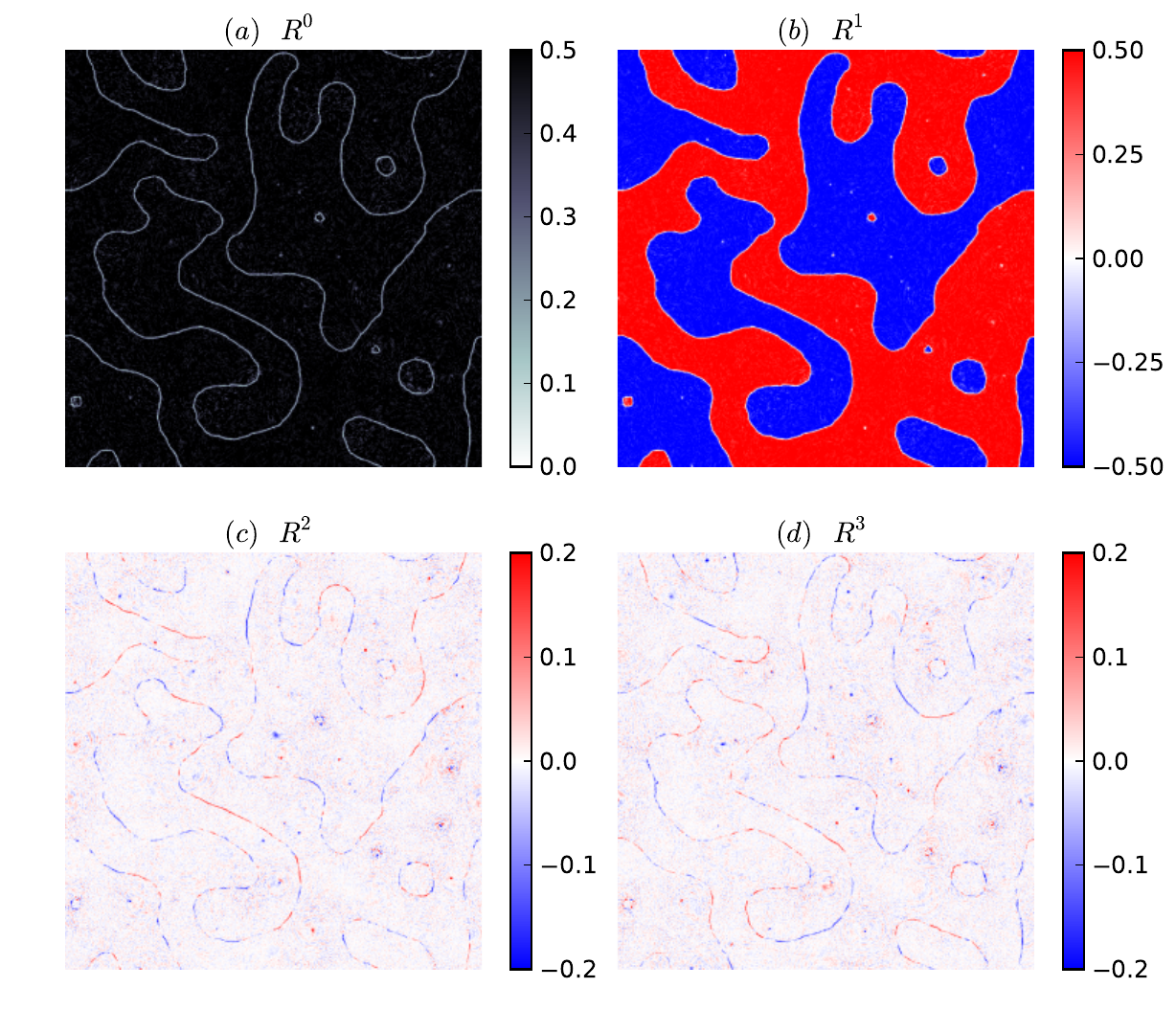}
    \caption{Spatial distributions of the components of $R^{\mu}$ in 2D simulations of the $Z_2$-symmetric 2HDM at $t=240$ for $M_H = M_A = M_{H^\pm} = 200\;\text{GeV}$, $\tan\beta = 0.85$ and $\cos(\alpha - \beta) = 1.0$. This timestep corresponds to the 3rd panel of Fig.~\ref{fig:Z2domains} and indeed the top-right panel is identical to the 3rd panel in Fig.~\ref{fig:Z2domains}.}
    \label{fig:Rspace}
\end{figure}

We have performed (2+1) dimensional simulations with $P=1024$ and $P=4096$ up to $t=\tau$ for the global field theory of the $Z_2$-symmetric 2HDM with Minkowksi metric, and the FRW metric in the radiation and matter dominated eras using the PRS algorithm.
The evolution a Minkowksi simulation is presented in Fig.~\ref{fig:Z2domains} along with the corresponding spatial variation of $R_\mu R^\mu$, which would be zero in the case of a neutral vacuum solution, $v_+\equiv 0$.
In these simulations domain walls do not form in the individual fields of the linear representation.
Rather, these fields evolve such that walls/condensates are manifest in the components of $R^\mu$.
Snapshots of the spatial variation of each of the components of $R^\mu$ during the same (2+1) dimensional simulation are presented in Fig.~\ref{fig:Rspace}.
Domains are mapped using the sign of $R^1$ as one would expect given the kink profiles of Fig.~\ref{fig:Z2FieldsFull2}.
The domain walls are manifest in the components of $R^\mu$ due to their invariance under a local unitary transformation of the doublets.
%One can think of this as requiring a local electroweak transformation to return the field configuration to that of Fig.~\ref{fig:Z2FieldsFull2} in the linear representation.

We find that there is a local violation of the neutral vacuum condition, $R^\mu R_\mu = 0$, on the domain walls in our simulations as shown in Fig.~\ref{fig:Z2domains};
a feature of the field dynamics not found in the minimum energy solutions, indicating that there is something more complicated going on than just evolution of a network of minimum energy configurations. We will return to this observation later in our discussion.

\begin{figure}
    \centering
    \hspace*{-0.71cm}
    \includegraphics[width=0.96\textwidth]{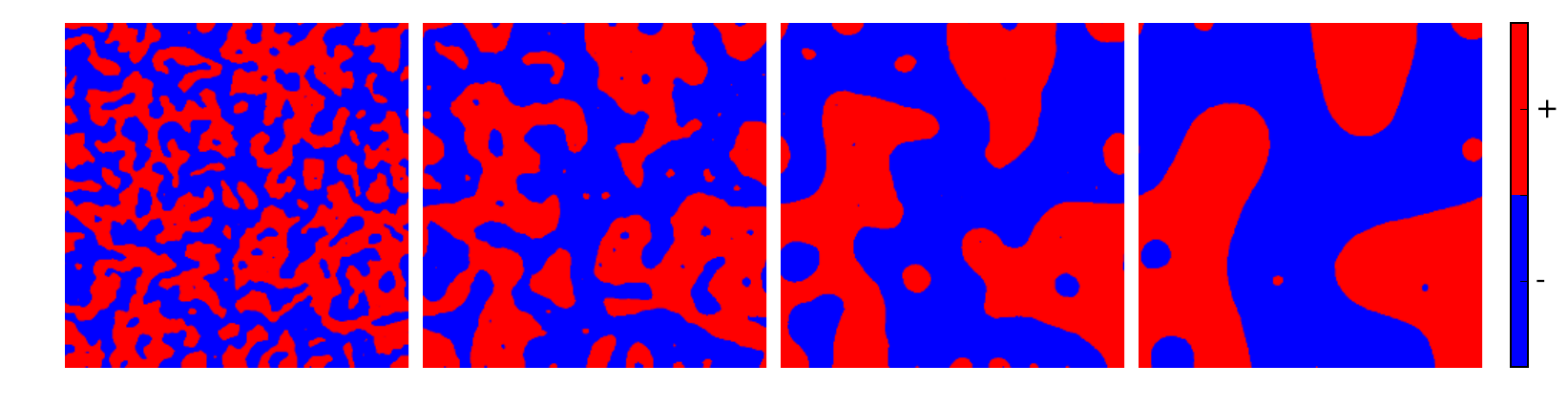} \\
    \includegraphics[width=\textwidth]{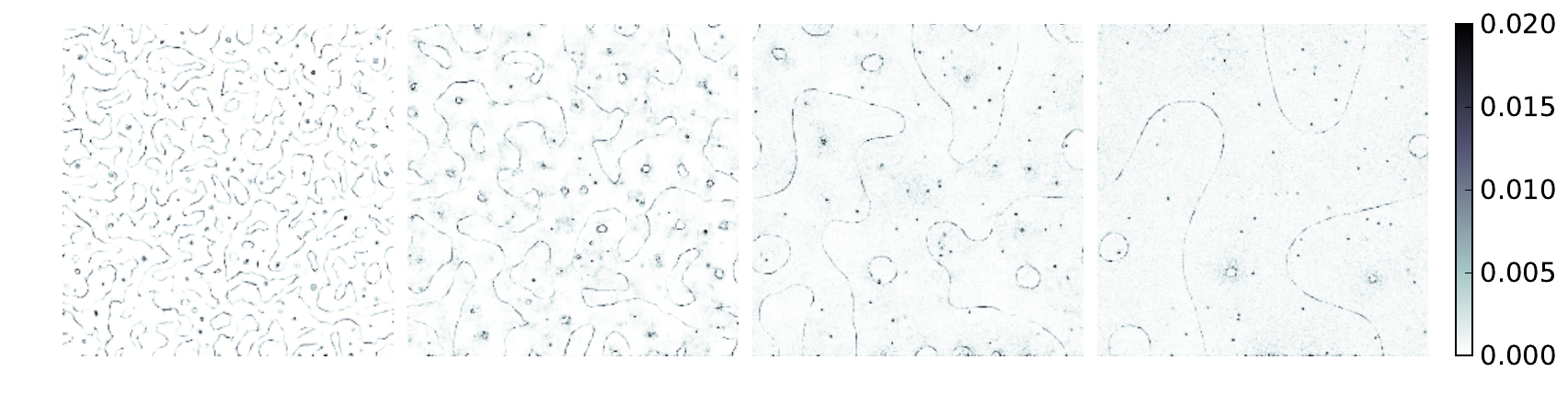}
    \caption{2D simulation of the evolution of domain walls in the $Z_2$-symmetric 2HDM with FRW metric in radiation dominated era. Parameters are $M_H = M_A = M_{H^\pm} = 200\;\text{GeV}$, $\tan\beta = 0.85$ and $\cos(\alpha - \beta) = 1.0$.
    Simulation was run for time, $t=480$ with temporal grid spacing, $\Delta t = 0.2$ and spatial grid size, $P = 1024$ with spacing, $\Delta x = 0.9$.
    Plots progress in time left-to-right and each plot is at double the timestep of the previous. The figures are presented using the same conventions as in Fig.~\ref{fig:Z2domains}.}
    \label{fig:Z2domainsPRS}
\end{figure}

Equivalent results for a domain wall network in a radiation dominated FRW universe are presented in Fig.~\ref{fig:Z2domainsPRS}. Again, we find a local violation of the neutral vacuum condition in the vicinity of the domain walls.

\begin{figure}
    \centering
    \hspace*{-0.92cm}
    \includegraphics[width=0.94\textwidth]{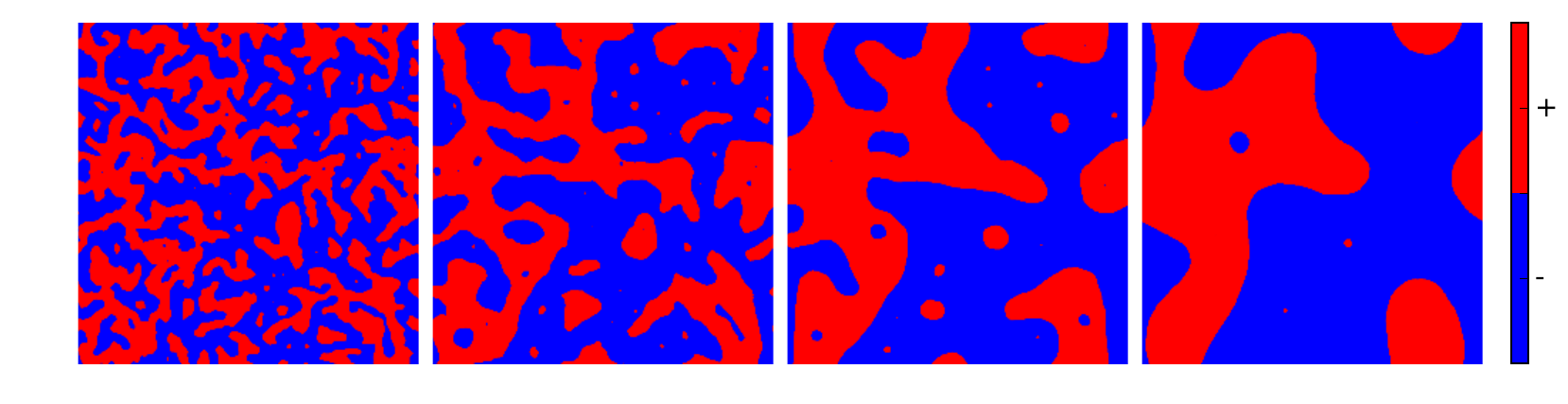} \\
    \includegraphics[width=\textwidth]{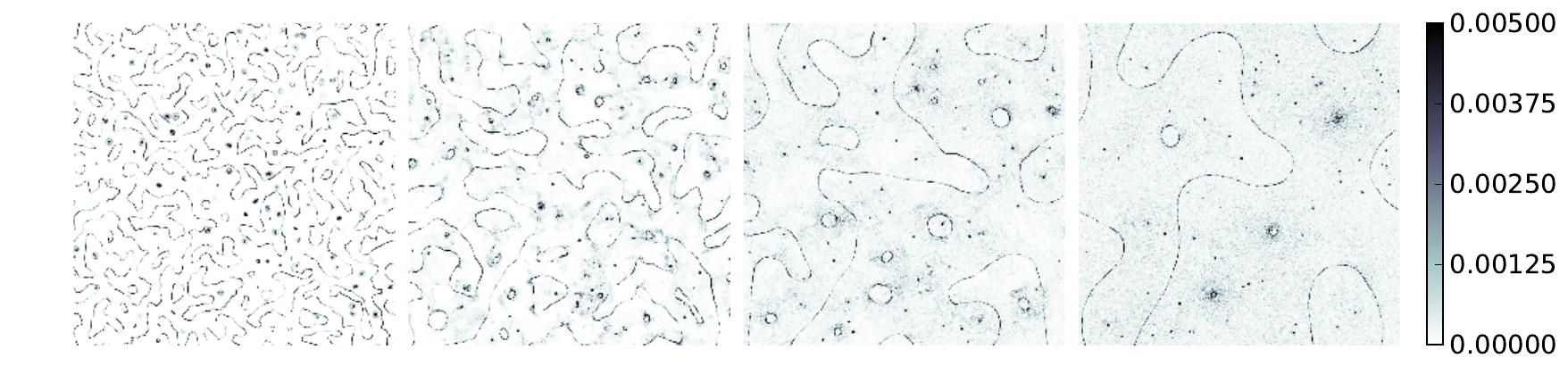}
    \caption{2D simulation of the evolution of domain walls in the $Z_2$-symmetric 2HDM with Minkowksi metric. Parameters are $M_H = M_A = M_{H^\pm} = 260\;\text{GeV}$, $\tan\beta = 0.85$ and $\cos(\alpha - \beta) = 1.0$.
    Simulation was run for time, $t=480$ with temporal grid spacing, $\Delta t = 0.2$ and spatial grid size, $P = 2048$ with spacing, $\Delta x = 0.45$. Note that the physical size of this is simulation is the same as in Figs.~\ref{fig:Z2domains} and \ref{fig:Z2domainsPRS}, but with higher resolution.
    Plots progress in time left-to-right and each plot is at double the timestep of the previous. The figures are presented using the same conventions as in Fig.~\ref{fig:Z2domains}.}
    \label{fig:Z2domains_260}
\end{figure}

The aforementioned dynamical features are manifest for a wide range of physical parameters. To illustrate this 
we consider a benchmark scenario motivated by the so-called Maximally Symmetric 2HDM~\cite{Dev:2014yca,Darvishi:2019ltl}, which has a
common heavy Higgs mass, $M_H = M_A = M_{H^\pm} = 260$ GeV. Unlike~~\cite{Dev:2014yca, Darvishi:2019ltl}, we assume here a Type-I Yukawa realization, so that the stringent constraints, mainly due to $B$-meson observables,
can be avoided~\cite{Arbey:2017gmh,Arbey:2019duh}.
For such a scenario, the domains and spatial variation of $R_\mu R^\mu$ are presented in Fig.~\ref{fig:Z2domains_260}.

\begin{figure}
    \centering
    \includegraphics[width=0.8\textwidth]{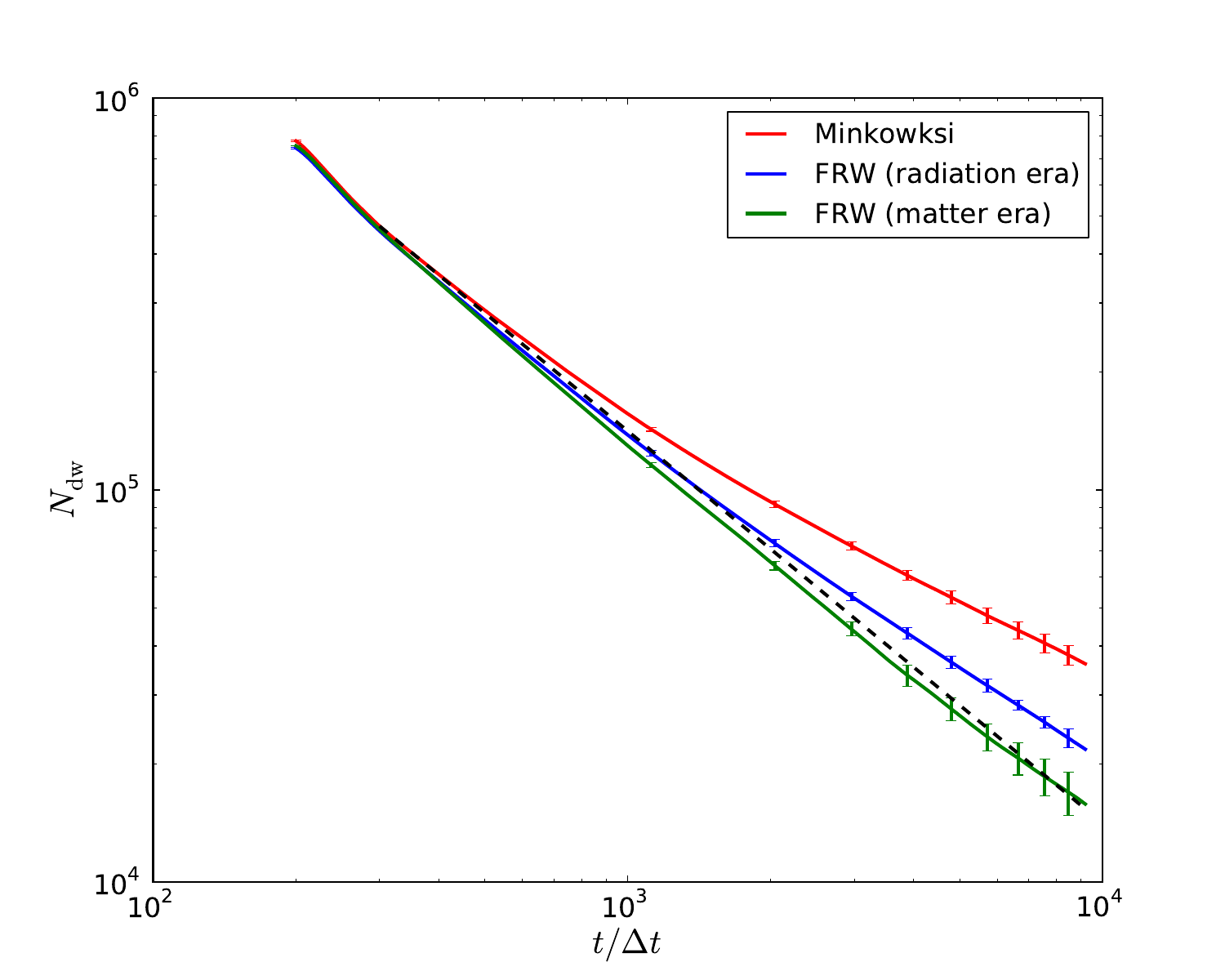}
    \caption{Evolution of the number of domain walls in 2D $Z_2$-symmetric 2HDM simulations averaged over 10 realizations for Minkowski, and FRW in radiation and matter dominated eras.
    Also plotted is the standard power law scaling for a domain wall network $\propto t^{-1}$.
    Parameters chosen were $M_H = M_A = M_{H^\pm} = 200\;\text{GeV}$, $\tan\beta = 0.85$ and $\cos(\alpha - \beta) = 1.0$.
    Simulations were run for time, $t=1840$ with temporal grid spacing, $\Delta t = 0.2$ and spatial grid size, $P = 4096$ with spacing, $\Delta x = 0.9$.
    Error bars, which are the standard deviation amongst the realizations, illustrate the numerical scatter between simulations.}
    \label{fig:Z2walls}
\end{figure}

The number of domain walls as a function of time in (2+1) dimensions with Minkowski and FRW metrics in radiation and matter dominated eras are presented in Fig.~\ref{fig:Z2walls}.
The time evolution of the number of domain walls in the system is obtained as an average over 10 realizations.
In both Minkowski space and an FRW radiation era we find domain walls in the $Z_2$-symmetric 2HDM do not scale in the standard way, whereas in the FRW matter era it is compatible with $N_{\rm dw}\propto t^{-1}$.
Fig.~\ref{fig:Z2walls} shows a a deviation from the $t^{-1}$ scaling found in simpler models such as  the Goldstone model \cite{Garagounis2003}.
In the 2HDM we find more domain walls at late times than one would naively expect.
This can be seen visually in that the number of walls from panel to panel in all of Figs.~\ref{fig:Z2domains}, \ref{fig:Z2domainsPRS} and \ref{fig:Z2domains_260} clearly does not decrease by a factor of two as one progresses from left to right.
This non-standard scaling is akin to the effect found in the so-called \textit{kinky vorton model} \cite{Pearson2009}.
In the kinky vorton model a Noether current produces a conserved charge which condenses on the domain walls.
The winding of the condensate field around the walls slows the collapse of the domain walls modifying their scaling behaviour. Our results suggest that there may be some extra interactions amongst the fields which prevents the walls from decaying as quickly as causality permits.

\begin{figure}
    \centering
    \includegraphics[width=\textwidth]{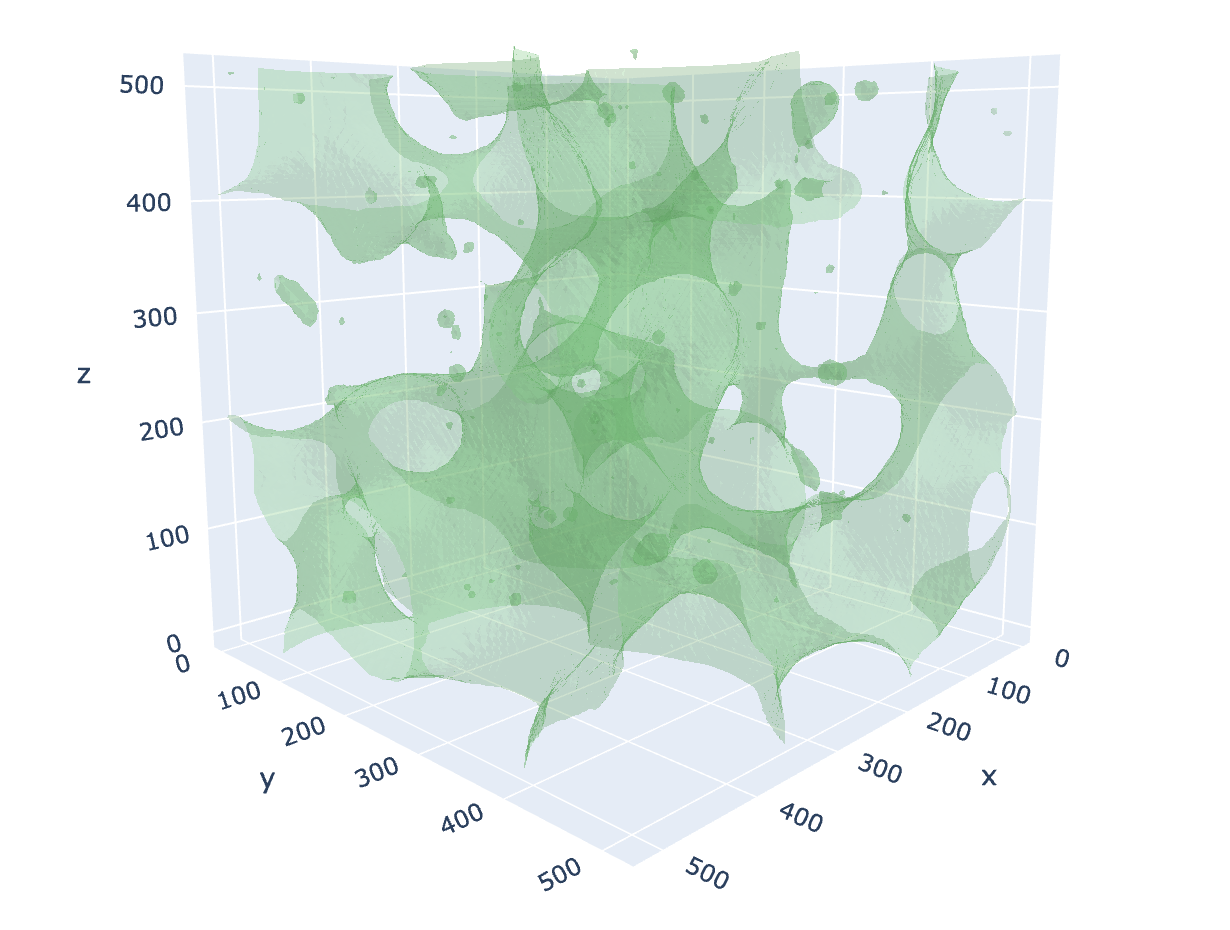}
    \caption{Isosurfaces of $R^1$ at $t=240$ showing domain walls in a 3D simulation of the $Z_2$-symmetric 2HDM. Parameters are $M_H = M_A = M_{H^\pm} = 200$ GeV, $\tan\beta=0.85$ and $\cos(\alpha-\beta)=1.0$. Simulations was run for time, $t=240$ with temporal grid spacing, $\Delta t = 0.2$ and spatial grid size, $P=512$ with spacing, $\Delta x = 0.9$.}
    \label{fig:isoZ2}
\end{figure}

\begin{figure}
    \centering
    \hspace*{-0.7cm}
    \includegraphics[width=0.96\textwidth]{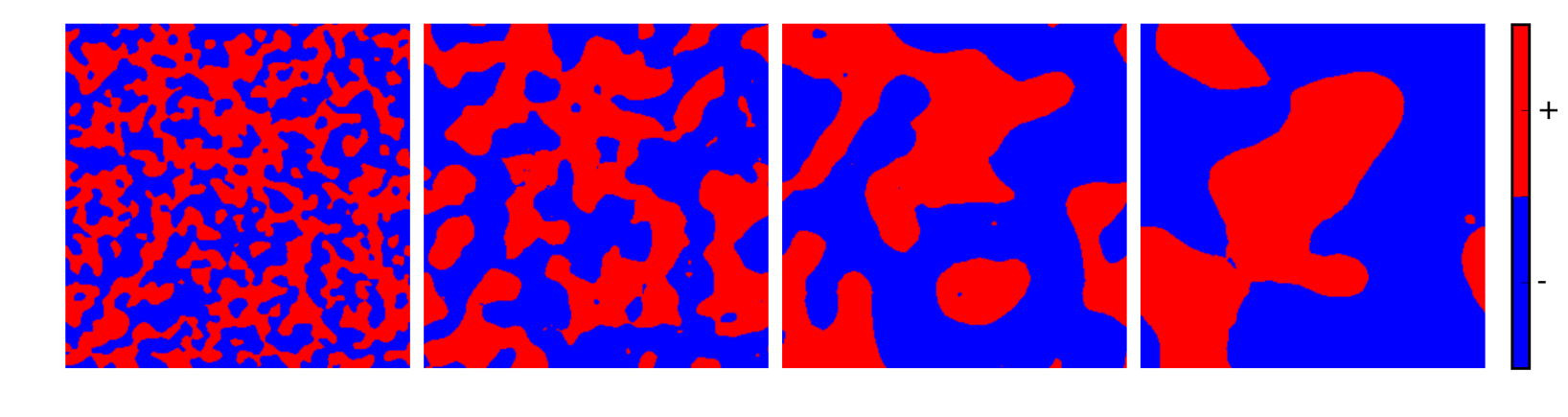} \\
    \includegraphics[width=\textwidth]{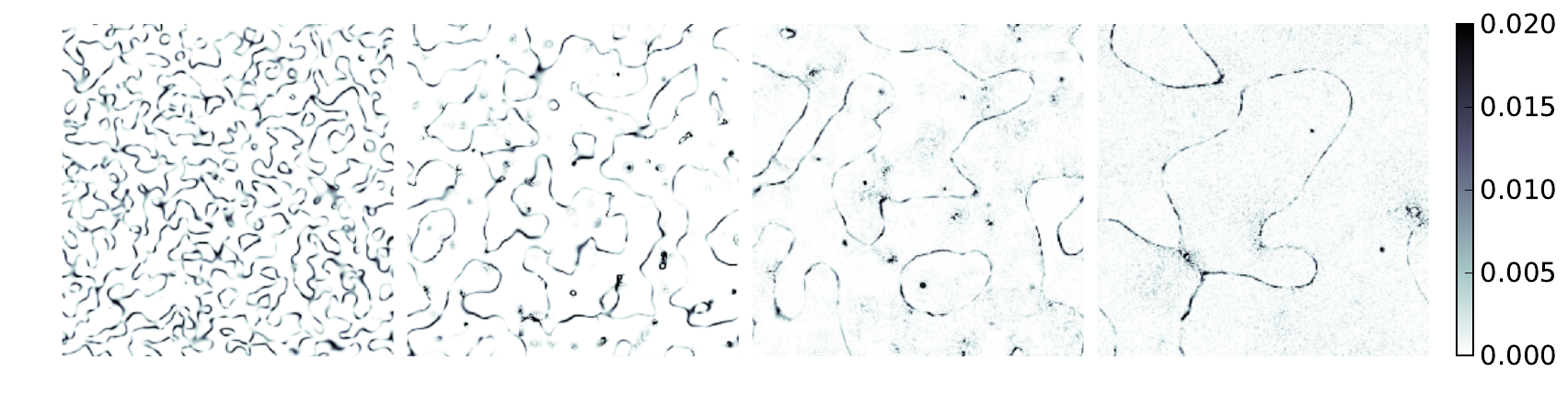}
    \caption{Slices through a 3D simulation of the evolution of domain walls in the $Z_2$-symmetric 2HDM with Minkowksi metric. Parameters are $M_H = M_A = M_{H^\pm} = 200\;\text{GeV}$, $\tan\beta = 0.85$ and $\cos(\alpha - \beta) = 1.0$.
    Simulation was run for time, $t=240$ with temporal grid spacing, $\Delta t = 0.2$ and spatial grid size, $P = 512$ with spacing, $\Delta x = 0.9$.
    Plots progress in time left-to-right and each plot is at double the timestep of the previous. The figures are presented using the same conventions as in Fig.~\ref{fig:Z2domains}.}
    \label{fig:Z2slice3D}
\end{figure}

\begin{figure}
    \centering
    \includegraphics[width=0.8\textwidth]{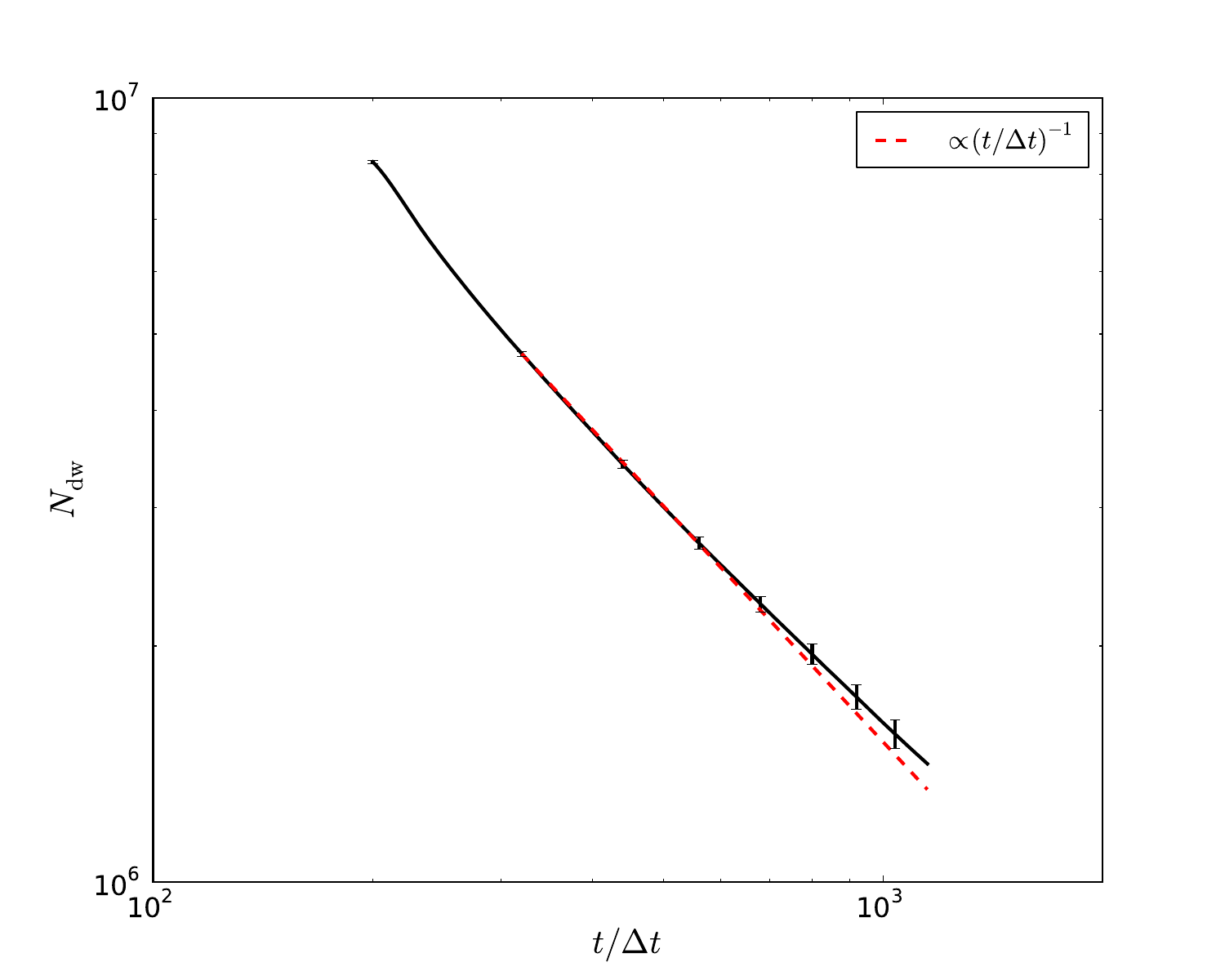}
    \caption{Evolution of the number of domain walls in 3D $Z_2$-symmetric 2HDM simulations averaged over 10 realisations with Minkowski metric.
    Also plotted is the standard power law scaling for a domain wall network.
    Parameters chosen were $M_H = M_A = M_{H^\pm} = 200\;\text{GeV}$, $\tan\beta = 0.85$ and $\cos(\alpha - \beta) = 1.0$.
    Simulations were run for time, $t=240$ with temporal grid spacing, $\Delta t = 0.2$ and spatial grid size, $P = 512$ with spacing, $\Delta x = 0.9$.
    Error bars present the numerical scatter via the standard deviation. There appears to be evidence for a deviation from the expected $\propto t^{-1}$ scaling as seen in (2+1) dimensions.}
    \label{fig:Z2scaling3D}
\end{figure}

For computational ease, the majority of our analysis has been performed in (2+1) dimensions.
Nonetheless, we have also performed simulations of domain walls in (3+1) dimensions to verify that the same dynamical features are manifest in those simulations as well.
A snapshot of a domain wall network from a (3+1) dimensional simulations of the $Z_2$-symmetric 2HDM is given in Fig.~\ref{fig:isoZ2}.
As in the case of (2+1) dimensions, we observe a local violation of the neutral vacuum condition in the vicinity of the domain walls as shown in Fig.~\ref{fig:Z2slice3D}.
We also find a deviation from the expected power law scaling as shown in Fig.~\ref{fig:Z2scaling3D}, albeit with significantly less dynamic range compared to the case of (2+1) dimensions.

\subsection{Kink Interactions}

The deviation from the expected power law scaling observed in Figs.~\ref{fig:Z2walls} and \ref{fig:Z2scaling3D} may suggest some interaction among the scalar fields which prevents the domain walls from scaling as fast as they otherwise would.
This could be attributed to the fact that the kink--anti-kink interaction can be attractive or repulsive depending where they form in the field configuration.
To investigate this we calculate the force between a kink and an anti-kink in (1+1) dimensions with a relative $\text{SU}(2)_L \times \text{U}(1)_Y$ global rotation between them.
The energy-momentum tensor for the 2HDM is given by
\begin{equation}
    {T^\mu}_\nu = \frac{\partial\mathcal{L}}{\partial\left(\partial_\mu\Phi_i\right)} \partial_\nu \Phi_i + \frac{\partial\mathcal{L}}{\partial\left(\partial^\nu\Phi_i\right)} \partial^\mu \Phi_i - {\delta^\mu}_\nu \mathcal{L}.
\end{equation}
Therefore, in (1+1) dimensions,
\begin{equation}
    {T^0}_1 = \dot{\Phi}_i^\dagger \Phi_i^\prime + {\Phi_i^\dagger}^\prime \dot{\Phi}_i
\end{equation}
where \textit{dots} and \textit{primes} denote temporal and spatial derivatives, respectively.
The momentum of the field configuration over some semi-infinite length is then
\begin{equation}
    P = \int_b^\infty \left(\dot{\Phi}_i^\dagger \Phi_i^\prime + {\Phi_i^\dagger}^\prime \dot{\Phi}_i\right) dx
\end{equation}
and the corresponding force
\begin{equation}\label{eq:force}
    F = \dot{P} = \int_b^\infty \left(\ddot{\Phi}_i^\dagger \Phi_i^\prime + \dot{\Phi}_i^\dagger \dot{\Phi}_i^\prime + \dot{\Phi}_i^{\dagger\prime} \dot{\Phi}_i + {\Phi_i^\dagger}^\prime \ddot{\Phi}_i\right) dx.
\end{equation}
Using the equations of motion to substitute for $\ddot{\Phi}_i$ and $\ddot{\Phi}_i^\dagger$ in \eqref{eq:force} and integrating, we find
\begin{equation}\label{eq:force_sub}
    F = \left[\left(\dot{\Phi}_i^\dagger\dot{\Phi}_i + \Phi_i^{\dagger\prime} \Phi_i^\prime\right) - V \right]_b^\infty.
\end{equation}

In order to evaluate the lower limit, $x = b$, we can derive asymptotic expressions for the kink and condensate.
For a static kink/condensate solution in one dimension, $\hat{\Phi}_i(x)$ (with $i=1,2$), the Euler-Lagrange equations reduce to
\begin{equation}\label{eq:ELeqstat}
    \Phi_i^{\prime\prime} = \frac{\partial V}{\partial \Phi_i^\dagger}.
\end{equation}
For a kink at $x = a$ and anti-kink at $x = -a$ we write the doublets as
\begin{equation}\label{eq:FieldsSmall}
    \Phi_1 = \Phi_1^0 + \delta\Phi_1, \qquad
    \Phi_2 = -\Phi_2^0 + \delta\Phi_2
\end{equation}
where $\delta\Phi_{1,2} \ll 1$ and $\Phi_{1,2}^0$ are the CP-preserving vacua of~\eqref{eq:NormalVac}.

We consider the force on a field configuration composed of a kink-anti kink pair with a relative $\text{SU}(2)_L \times \text{U}(1)_Y$ transformation between the kink and anti kink as follows
\begin{equation}
  \begin{split}
    \Phi_1(x) =&\ \hat{\Phi}_1(x-a) + U \left[\hat{\Phi}_1(x+a) - \Phi_1^0\right],\\
    \Phi_2(x) =&\ \hat{\Phi}_2(x-a) - U \left[\hat{\Phi}_2(x+a) - \Phi_2^0\right],
  \end{split}
\end{equation}
where $\hat{\Phi}_{1,2}(x)$ are static solutions\footnote{Note, the relative sign between the expressions for the doublets is such that one describes a kink-anti kink pair and the other the corresponding condensate profile, c.f. Fig.~\ref{fig:Z2FieldsFull2}} and
\begin{equation}
    U = e^{i\theta}\left(\begin{matrix}
    \bar{A} & B \\
    -\Bar{B} & A
    \end{matrix}\right)
\end{equation}
is a global $\text{SU}(2)_L \times \text{U}(1)_Y$ matrix.
It should be noted that this is an approximate solution, becoming exact in the limit that $U$ becomes an identity matrix.
Nonetheless, this field configuration has the desired profile over the range of integration. 
For $-a \ll b \ll a$, we have $\Hat{\Phi}_i(x + a) - \Phi_i^0 \simeq 0$ and, hence, $\Hat{\Phi}_i^\prime(x + a) \simeq 0$,
allowing us to linearize \eqref{eq:force_sub} in these small quantities.
Firstly, we can linearize the spatial derivatives in \eqref{eq:force_sub} as
\begin{equation}
    \Phi_1^{\dagger\prime}\Phi_1^\prime \simeq \hat{\Phi}_{1,-}^{\dagger\prime}\hat{\Phi}_{1,-}^\prime + \hat{\Phi}_{1,-}^{\dagger\prime} U \hat{\Phi}_{1,+}^\prime + \hat{\Phi}_{1,+}^{\dagger\prime} U^\dagger \hat{\Phi}_{1,-}^\prime,
\end{equation}
\begin{equation}
    \Phi_2^{\dagger\prime}\Phi_2^\prime \simeq \hat{\Phi}_{2,-}^{\dagger\prime}\hat{\Phi}_{2,-}^\prime - \hat{\Phi}_{2,-}^{\dagger\prime} U \hat{\Phi}_{2,+}^\prime - \hat{\Phi}_{2,+}^{\dagger\prime} U^\dagger \hat{\Phi}_{2,-}^\prime,
\end{equation}
where we have defined the short-hand notation $\hat{\Phi}_{i,\pm} = \hat{\Phi}_i(x \pm a)$.
The 2HDM potential of \eqref{eq:Z2Potential} linearizes as
\begin{align}
        V(\Phi_1,\Phi_2) =&\ 
        V(\hat{\Phi}_{1,-},\hat{\Phi}_{2,-})\nonumber\\[2mm]
        &+ (\hat{\Phi}_{1,+}^\dagger - \Phi_1^{0\dagger})
        U^\dagger \frac{\partial V}{\partial \Phi_1^\dagger}(\hat{\Phi}_{1,-},\hat{\Phi}_{2,-})
        + \frac{\partial V}{\partial \Phi_1}(\hat{\Phi}_{1,-},\hat{\Phi}_{2,-})
        U(\hat{\Phi}_{1,+} - \Phi_1^{0})\nonumber\\
        &- (\hat{\Phi}_{2,+}^\dagger 
        - \Phi_2^{0\dagger})U^\dagger \frac{\partial V}{\partial \Phi_2^\dagger}(\hat{\Phi}_{1,-},\hat{\Phi}_{2,-})
        - \frac{\partial V}{\partial \Phi_2}(\hat{\Phi}_{1,-},\hat{\Phi}_{2,-})
        U(\hat{\Phi}_{2,+} - \Phi_2^{0}).
\end{align}

Hence, the linearized version of the force on the field configuration can be written as,
\begin{equation}\label{eq:forcemin}
    F = \left[ \hat{\Phi}_{1,-}^{\dagger\prime}U\hat{\Phi}_{1,+}^\prime - \hat{\Phi}_{1,-}^{\dagger\prime\prime} U (\hat{\Phi}_{1,+} - \Phi_1^0) - \hat{\Phi}_{2,-}^{\dagger\prime}U\hat{\Phi}_{2,+}^\prime + \hat{\Phi}_{2,-}^{\dagger\prime\prime} U (\hat{\Phi}_{2,+} - \Phi_2^0) + \text{h.c.} \right]_b^\infty.
\end{equation}
Since the static solutions are asymptotically flat, their gradients vanish at infinity. Therefore, there is no contribution from the upper limit of \eqref{eq:forcemin}. Furthermore, substituting the explicit form of the doublets, we reduce the above expression to
\begin{equation}
    F = \frac{1}{2}(e^{i\theta}A+e^{-i\theta}\Bar{A})\left[ v_{1,-}^\prime v_{1,+}^\prime - v_{1,-}^{\prime\prime}(v_{1,+} - v_1^0) - v_{2,-}^\prime v_{2,+}^\prime + v_{2,-}^{\prime\prime}(v_{2,+} - v_2^0) \right]_b^\infty.
\end{equation}
In order to evaluate the lower limit, let us define the asymptotic expressions
\begin{align}
        v_{1,-} \equiv&\ v_1(x-a) \simeq v_1^0 - \rho e^{\mu(x-a)}, \qquad 
        &v_{1,+} \equiv 
        v_1(x+a) \simeq v_1^0 - \rho e^{-\mu(x+a)}\,,\nonumber\\
        v_{2,-} \equiv&\ v_2(x-a) \simeq -v_2^0 + \sigma e^{\nu(x-a)}, \qquad 
        &v_{2,+} \equiv v_2(x+a) 
        \simeq v_2^0 - \sigma e^{-\nu(x-a)},
\end{align}
where $\mu = \sqrt{2} \mu_1$, $\nu = \sqrt{2} \mu_2$ and $\rho$ and $\sigma$ are normalization factors that are undetermined at this level of approximation.
Using these asymptotic expressions for the kink solutions we can now write the force as
\begin{equation}
    F = \frac{d E_{\text{int}}}{d R} = (e^{i\theta}A+e^{-i\theta}\Bar{A})\left(\rho^2\mu^2 e^{-\mu R} + \sigma^2\nu^2 e^{-\nu R}\right)
\end{equation}
where $R = 2a$ is the kink separation. Therefore, the interaction energy is
\begin{equation}
    E_{\text{int}} = -(e^{i\theta}A+e^{-i\theta}\Bar{A})\left(\rho^2\mu e^{-\mu R} + \sigma^2 \nu e^{-\nu R}\right).
\end{equation}
We see that the group parameters, $A$ and $\theta$, are present in the expression.
This indicates that the sign of the force will change depending on the values of the group parameters.
As such, there can be repulsive as well as attractive interactions between domain walls in different parts of the field configuration.
Note that for $U=I$ (where $\theta=0$ and $A=1$) we recover the usual negative interaction energy signifying an attractive force between the kink and anti kink.

Now imagine an approximately circular wall evolving as part of the wall network in (2+1) dimensions (or an approximately spherical wall in (3+1) dimensions). The standard picture would be for the walls to collapse as fast as causality would allow under their own tension - this is what leads to the expectation that $N_{\rm dw}\propto t^{-1}$. However, in this case we have shown that there are extra forces associated with the relative phases of the vacuum which could be, and indeed will be different on each part of the wall. These forces will clearly interfere with the dynamics of the wall network and we can expect that the scaling behaviour will be modified.  

\subsection{Relatively Gauge-Rotated Vacua}

Let us now consider the effect of introducing a relative electroweak transformation between vacua on either side of a kink obtained via gradient flow. This is something which is not possible in the case of the standard domain wall solution in the discrete Goldstone model, but which is manifest in the 2HDMs with discrete symmetries, and would be expected in simulations from random initial conditions. In order to achieve this, we introduce a relative transformation between the vacuum field configuration at the boundaries and impose Dirichlet boundary conditions on the gradient flow,
\begin{equation}
    \Phi_i(-\infty) = \frac{1}{\sqrt{2}}\left(\begin{matrix}
    0 \\
    \pm v_i
    \end{matrix}\right), \quad
    \Phi_i(+\infty) = \frac{1}{\sqrt{2}}U\left(\begin{matrix}
    0 \\
    \mp v_i
    \end{matrix}\right),
\end{equation}
where $U$ has the form of \eqref{eq:EWGT}.
If we consider the effect of each group parameters separately, we see that a non-zero $\gamma_1$ excites the upper components of the CP-preserving vacua, while $\theta$ and $\gamma_2$ excite the imaginary part of the lower components.

\begin{figure}
    \centering
    \includegraphics[width=\textwidth]{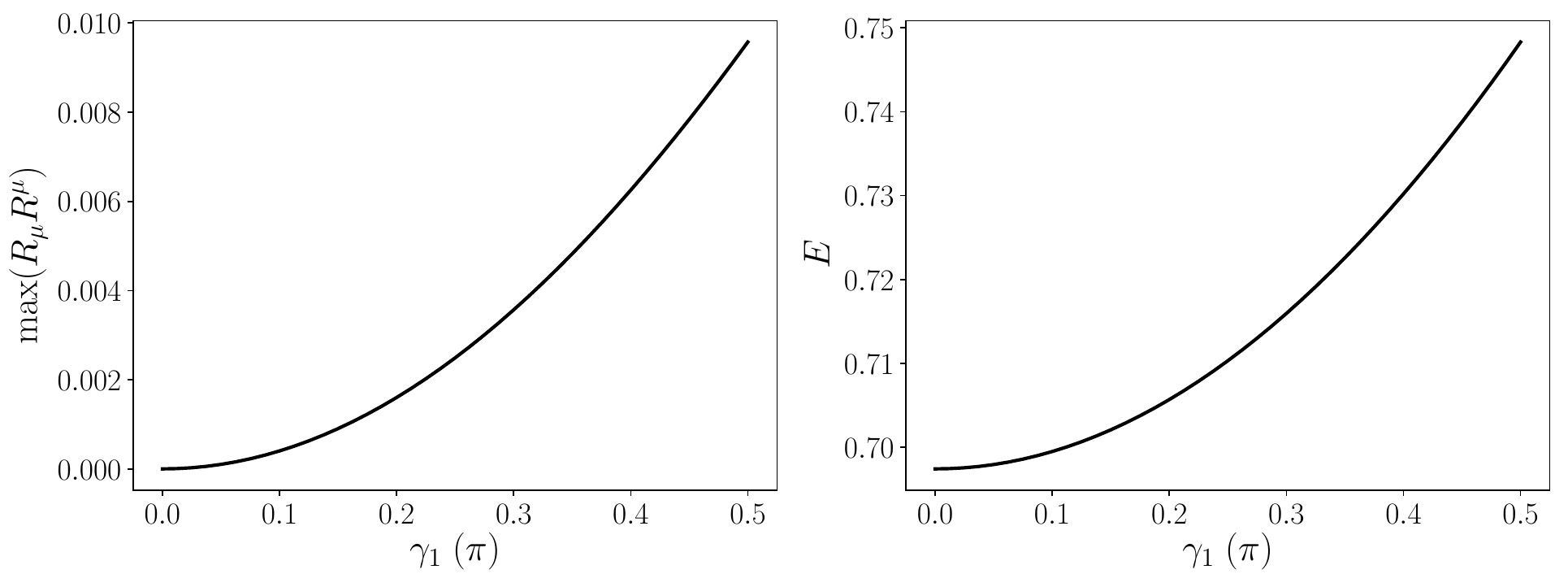}
    \caption{Peak value of $R_\mu R^\mu$ for long-time gradient flow solutions as a function of $\gamma_1$ at the right-hand boundary (left) and the corresponding energy per unit area of the solutions (right). The solution with $\gamma_1$ is the minimum energy solution previously calculated with $E=0.697$ and $R^\mu R_\mu\equiv 0$.}
    \label{fig:gamma_sweep}
\end{figure}

In order to investigate the impact of $U\ne I$, we have first calculated solutions for a range of the parameter $\gamma_1$ keeping $\gamma_2=\gamma_3=\theta=0$ fixed throughout.
The peak value of $R_\mu R^\mu$ as a function of $\gamma_1$ is presented in Fig.~\ref{fig:gamma_sweep} along with the corresponding energy per unit area of the solutions.
The solution with lowest energy is indeed the solution where there is no relative transformation between the boundaries.
For this solution we observe no violation of the neutral vacuum condition as expected.

\begin{figure}
    \centering
    \includegraphics[width=\textwidth]{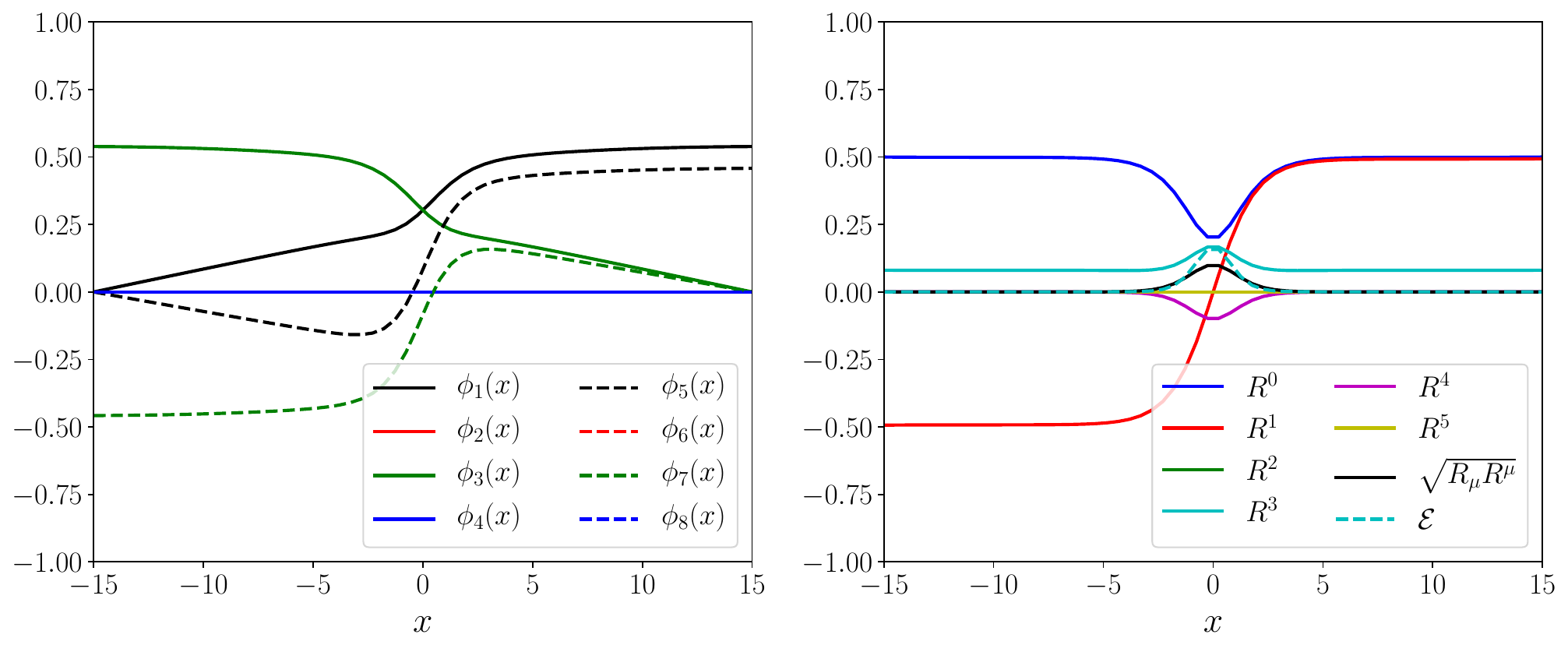}
    \caption{Kink solution with $\gamma_1=\pi/2$ at the right hand boundary (left) and corresponding $R$-space profiles for $M_H = M_A = M_{H^\pm} = 200 \text{GeV}$, $\tan\beta=0.85$ and $\cos(\alpha-\beta) = 1.0$. Note that there is no obvious wall solution in any of the fields $\phi_1,..,\phi_8$, but there is clearly a kink in the profile of $R^1$. Moreover, in this case $R^\mu R_\mu$ is non-zero at the position of the kink, and $R^4$ is also non-zero, but negative.  Since $\theta=0$, $R^4$ is non-zero while $R^5 = 0$ (see \eqref{eq:RA}) and we also have that  $\sqrt{R_\mu R^\mu} = \left|R^4\right|$.}
    \label{fig:gamma_sol}
\end{figure}

However, for any other value of  $\gamma_1$ we obtain higher-energy solutions where $R_\mu R^\mu$ is non-zero and peaked locally at the position the kink as seen in our simulations from random initial conditions. An example solution for $\gamma_1 = \pi/2$ and the corresponding $R$-space profiles are presented in Fig.~\ref{fig:gamma_sol}.
We note that the neutral vacuum condition is respected only when the upper components of the doublets are zero at the rotated boundary, i.e. in the absence of $v_+$.
We also observe an increase in the energy per unit area as in Fig.~\ref{fig:gamma_sweep} for solution with transformations involving the other group parameters, $\gamma_2$, $\gamma_3$ and $\theta$.

\begin{figure}
    \centering
    \includegraphics[width=\textwidth]{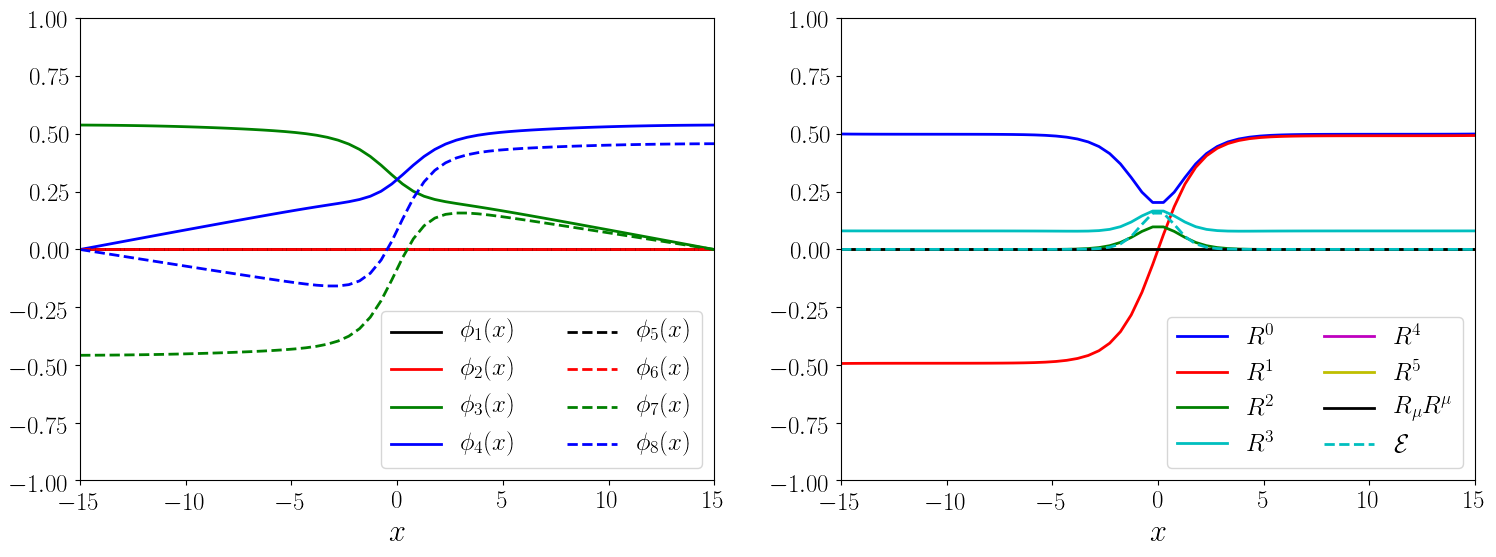}
    \caption{Kink solution with $\theta=\pi/2$ at the right hand boundary (left) and corresponding $R$-space profiles for $M_H = M_A = M_{H^\pm} = 200 \text{GeV}$, $\tan\beta=0.85$ and $\cos(\alpha-\beta) = 1.0$. Note that there is no obvious wall solution in any of the fields $\phi_1,..,\phi_8$, but there is clear kink in the profile of $R^1$. Moreover, in this case $R^2$ is non-zero at the position of the kink.}
    \label{fig:theta_sol}
\end{figure}

\begin{figure}
    \centering
    \includegraphics[width=\textwidth]{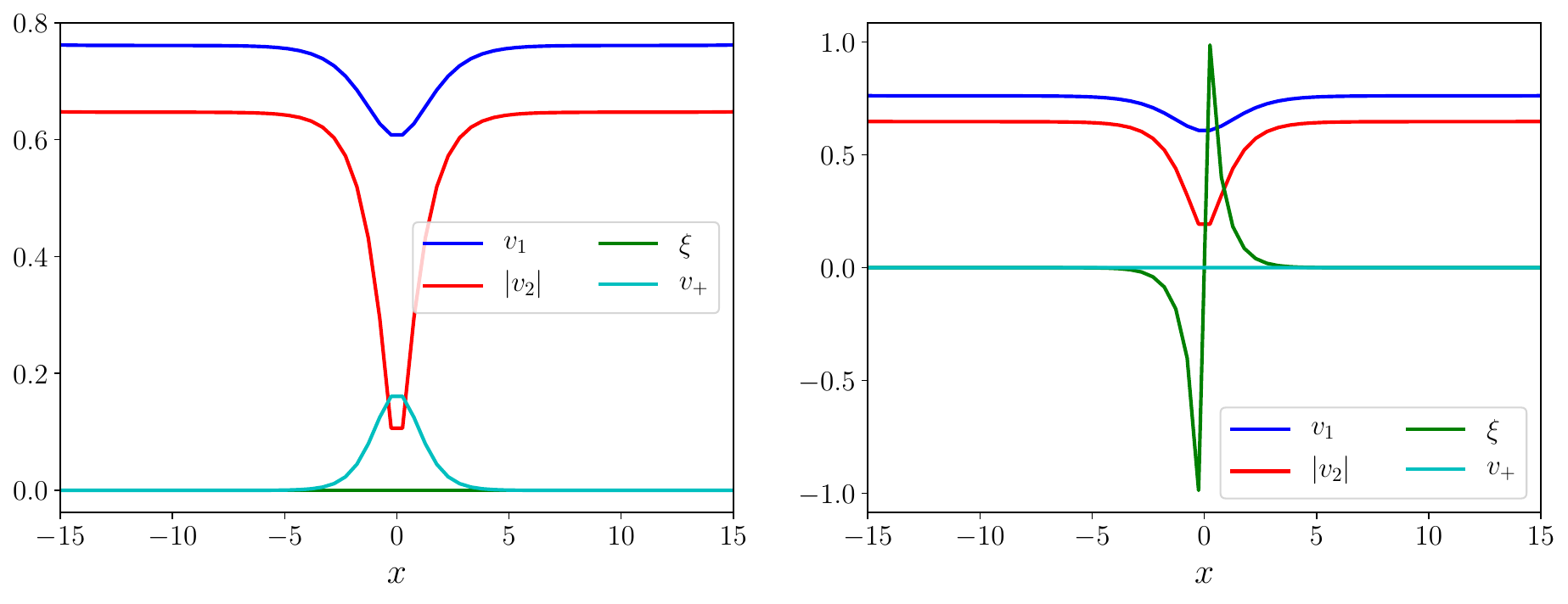}
    \caption{Vacuum manifold for a minimum energy solution with $\gamma_1=\pi/2$ (left) and $\theta = \pi/2$ (right) at the right hand boundary. Parameters chosen were $M_H = M_A = M_{H^\pm} = 200 \; \text{GeV}$, $\tan\beta=0.85$ and $\cos(\alpha-\beta) = 1.0$. In both cases all other electroweak group parameters are set to zero.
    %{\bf Note issue with right-hand panel}
    }
    \label{fig:gamma_angles}
\end{figure}

We have also investigated solutions when $\theta$ and $\gamma_2\ne 0$. Since they both excite the same components of the field configuration ($\phi_4$ and $\phi_8$) with opposite sign, we only present results for an example solution for $\theta = \pi/2$ with $\gamma_1=\gamma_2=\gamma_3=0$ which is presented in Fig.~\ref{fig:theta_sol}.
In Figs. \ref{fig:gamma_sol} and \ref{fig:theta_sol} we see that the vacuum values of the components $R^A$ remain unchanged, however, solutions with a relative transformation between the boundaries have features around the kink not observed in the minimum energy solution of Fig.~\ref{fig:Z2FieldsFull2}.
Specifically, a non-zero $\gamma_1$ leads to a solutions with non-zero $R^4$ and $R^5$ on the kink and hence a violation of the neutral vacuum condition.
For non-zero $\theta$ (or $\gamma_2$) we obtain solutions with a non-zero $R^2$ around the kink.
Corresponding plots of the vacuum manifold parameters are given in Fig.~\ref{fig:gamma_angles}.
For non-zero $\gamma_1$ we obtain a non-zero $v_+$ peaked at the kink indicating a local violation of the neutral vacuum condition.
For non-zero $\theta$ we observe a non-zero $\xi$ around the kink.

\begin{figure}
    \centering
    \includegraphics[width=\textwidth]{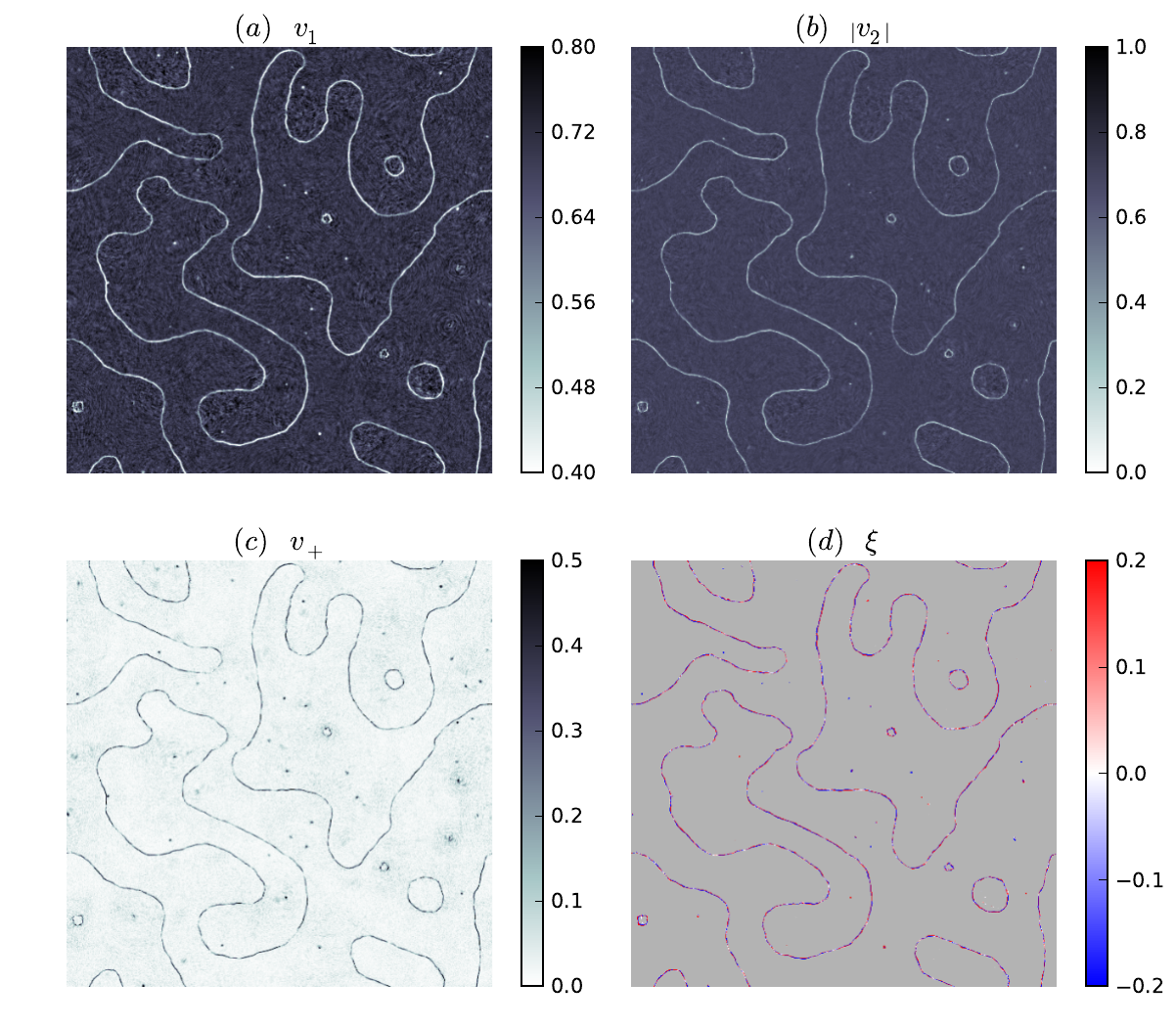}
    \caption{Results of a 2D simulation in the $Z_2$-symmetric 2HDM at $t = 240$ for $M_H = M_A = M_{H^\pm} = 200 \text{GeV}$, $\tan\beta=0.85$ and $\cos(\alpha-\beta) = 1.0$ showing the spatial variation of the vacuum manifold parameters, $v_1, v_2, \xi$ and $v_+$. The behaviour of the vacuum manifold parameters are consistent with those obtained for the kinks with rotated vacua. In particular, we observe the local minimum in $v_1$ in the vicinity of the domain walls; the non-zero $v_+$ on the domain walls corresponding to the local violation of the neutral vacuum condition and the transverse variation of $\xi$ over the domain walls. Note that one can only calculate $|v_2|$ from the fields of the linear representation and hence in (b) we see only the modulus of the kink solution and also that $\xi$ is plotted only in the vicinity of the domain walls.}
    \label{fig:vac_man}
\end{figure}

\begin{figure}
    \centering
    \includegraphics[width=\textwidth]{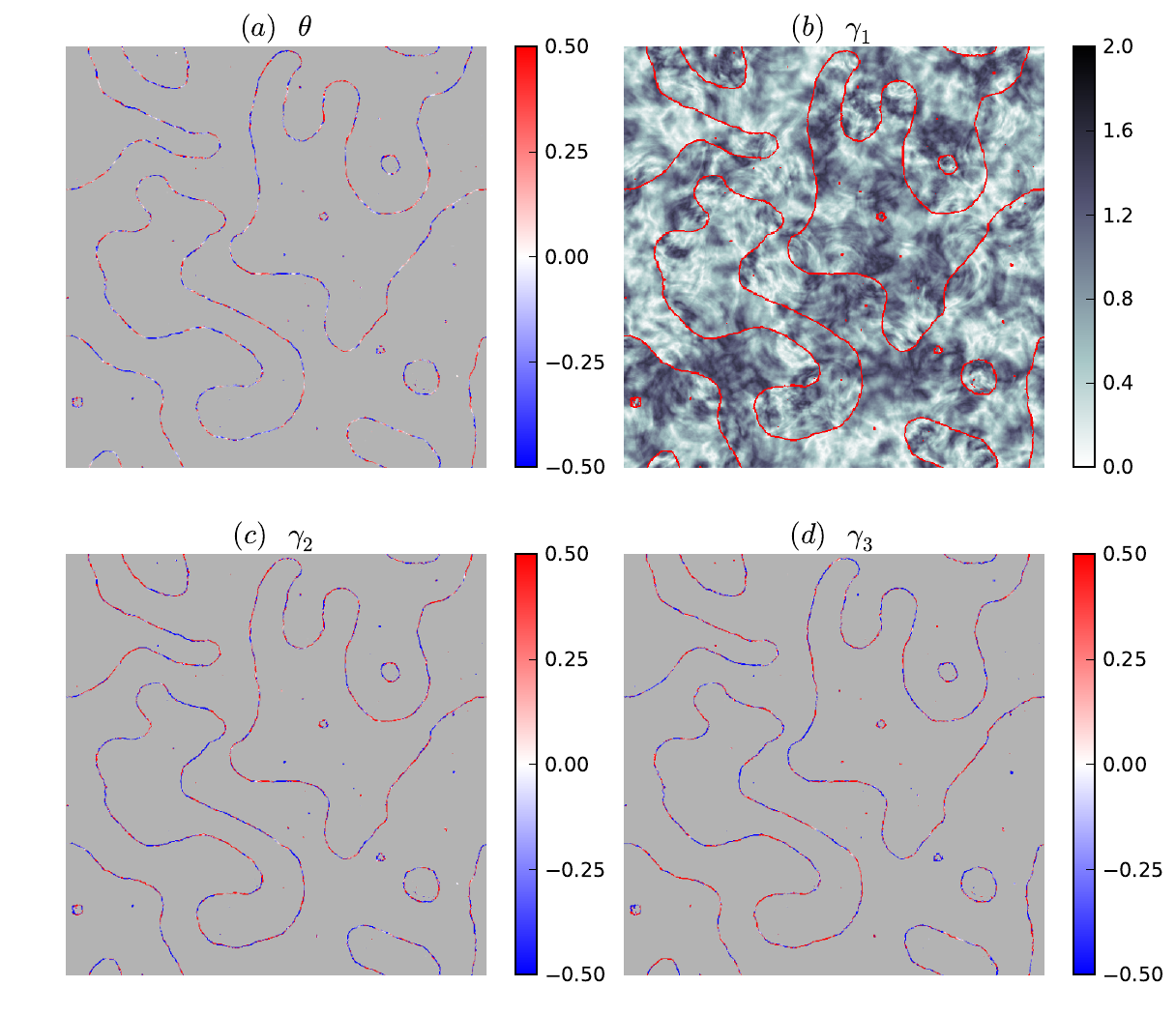}
    \caption{Results of a 2D simulation in the $Z_2$-symmetric 2HDM at $t = 240$ for $M_H = M_A = M_{H^\pm} = 200 \text{GeV}$, $\tan\beta=0.85$ and $\cos(\alpha-\beta) = 1.0$: (a) group parameter $\theta$ plotted in the vicinity of the domain walls; (b) group parameter $\gamma_1$ overlaid with red/light line showing location of domain walls; (c) group parameter $\gamma_2$ plotted in the vicinity of the domain walls and (d) group parameters $\gamma_3$ plotted in the vicinity of the domain walls. (a), (b) and (c) all show a winding of the respective group parameter around the domain walls.}
    \label{fig:angles}
\end{figure}

Spatial variation of the vacuum manifold parameters and group parameters from a (2+1) dimensional simulation are presented in Figs. \ref{fig:vac_man} and \ref{fig:angles}, respectively.
The profiles of the vacuum manifold parameters are consistent with those obtained for the kinks with rotated vacua.
Specifically, the peak in $v_+$ and the profile of $\xi$ seen in Fig.~\ref{fig:gamma_angles} are manifest in the simulation results.
We also observe a winding of the parameters $\theta, \gamma_2$ and $\gamma_3$ around the domain walls.
%and a variation of $\gamma_1$ across the domain walls as seen in Fig.~\ref{fig:gamma_angles}.
If these features are indeed akin to the kinky vorton model, one would expect there to be some Noether-like charge winding around the domain walls.

\begin{figure}
    \centering
    \includegraphics[width=\textwidth]{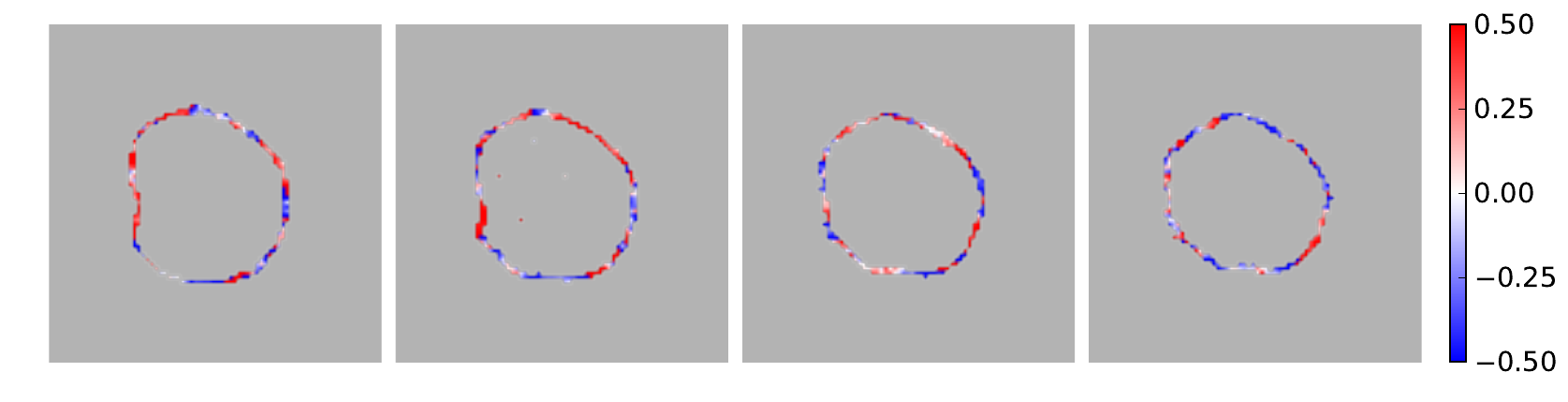}
    \caption{Late time temporal evolution of the electroweak group parameter, $\theta$, in the vicinity of a domain wall in the $Z_2$-symmetric 2HDM for for $M_H = M_A = M_{H^\pm} = 200 \text{GeV}$, $\tan\beta=0.85$ and $\cos(\alpha-\beta) = 1.0$.}
    \label{fig:theta_winding}
\end{figure}

A plot showing the time evolution of $\theta$ is presented in Fig.~\ref{fig:theta_winding}.
We see that there is some clockwise winding of the electroweak angle, $\theta$, around the domain wall.
Note that Fig.~\ref{fig:theta_winding} shows a select spatial region containing a circular domain wall at late times in our simulations.

% CP1 Symmetry
\section{CP1 Symmetry}
\label{sec:CP1}

A CP1 transformation\footnote{Note that CP1 represents the usual CP symmetry.} of the Higgs doublets is given by
\begin{equation}
    \Phi_1 \rightarrow \Phi_1^*, \quad
    \Phi_2 \rightarrow \Phi_2^*.
\end{equation}
Using the parameter restrictions of Table~\ref{tab:2HDMConstraints}, the CP-symmetric 2HDM potential can be written as
\begin{align}
  \label{eq:CP1Potential}
    V =& -\mu_1^2 (\Phi_1^\dagger\Phi_1) - \mu_2^2 (\Phi_2^\dagger\Phi_2) - 2m_{12}^2\text{Re}\left(\Phi_1^\dagger\Phi_2\right) + \lambda_1 (\Phi_1^\dagger\Phi_1)^2 + \lambda_2 (\Phi_2^\dagger\Phi_2)^2 + \lambda_3(\Phi_1^\dagger\Phi_1)(\Phi_2^\dagger\Phi_2)\nonumber\\
    &+ \lambda_{45}\left[\text{Re}\left(\Phi_1^\dagger\Phi_2\right)\right]^2 + \Bar{\lambda}_{45}\left[\text{Im}\left(\Phi_1^\dagger\Phi_2\right)\right]^2 + 2\text{Re}\left(\Phi_1^\dagger\Phi_2\right)\left(\lambda_6 (\Phi_1^\dagger\Phi_1) + \lambda_7 (\Phi_2^\dagger\Phi_2)\right).\quad
\end{align}

For the CP-breaking vacuum \eqref{eq:CPVac}, the VEVs are calculated in terms of the potential parameters in \cite{Brawn2011}. The final results are given by
\begin{equation}\label{eq:CP1v1}
    v_1^2 = \frac{2\left(2\lambda_2\lambda_5 - \lambda_7^2\right)\mu_1^2 + 2\left(\lambda_6\lambda_7 - \Bar{\lambda}_{345}\lambda_5\right)\mu_2^2 + 2\left(\Bar{\lambda}_{345}\lambda_7 - 2\lambda_2\lambda_6 \right)m_{12}^2}{\lambda_5\left(4\lambda_1\lambda_2 - \Bar{\lambda}_{345}^2\right) - 2\lambda_1\lambda_7^2 - 2\lambda_2\lambda_6^2 + 2\Bar{\lambda}_{345}\lambda_6\lambda_7},
\end{equation}
\begin{equation}\label{eq:CP1v2}
    v_2^2 = \frac{2\left(2\lambda_1\lambda_5 - \lambda_6^2\right)\mu_2^2 + 2\left(\lambda_6\lambda_7 - \Bar{\lambda}_{345}\lambda_5\right)\mu_1^2 + 2\left(\Bar{\lambda}_{345}\lambda_6 - 2\lambda_1\lambda_7 \right)m_{12}^2}{\lambda_5\left(4\lambda_1\lambda_2 - \Bar{\lambda}_{345}^2\right) - 2\lambda_1\lambda_7^2 - 2\lambda_2\lambda_6^2 + 2\Bar{\lambda}_{345}\lambda_6\lambda_7},
\end{equation}
and
\begin{equation}\label{eq:CP1xi}
    \xi = \arccos\left(\frac{2m_{12}^2 - \lambda_6 v_1^2 - \lambda_7 v_2^2}{2\lambda_5 v_1 v_2}\right).
\end{equation}

In order to find the masses of the physical states, we must again calculate the eigenvalues of the Hessian matrix.
We use the minimization conditions,
\begin{eqnarray}\label{eq:MuMins}
    \mu_1^2 &=& \lambda_1 v_1^2 + \frac{1}{2}\bar{\lambda}_{345} v_2^2 + \lambda_6 v_1 v_2 c_\xi,\\
    \mu_2^2 &=& \lambda_2 v_2^2 + \frac{1}{2}\bar{\lambda}_{345} v_1^2 + \lambda_7 v_1 v_2 c_\xi
\end{eqnarray}
and
\begin{equation}\label{eq:m12Min}
    m_{12}^2 = \lambda_5 v_1 v_2 c_\xi + \frac{1}{2}\lambda_6 v_1^2 + \frac{1}{2}\lambda_7 v_2^2\,,
\end{equation}
in order to further simplify the mass matrices given in~\cite{Pilaftsis1999}.
In this way, the charged sector mass matrix simplifies to
\begin{equation}\label{eq:CP1ChargedMass}
	\mathcal{M}_{H^\pm}^2 = - \frac{1}{2}\bar{\lambda}_{45} \left(\begin{matrix}
	v_2^2 & -v_1v_2 e^{i\xi} \\
	-v_1v_2e^{-i\xi} & v_1^2
	\end{matrix}\right).
\end{equation}
As expected, the charged mass matrix \eqref{eq:CP1ChargedMass} has one zero eigenvalue corresponding to the Goldstone bosons, $G^\pm$, and the mass of the charged Higgs bosons is
\begin{equation}\label{eq:CP1chargemass}
	M_{H^\pm}^2 = -\frac{1}{2}\bar{\lambda}_{45}v_{\text{SM}}^2.
\end{equation}
Unlike in the $Z_2$ case, the CP1-symmetric 2HDM has a non-zero phase, $\xi$, and the mass matrices for the CP-even and CP-odd scalars do not decouple as the model is not explicitly CP-conserving.
Working in the $\{\varphi_1,\varphi_2,A,G^0\}$ basis, the $G^0$ axis decouples from the other 3 and, as one would expect, has a zero eigenvalue.
Hence, we obtain a symmetric $3\times 3$ neutral mass matrix, $\mathcal{M}_N^2$, with elements,
\begin{eqnarray}
  \label{eq:calMNCP1}
    (\mathcal{M}_N^2)_{11} &=& 2 \lambda _1 v_1^2 + \lambda _5 v_2^2 c_\xi^2 + 2 \lambda _6 v_1 v_2 c_\xi\,,\nonumber\\
    (\mathcal{M}_N^2)_{12} &=& (\lambda _{34} -\lambda _5 s_\xi^2) v_1 v_2 + (\lambda _6 v_1^2 + \lambda _7 v_2^2) c_\xi\nonumber\,,\\
    (\mathcal{M}_N^2)_{13} &=& -s_\xi v_\text{SM} \left(\lambda _5 v_2 c_\xi + \lambda _6 v_1\right)\,,\\
    (\mathcal{M}_N^2)_{22} &=& 2 \lambda _2 v_2^2 + \lambda _5 v_1^2 c_\xi^2 + 2 \lambda _7 v_1 v_2 c_\xi\,,\nonumber\\
    (\mathcal{M}_N^2)_{23} &=& -s_\xi v_\text{SM} \left(\lambda _5 v_1 c_\xi + \lambda _7 v_2\right)\,,\nonumber\\
    (\mathcal{M}_N^2)_{33} &=& \lambda _5 v_\text{SM}^2 s_\xi^2\,.\nonumber
\end{eqnarray}
The masses for the three neutral Higgs bosons, $M_{H_{1,2,3}}$, are obtained as 
roots of a cubic characteristic equation, exactly as done in~\cite{Pilaftsis1999}.
Which mass eigenstate ($i=1,2,3$) corresponds to which Higgs particle is then determined by its relative coupling to the $Z^0$ boson,
\begin{equation}
    g_{H_i ZZ} = \cos\beta O_{1i} + \sin\beta O_{2i},
\end{equation}
where $O$ is an orthogonal 3$\times$3-dimensional matrix that diagonalizes the scalar mass matrix~$\mathcal{M}_N^2$ given
in~\eqref{eq:calMNCP1}.
In particular, the SM-like scalar will be that which has 
$g_{H_i ZZ}$ close to~1.

Ideally, to simulate the CP1-symmetric 2HDM we would obtain expressions for the scalar masses, $M_{H_{1,2,3}}$, of \eqref{eq:calMNCP1} and invert this system of equations to obtain expressions for the parameters of the scalar potential in terms of the physical observables as was performed for the $Z_2$-symmetric 2HDM in Appendix~\ref{sec:Rescaling}.
While this calculation is achievable in the $Z_2$ case, the CP1-symmetric 2HDM potential is significantly more complicated due to the mixing of CP-even/odd states requiring a more complicated system of mixing angles \cite{Osland:2008aw} to diagonalize \eqref{eq:calMNCP1}.
Nonetheless, the values of the neutral scalar mass eigenvalues can be obtained numerically for a given set of potential parameters.
To achieve this we first simplify \eqref{eq:calMNCP1} by working in the reduced basis of Table~\ref{tab:Diag_Red_Constraints} with additional restrictions $\lambda_4 = \lambda_6 = 0$.
Under these additional restrictions the quartic coupling parameter $\lambda_5$ fixes the charged Higgs mass as seen in \eqref{eq:CP1chargemass}.
To obtain maximal spontaneous CP violation, we specify $\xi = \pi/4$ and choose $\tan\beta$ with $v_1 = c_\beta v_\text{SM}$ and $v_2 = s_\beta v_\text{SM}$ allowing us to infer the quadratic parameters, $\mu_1^2$, $\mu_2^2$ and $m_{12}^2$, from the minimization conditions \eqref{eq:MuMins}--\eqref{eq:m12Min}.
We choose $\lambda_3$ such that the mass matrix element $(\mathcal{M}_N^2)_{12}$ of \eqref{eq:calMNCP1} is small to achieve SM alignment.
The remaining parameters $\lambda_1$ and $\lambda_2$ are used to fix the scalar masses, $M_{H_{1,2,3}}$.

\subsection{Kinks Solutions}

As for the $Z_2$-symmetric 2HDM, we present a generalized treatment of kinks in the CP1-symmetric 2HDM from \cite{Brawn2011}.
We obtain relaxed solutions from the general field configuration and general vacuum ansatz \eqref{eq:ChargedVac}.

\begin{figure}
    \centering
    \includegraphics[width=\textwidth]{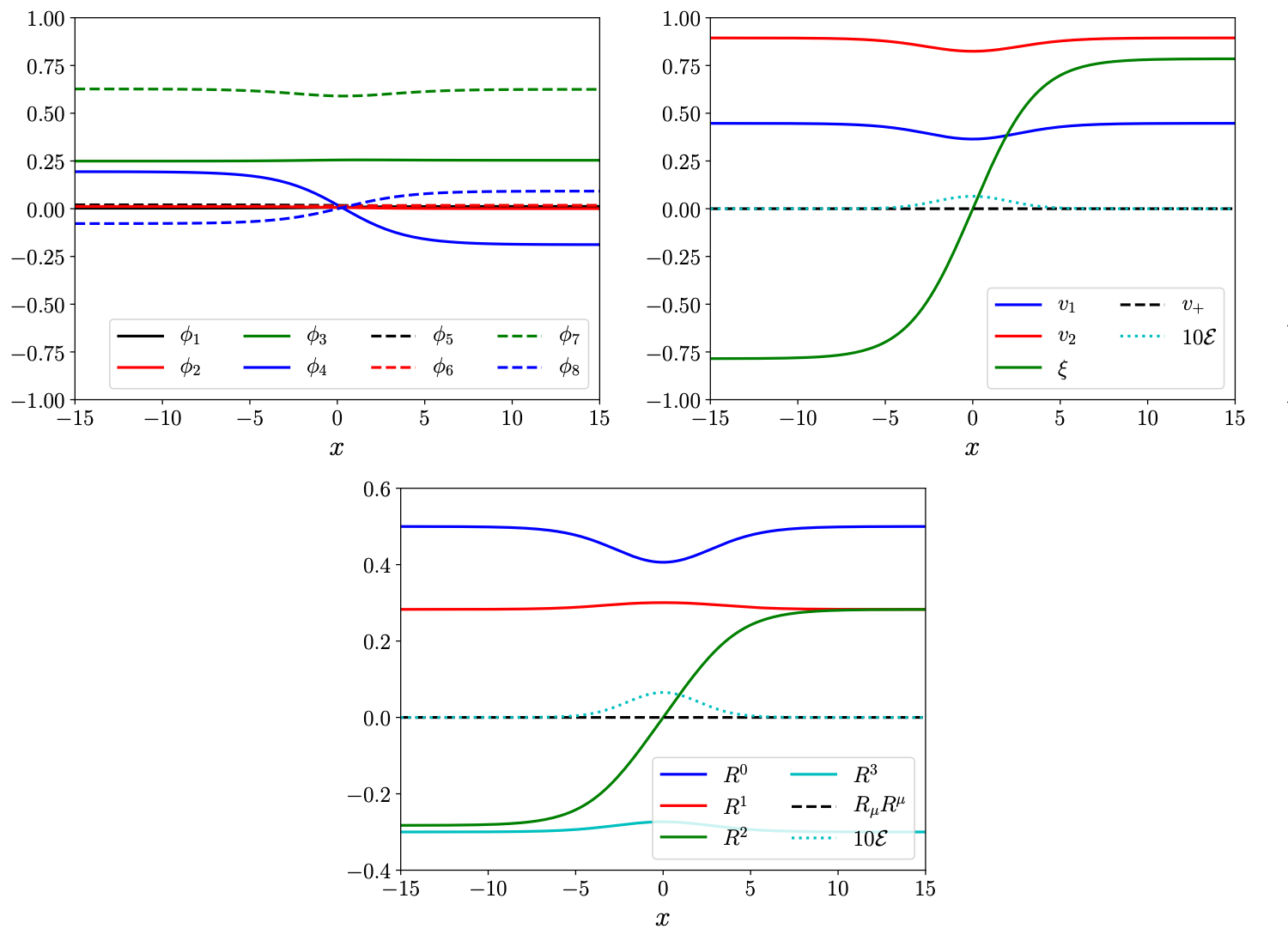}
    \caption{Kink solution of the CP1-symmetric 2HDM in the linear representation (left) and vacuum manifold parametrization (right) along with corresponding $R$-space profiles in the reduced basis of Table~\ref{tab:Diag_Red_Constraints} with $\lambda_4 = \lambda_6 = 0$. Note that the $R$-space profiles are identical for the solutions relaxed in the two different representations. Parameters chosen correspond to $ M_1 = $~125 GeV, $ M_2 = $~307 GeV, $ M_3 = $~455 GeV, $ M_{H^\pm} = $~300 GeV, $\xi = \pi/4$ and $\tan\beta = 2.0$. Total energy of the solution is 0.064.}%184474345.}
    \label{fig:CP1kinkphi}
\end{figure}

Inserting the general vacuum, \eqref{eq:ChargedVac}, into the potential, \eqref{eq:CP1Potential}, one finds the following expression for the energy per unit area for the CP1-symmetric 2HDM:
\begin{equation}
\begin{split}
\mathcal{E} =& \frac{1}{2}\left(\frac{d v_1}{d x}\right)^2 + \frac{1}{2}\left(\frac{d v_2}{d x}\right)^2 + \frac{1}{2}\left(\frac{d v_+}{d x}\right)^2 + \frac{1}{2} v_2^2 \left(\frac{d \xi}{d x}\right)^2 - \frac{1}{2} \mu_1^2 v_1^2 - \frac{1}{2} \mu_2^2 \left(v_2^2 + v_+^2\right) \\
& - m_{12}^2 v_1 v_2 c_\xi + \frac{1}{4} \lambda_1 v_1^4 + \frac{1}{4} \lambda_2 \left(v_2^2 + v_+^2 \right)^2 + \frac{1}{4}\lambda_3 v_1^2 v_+^2 + \frac{1}{4}\left(\lambda_{34} + \lambda_5 c_{2\xi}\right)v_1^2 v_2^2 \\
&+ \frac{1}{2}\lambda_6 v_1^3 v_2 c_\xi + \frac{1}{2}\lambda_7 v_1 v_2 \left(v_2^2 + v_+^2\right)c_\xi.
\end{split}
\end{equation}
The resulting gradient flow equations are
\begin{align}
   \label{eq:CP1VEVs}
%	\begin{equation}
%	\begin{split}
	\frac{\partial v_1}{\partial t} =&\ \frac{\partial^2 v_1}{\partial x^2} + \mu_1^2 v_1 + m_{12}^2 v_2 c_\xi - \lambda_1 v_1^3 - \frac{1}{2}\lambda_3 v_1 v_+^2 - \frac{1}{2}\left(\lambda_{34} + \lambda_5 c_{2\xi}\right) v_1 v_2^2 \nonumber\\
	&- \frac{3}{2}\lambda_6 v_1^2 v_2 c_\xi - \frac{1}{2}\lambda_7 v_2 \left(v_2^2 + v_+^2\right) c_\xi,\nonumber\\[3mm]
%	\end{split}
%	\end{equation}
%	\begin{equation}
%	\begin{split}
	\frac{\partial v_2}{\partial t} =&\ \frac{\partial^2 v_2}{\partial x^2} - v_2\left(\frac{\partial \xi}{\partial x}\right)^2 + \mu_2^2 v_2 + m_{12}^2 v_1 c_\xi - \lambda_2 v_2\left(v_2^2 + v_+^2\right) - \frac{1}{2}\left(\lambda_{34} + \lambda_5 c_{2\xi}\right) v_1^2 v_2 \nonumber\\
	& - \frac{1}{2}\lambda_6 v_1^3 c_\xi - \frac{1}{2}\lambda_7 v_1 \left(3v_2^2 + v_+^2\right) c_\xi,\nonumber\\[3mm]
%	\end{split}
%	\end{equation}
%	\begin{equation}
	\frac{\partial v_+}{\partial t} =&\ \frac{\partial^2 v_+}{\partial x^2} + \mu_2^2 v_+ - \lambda_2 v_+\left(v_2^2 + v_+^2\right) - \frac{1}{2}\lambda_3 v_1^2 v_+ - \lambda_7 v_1 v_2 v_+ c_\xi,\\[3mm]
%	\end{equation}
%	\begin{equation}
%	\begin{split}
	\frac{\partial \xi}{\partial t} =&\ v_2^2\frac{\partial^2 \xi}{\partial x^2} + 2v_2 \left(\frac{\partial v_2}{\partial x}\right)\left(\frac{\partial \xi}{\partial x}\right) - m_{12}^2 v_1 v_2 s_\xi + \frac{1}{2}\lambda_5 v_1^2 v_2^2 s_{2\xi}  \nonumber\\
	& + \frac{1}{2}\lambda_6 v_1^3 v_2 s_\xi + \frac{1}{2}\lambda_7 v_1 v_2 \left(v_2^2 + v_+^2\right)s_\xi.\nonumber
%	\end{split}
%	\end{equation}
\end{align}
These gradient flow equations simplify in the reduced basis with $\lambda_4 = \lambda_6 = 0$ for which we obtain physical parameters.
The relaxed solutions of this system of equations in the reduced basis are given in Fig.~\ref{fig:CP1kinkphi} in both the linear representation of the Higgs doublets and for the vacuum manifold parameters, $v_1$, $v_2$, $v_+$ and $\xi$. 
It can be seen that a kink forms in $R^2$ and correspondingly in $\xi$.
These two systems of equations are equivalent having the same energy per unit area and profiles for $R^\mu$.
This is the case for all model parameters we have considered including arbitrary parameters sets in the non-reduced basis of Table~\ref{tab:2HDMConstraints}.
Therefore, as in the $Z_2$ case, there are a myriad of equivalent field configurations related by global electroweak transformations.
One such solution was obtained in \cite{Brawn2011} using the restricted ansatz $v_+ \equiv 0$.
The minimum energy solution to \eqref{eq:CP1VEVs} indeed has this property.
For both systems of equations a kink forms in $R^2$ and components of $R^\mu$ have asymptotic values given by \eqref{eq:RVectorVac}.
The quantity $R^2$ is odd under a CP1 transformation whilst $R^0$, $R^1$ and $R^3$ are even.
Hence, we may define the gauge-invariant topological current,
\begin{equation}
    J^\mu = \frac{1}{2\langle R^2\rangle}\varepsilon^{\mu\nu}\partial_\nu \left[-i\left(\Phi_1^\dagger\Phi_2 - \Phi_2^\dagger\Phi_1\right)\right] = \frac{1}{2\langle R^2\rangle}\varepsilon^{\mu\nu}\partial_\nu R^2,
\end{equation}
and corresponding topological charge,
\begin{equation}
    Q = \int_{-\infty}^\infty dx J^0 = \frac{1}{2\langle R^2 \rangle}\left[R^2(\infty) - R^2(-\infty)\right].
\end{equation}
Again, $Q=1$ corresponds to a kink and $Q=-1$ to an anti kink.

\subsection{Dynamical Simulations from Random Initial Conditions}

\begin{figure}
    \centering
    \hspace*{-0.71cm}
    \includegraphics[width=0.96\textwidth]{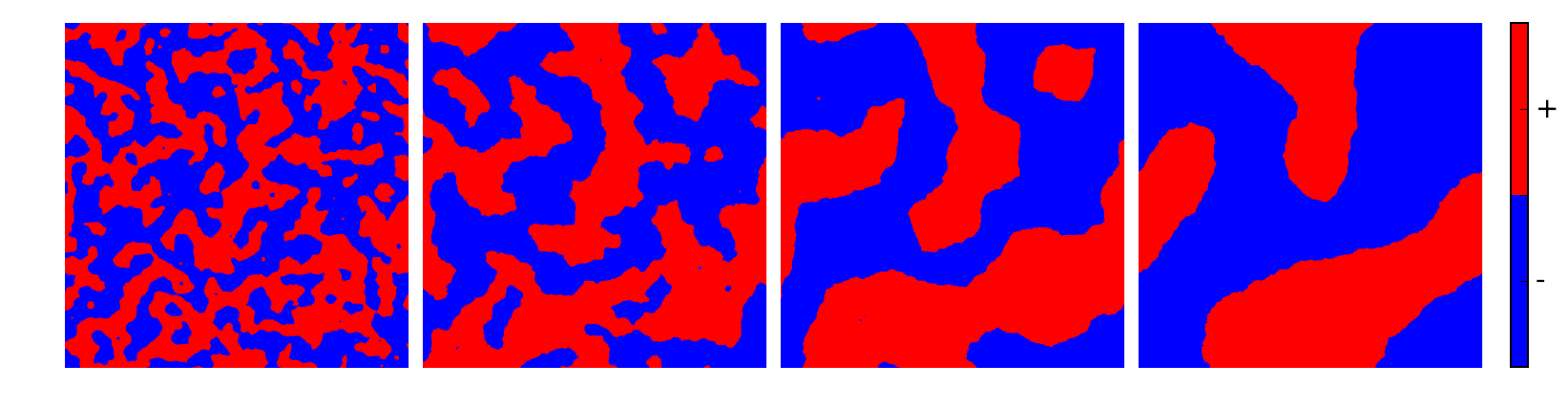} \\
    \includegraphics[width=\textwidth]{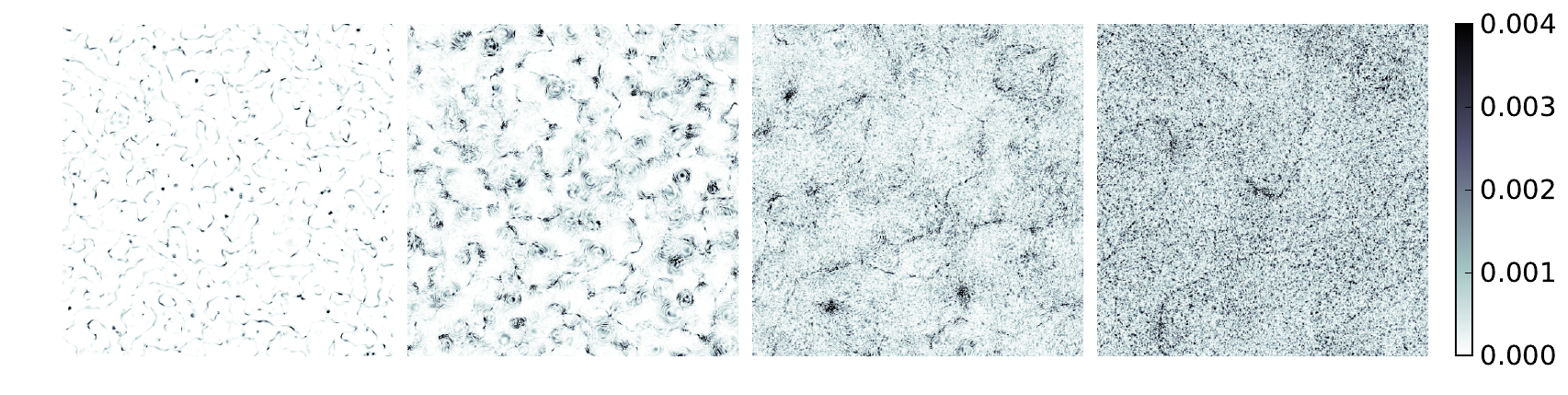}
    \caption{2D simulation of the evolution of domain walls in the CP1-symmetric 2HDM with Minkowski metric. Parameters chosen correspond to $ M_1 = $~125 GeV, $ M_2 = $~307 GeV, $ M_3 = $~455 GeV, $ M_{H^\pm} = $~300 GeV, $\xi = \pi/4$ and $\tan\beta = 2.0$.
    Simulation was run for time, $t=480$ with temporal grid spacing, $\Delta t = 0.2$ and spatial grid size, $P = 1024$ with spacing, $\Delta x = 0.9$.
    Plots progress in time left-to-right and each plot is at double the timestep of the previous.
    The upper figures present the sign of $R^2$ and the lower ones show the spatial variation of $v_+^2$.}
    \label{fig:CP1domainsphys}
\end{figure}

We have performed (2+1) dimensional simulations with $P = 1024$ and 4096 up to $t = \tau$ for the global field theory of the CP1-symmetric 2HDM with Minkowski, and FRW metric in the radiation and matter dominated eras using the PRS algorithm.
The evolution of a Minkowski simulation is presented in Fig.~\ref{fig:CP1domainsphys} along with the corresponding spatial variation of $v_+$, which would be zero in the case of a neutral vacuum solution, $v_+ \equiv 0$.
The domains are mapped according to the sign of $R^2$ as one would expect given the kink solutions in Fig.~\ref{fig:CP1kinkphi}.

\begin{figure}
    \centering
    \hspace*{-0.71cm}
    \includegraphics[width=0.96\textwidth]{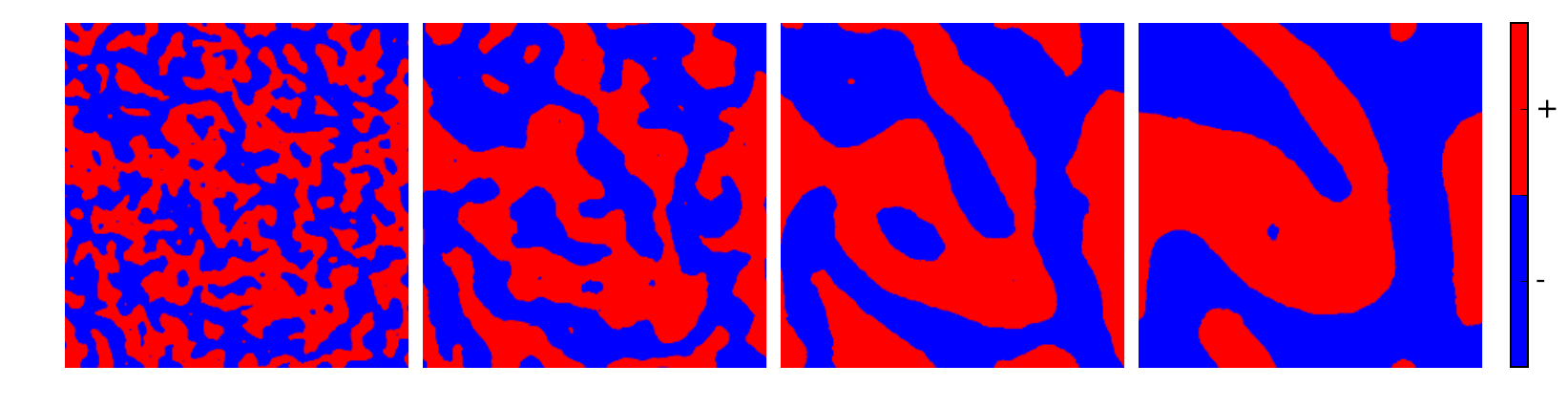} \\
    \includegraphics[width=\textwidth]{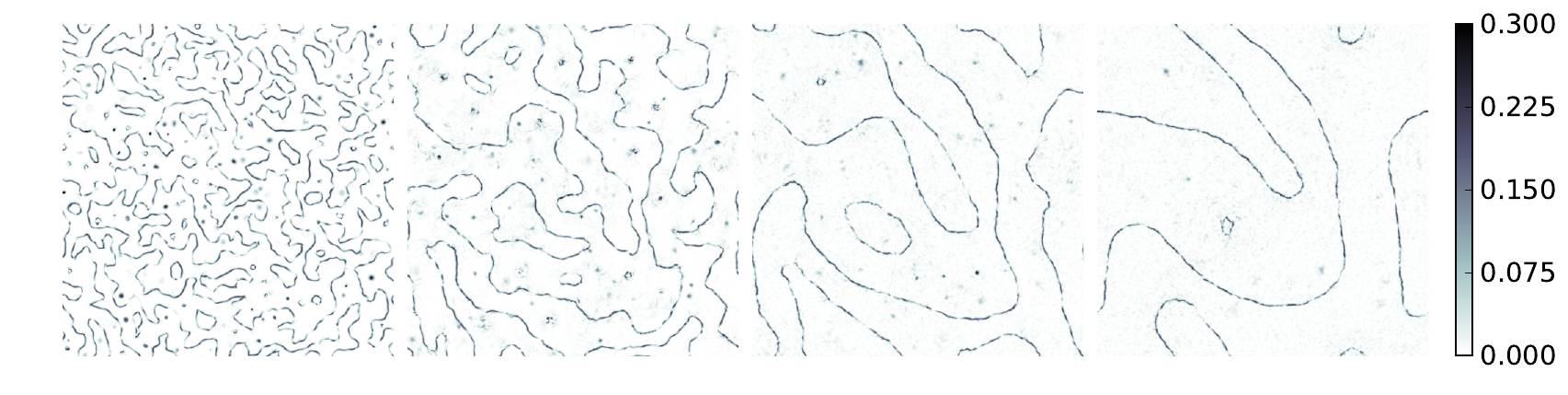}
    \caption{2D simulation of the evolution of domain walls in the CP1-symmetric 2HDM with Minkowski metric. Parameters are $\{\mu_1^2,\mu_2^2,m_{12}^2,\lambda_1,\lambda_2,\lambda_3,\lambda_4,\lambda_5,\lambda_6,\lambda_7\} = \{1,1,0.1,1,1,2,1.5,2,0,0\}$.
    Simulation was run for time, $t=480$ with temporal grid spacing, $\Delta t = 0.2$ and spatial grid size, $P = 1024$ with spacing, $\Delta x = 0.9$.
    Plots progress in time left-to-right and each plot is at double the timestep of the previous.
    The upper figures present the sign of $R^2$ and the lower ones show the spatial variation of $R_\mu R^\mu$.
    Clearly $R_\mu R^\mu \ne 0$ on the walls where $R^2$ changes sign.}
    \label{fig:CP1domains}
\end{figure}

As in the $Z_2$ case, we find a local violation of the neutral vacuum condition, $R_\mu R^\mu = 0$, on the domain walls in our simulations as shown in Fig.~\ref{fig:CP1domainsphys}.
The magnitude of this violation is small for physical parameter sets we have considered.
Nonetheless, the violation is present and is qualitatively similar to that found in the $Z_2$-symmetric 2HDM.
For clarity of presentation we have also performed simulations in the reduced basis for an arbitrary set of scalar potential parameters as presented in Fig.~\ref{fig:CP1domains}.
In this case the violation of the neutral vacuum on the domain walls which emerges in our simulations can be more easily seen.
For comparison to Fig.~\ref{fig:CP1domainsphys}, the corresponding physical observables for the parameters chosen in Fig.~\ref{fig:CP1domains} are $ M_1 = $~123 GeV, $ M_2 = $~317 GeV, $ M_3 = $~354 GeV, $ M_{H^\pm} = $~123 GeV, $\xi = 1.48\;\text{rad}$ and $\tan\beta = 1.0$.
It should be reiterated that the qualitative features presented here are general to all parameter sets we have considered.

\begin{figure}
    \centering
    \includegraphics[width=0.8\textwidth]{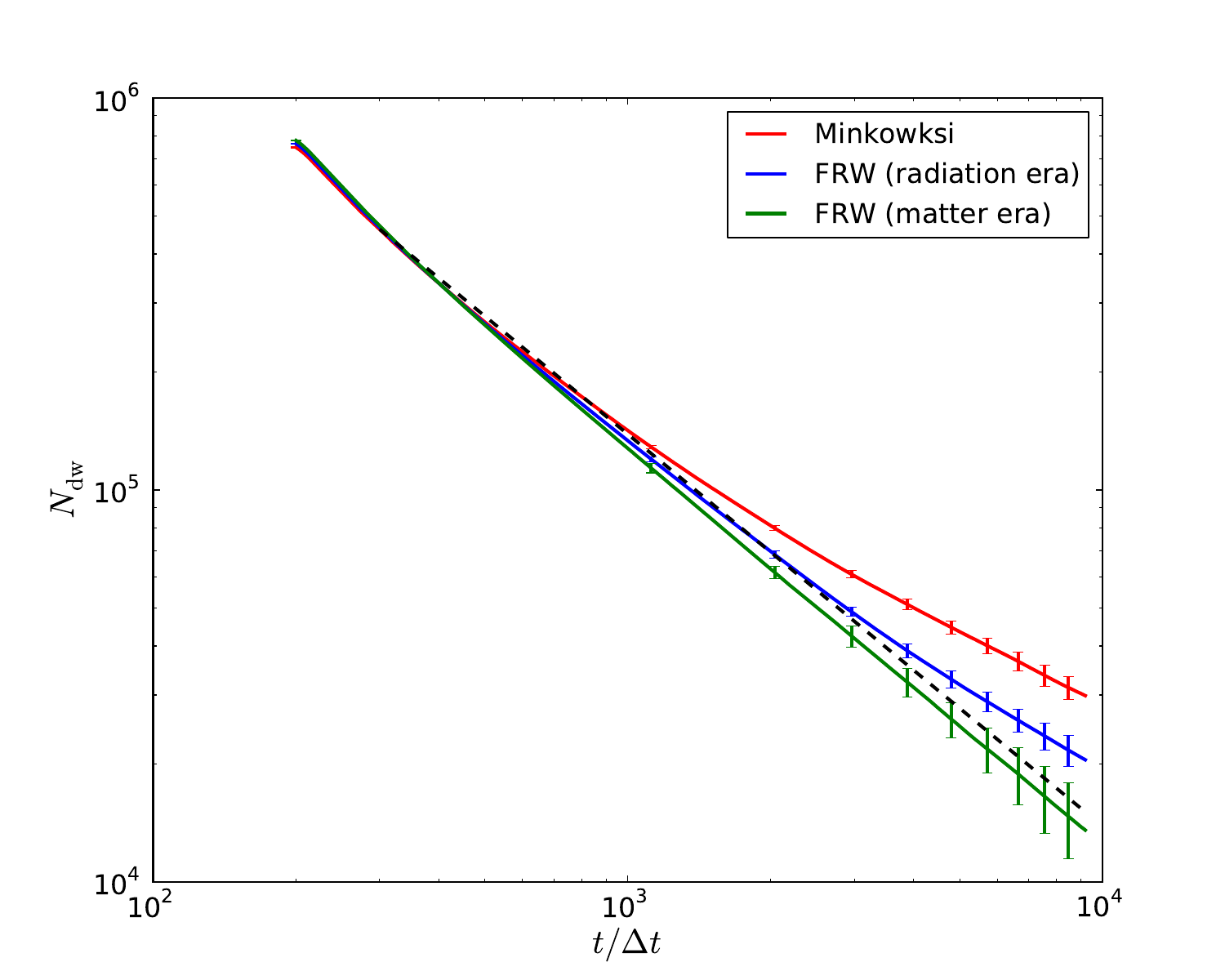}
    \caption{Evolution of the number of domain walls in 2D CP1-symmetric 2HDM simulations averaged over 10 realizations for Minkowski, and FRW in radiation and matter dominated eras.
    Also plotted is the standard power law scaling for a domain wall network $\propto t^{-1}$.
    Parameters chosen were $\{\mu_1^2,\mu_2^2,m_{12}^2,\lambda_1,\lambda_2,\lambda_3,\lambda_4,\lambda_5,\lambda_6,\lambda_7\} = \{1,1,0.1,1,1,2,1.5,2,0,0\}$.
    Simulations were run for time, $t=1840$ with temporal grid spacing, $\Delta t = 0.2$ and spatial grid size, $P = 4096$ with spacing, $\Delta x = 0.9$.
    Error bars, which are the standard deviation amongst the realizations, illustrates the numerical scatter between simulations.}
    \label{fig:CP1walls}
\end{figure}

The number of domain walls as a function of time in (2+1) dimensions with Minkowski and FRW metrics in radiation and matter dominated eras are presented in Fig.~\ref{fig:CP1walls}.
The time evolution of the number of domain walls is again obtained as an average over 10 realizations.
As in the $Z_2$-symmetric case, we find that in both Minkowski space and FRW radiation era domain walls do not follow standard scaling, whereas the FRW matter era is compatible with $N_\text{dw} \propto t^{-1}$.

% CP2 Symmetry
\section{CP2 Symmetry}
\label{sec:CP2}

A CP2 transformation of the Higgs doublets is given by~\cite{Ivanov2007,Ferreira:2009wh,Brawn2011}
\begin{equation}
    \Phi_1 \rightarrow \Phi_2^*,\quad
    \Phi_2 \rightarrow -\Phi_1^*.
\end{equation}
Using the parameter constraints of Table~\ref{tab:2HDMConstraints}, the CP2-symmetric 2HDM potential can be written as
\begin{align}
  \label{eq:CP2Potential}
    V &= -\mu_1^2\left(\Phi_1^\dagger\Phi_1 + \Phi_2^\dagger\Phi_2\right) + \lambda_1 \left[(\Phi_1^\dagger\Phi_1)^2 + (\Phi_2^\dagger\Phi_2)^2\right] + \lambda_3 (\Phi_1^\dagger\Phi_1)(\Phi_2^\dagger\Phi_2)\nonumber\\
    &\quad + \left(\lambda_{4} + R_5\right)\left[\text{Re}\left(\Phi_1^\dagger\Phi_2\right)\right]^2 
    + \left(\lambda_4 - R_5\right)\left[\text{Im}\left(\Phi_1^\dagger\Phi_2\right)\right]^2 \\
    &\quad - 2 I_5 \text{Re}\left(\Phi_1^\dagger\Phi_2\right)\text{Im}\left(\Phi_1^\dagger\Phi_2\right) + 2\left(\Phi_1^\dagger\Phi_2 - \Phi_2^\dagger\Phi_1\right)\left[ R_6 \text{Re}\left(\Phi_1^\dagger\Phi_2\right) - I_6\text{Im}\left(\Phi_1^\dagger\Phi_2\right)\right]\,,\nonumber
\end{align}
where we have introduced the notation $\text{Re}\left(\lambda_i\right) = R_i$ and $\text{Im}\left(\lambda_i\right) = I_i$.
The minimization conditions \cite{Pilaftsis1999} for the CP2-symmetric potential are
\begin{equation}
    \begin{split}
        \mu_1^2 & = \lambda_1 v_1^2 + \frac{1}{2}\left[\lambda_{34}+\text{Re}(\lambda_5 e^{2i\xi})\right]v_2^2 + \frac{1}{2}\text{Re}(\lambda_6 e^{i\xi}) (3v_1^2 - v_2^2)\tan\beta \\
        & = \lambda_1 v_2^2 + \frac{1}{2}\left[\lambda_{34}+\text{Re}(\lambda_5 e^{2i\xi})\right]v_1^2 - \frac{1}{2}\text{Re}(\lambda_6 e^{i\xi})(3v_2^2 - v_1^2)\cot\beta
    \end{split}
\end{equation}
and
\begin{equation}
    \text{Im}(\lambda_5e^{2i\xi}) v_1v_2 + \text{Im}(\lambda_6 e^{i\xi})(v_1^2 - v_2^2) = 0.
\end{equation}
Making use of these expressions, the elements of the charged mass matrix can be written as
\begin{equation}
  \begin{split}
    (\mathcal{M}_C^2)_{11} & = -\frac{1}{2}\left[\lambda_4 + \text{Re}(\lambda_5 e^{2i\xi})\right] v_2^2 - \frac{1}{2}\text{Re}(\lambda_6 e^{i\xi})(v_1^2 - v_2^2)\tan\beta\,,\\
    (\mathcal{M}_C^2)_{21} & = (\mathcal{M}_C^2)_{12}^* = \frac{1}{2}\left(\lambda_4 + \lambda_5e^{2i\xi}\right)v_1v_2 + \frac{1}{2}\lambda_6 e^{i\xi}(v_1^2-v_2^2)\,,\\
    (\mathcal{M}_C^2)_{22} & = -\frac{1}{2}\left[\lambda_4 + \text{Re}(\lambda_5 e^{2i\xi})\right] v_1^2 - \frac{1}{2}\text{Re}(\lambda_6 e^{i\xi})(v_1^2 - v_2^2)\cot\beta\,,
  \end{split}
\end{equation}
which has one zero eigenvalue for the Goldstone bosons, $G^\pm$, and a charged Higgs mass,
\begin{equation}
    M_{H^\pm}^2 = -\frac{1}{2}\left[\lambda_4 + \text{Re}(\lambda_5 e^{2i\xi}) + \text{Re}(\lambda_6e^{i\xi})(\cot\beta - \tan\beta)\right]v_\text{SM}^2.
\end{equation}
Moreover, the elements of the $3\times3$ squared mass matrix read
\begin{align}
    (\mathcal{M}_N^2)_{11} & = 2 \lambda _1 v_1^2+\frac{1}{2} \text{Re}\left(\lambda _6 e^{i \xi }\right) \left(3 v_1^2+v_2^2\right) \tan\beta\,,\nonumber\\
    (\mathcal{M}_N^2)_{12} & = \left[\lambda _{34} + \text{Re}\left(\lambda _5 e^{2 i \xi }\right)\right] v_1 v_2+\frac{3}{2} \text{Re}\left(\lambda _6 e^{i \xi }\right) \left(v_1^2-v_2^2\right)\,,\nonumber\\
    (\mathcal{M}_N^2)_{13} & = -\frac{1}{2} \left[\text{Im}\left(\lambda _5 e^{2 i \xi }\right) v_2 + 2 \text{Im}\left(\lambda _6 e^{i \xi }\right) v_1\right] v_{\text{SM}}\,,\nonumber\\
    (\mathcal{M}_N^2)_{22} & = 2 \lambda _1 v_2^2-\frac{1}{2} \text{Re}\left(\lambda _6 e^{i \xi }\right) \left(v_1^2+3 v_2^2\right) \cot\beta\, ,\\
    (\mathcal{M}_N^2)_{23} & = -\frac{1}{2} \left[\text{Im}\left(\lambda _5 e^{2 i \xi }\right) v_1-2 \text{Im}\left(\lambda _6 e^{i \xi }\right) v_2\right] v_{\text{SM}}\, ,\nonumber\\
    (\mathcal{M}_N^2)_{33} & =  \frac{1}{2} \left\{2\lambda _1 \left(s_\beta v_1^2+c_\beta v_2^2\right) +\left[\lambda _{34}-\text{Re}\left(\lambda _5 e^{2 i \xi }\right)\right] \left(c_\beta v_1^2+s_\beta v_2^2\right)\right.\nonumber\\
    &\left.\quad -4 \text{Re}\left(\lambda _5 e^{2 i \xi }\right) s_\beta c_\beta v_1 v_2 - \text{Re}\left(\lambda _6 e^{i \xi }\right) \left(2 c_{2\beta} v_1 v_2+s_{2\beta} \left(v_1^2-v_2^2\right)\right) -2 \mu _1^2\right\}.\quad\nonumber
  \end{align}
As in the CP1 case, the three squared mass eigenvalues of the above matrix, $M^2_{h,H}$ and $M^2_A$, are obtained by 
solving a cubic characteristic equation.
However, in the diagonally reduced basis of Table \ref{tab:Diag_Red_Constraints}, all scalar masses 
of the CP2-symmetric 2HDM take on a simple form given by
\begin{eqnarray}
        M_{h,H}^2 &=& \left(\lambda_1 \mp \frac{1}{2}\tilde{\lambda}_{345}\right) v_{\text{SM}}^2\,,\\
         M_A^2 &=& \left|\lambda_5\right| v_{\text{SM}}^2\,,\\
        M_{H^\pm}^2 &=& -\frac{1}{2}\left(\lambda_4 - \left|\lambda_5\right|\right) v_{\text{SM}}^2\,.
\end{eqnarray}
Thus, following a rescaling for dimensionless energy, we can write the CP2-symmetric 2HDM in the diagonally reduced basis in terms of the Higgs masses,
\begin{align}\label{eq:DRCP2potential}
        \hat{V} &= -\frac{1}{2}\left(\hat{\Phi}^\dagger_1\hat{\Phi}_1 + \hat{\Phi}^\dagger_2\hat{\Phi}_2\right) + \frac{1 + \hat{M}_H^2}{2}\left[(\hat{\Phi}^\dagger_1\hat{\Phi}_1)^2 + (\hat{\Phi}^\dagger_2\hat{\Phi}_2)^2\right]\nonumber\\
        &\quad + \left(1 - \hat{M}_H^2 + 2\hat{M}_{H^\pm}^2\right)(\hat{\Phi}^\dagger_1\hat{\Phi}_1)(\hat{\Phi}^\dagger_2\hat{\Phi}_2)
        - 2\hat{M}_{H^\pm}^2\left[\text{Re}\left(\hat{\Phi}_1^\dagger\hat{\Phi}_2\right)\right]^2\nonumber\\
        &\quad + 2\left(\hat{M}_A^2 - \hat{M}_{H^\pm}^2\right)\left[\text{Im}\left(\hat{\Phi}_1^\dagger\hat{\Phi}_2\right)\right]^2.
\end{align}
Notice that in the diagonally reduced basis, it is easier to see that CP2 becomes a restricted scenario of the $Z_2$ symmetry with $\tan\beta = 1$ and $g_{hZZ} = 1$~\cite{Darvishi:2020teg}. Therefore, without loss of generality, in all our numerical results presented in the following sections, we perform our simulations in the diagonally reduced parameters space of the CP2-symmetric 2HDM.

\subsection{Kinks Solutions}

As for the $Z_2$ and CP1 cases, we have performed a generalized treatment of kinks in the diagonally reduced CP2-symmetric 2HDM.
We obtain relaxed solutions from the general field configuration and general vacuum ansatz, \eqref{eq:ChargedVac}. 

\begin{figure}
    \centering
    \includegraphics[width=\textwidth]{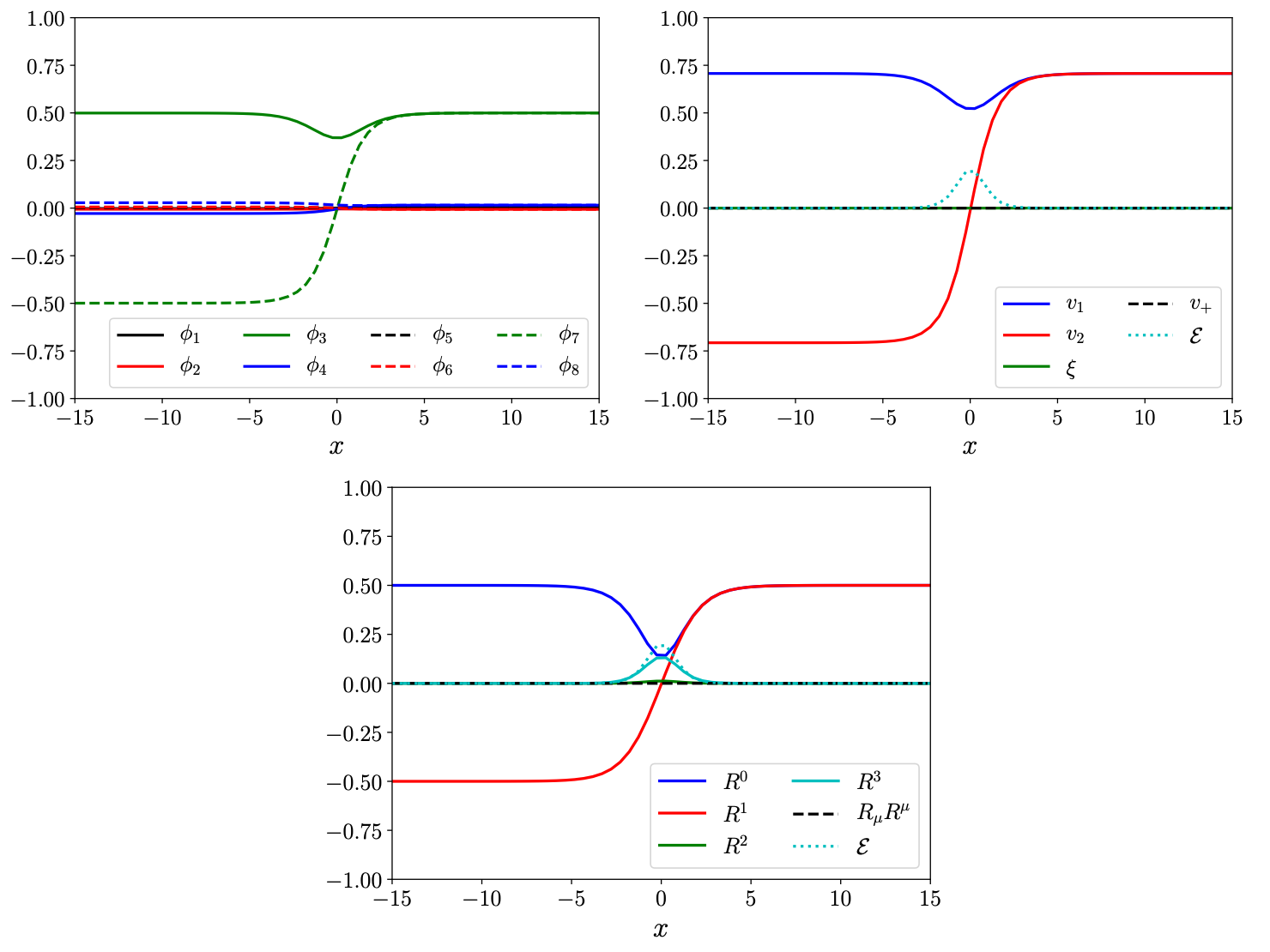}
    \caption{Kink solution of the diagonally reduced CP2-symmetric 2HDM in the linear representation (top, left) and vacuum manifold parameterization (top, right) along with corresponding $R$-space profiles (bottom). Parameters chosen were $M_H = M_A = M_{H^\pm} = 250$ GeV. Total energy of the solution is 0.858.}%154852109.}
    \label{fig:CP2kinkphi}
\end{figure}

Inserting the general vacuum, \eqref{eq:ChargedVac}, into the potential, \eqref{eq:CP2Potential}, one finds the energy per unit area for the diagonally reduced CP2-symmetric 2HDM:
\begin{equation}
\begin{split}
\mathcal{E} = &\frac{1}{2}\left(\frac{d v_1}{d x}\right)^2 + \frac{1}{2}\left(\frac{d v_2}{d x}\right)^2 + \frac{1}{2}\left(\frac{d v_+}{d x}\right)^2 + \frac{1}{2} v_2^2 \left(\frac{d \xi}{d x}\right)^2 - \frac{1}{2}\mu_1^2\left(v_1^2 + v_2^2 + v_+^2\right) \\
&+ \frac{1}{4}\lambda_1 \left[v_1^4 + \left(v_2^2 + v_+^2\right)^2\right] + \frac{1}{4} \lambda_3 v_1^2 v_+^2 + \frac{1}{4}\left(\lambda_{34} + R_5 c_{2\xi}\right) v_1^2 v_2^2 \\
%&- \frac{1}{4} I_5 v_1^2 v_2^2 s_{2\xi} + \frac{1}{2}v_1 v_2 \left(v_1^2 - v_2^2 - v_+^2\right)\left(R_6 c_\xi - I_6 s_\xi\right).
\end{split}
\end{equation}
The resulting gradient flow equations are
\begin{equation}
	\begin{split}
	\frac{\partial v_1}{\partial t} =&\ \frac{\partial^2 v_1}{\partial x^2} + \mu_1^2 v_1 - \lambda_1 v_1^3 - \frac{1}{2} \lambda_3 v_1 v_+^2 - \frac{1}{2}\left(\lambda_{34} + \lambda_5 c_{2\xi}\right) v_1 v_2^2, \\
%	\end{split}
%	\end{equation}
%	\begin{equation}
%	\begin{split}
	\frac{\partial v_2}{\partial t} =&\ \frac{\partial^2 v_2}{\partial x^2} - v_2\left(\frac{d \xi}{d x}\right)^2 + \mu_1^2 v_2 - \lambda_1 v_2\left(v_2^2 + v_+^2\right) - \frac{1}{2}\left(\lambda_{34} + \lambda_5 c_{2\xi}\right)v_1^2 v_2, \\
%	\end{split}
%	\end{equation}
%	\begin{equation}
%	\begin{split}
	\frac{\partial v_+}{\partial t} =&\ \frac{\partial^2 v_+}{\partial x^2} + \mu_1^2 v_+ - \lambda_1 v_+\left(v_2^2 + v_+^2\right) - \frac{1}{2}\lambda_3 v_1^2 v_+,  \\
%	\end{split}
%	\end{equation}
%	\begin{equation}
%	\begin{split}
	\frac{\partial \xi}{\partial t} =&\ v_2^2\frac{\partial^2 \xi}{\partial x^2} + 2v_2 \left(\frac{\partial v_2}{\partial x}\right)\left(\frac{\partial \xi}{\partial x}\right) + \frac{1}{2}\lambda_5 v_1^2 v_2 ^2 s_{2\xi}.
	\end{split}
\end{equation}
In Fig.~\ref{fig:CP2kinkphi}, we present the relaxed solutions both in the linear and bilinear field-space representations. In the latter representation, we see the formation of a kink along the $R^1$ direction.

Under a CP2 transformation, the quantities $R^{1,2,3}$ are odd in a general weak basis, 
whilst $R^0$ is even.
To distinguish CP2 from the other cases, $Z_2$ and CP1, one could in principle define the gauge-invariant topological current by means of $R^3$, i.e.
\begin{equation}
  \label{eq:CP2Jmu}
    J^\mu = \frac{1}{2\langle R^3\rangle}\varepsilon^{\mu\nu}\partial_\nu \left(\Phi_1^\dagger\Phi_1 - \Phi_2^\dagger\Phi_2\right)\: =\: \frac{1}{2\langle R^3\rangle}\varepsilon^{\mu\nu}\partial_\nu R^3,
\end{equation}
and the respective topological charge,
\begin{equation}
  \label{eq:CP2Q}
    Q = \int_{-\infty}^\infty dx J^0 = \frac{1}{2\langle R^3 \rangle}\left[R^3(\infty) - R^3(-\infty)\right].
\end{equation}
As before, we stipulate that $Q=1$ describes a kink and $Q=-1$ an anti-kink. However, as we will see in the next section, in the diagonally reduced basis, we have that $R^{2,3} (\pm \infty) = 0$, since the 
permutation symmetry $S_2$, 
under the exchange $\Phi_1 \leftrightarrow \Phi_2$, remains unbroken. As a consequence,
the definitions for the topological current and 
charge in~\eqref{eq:CP2Jmu} and~\eqref{eq:CP2Q} become singular. For this reason, we will
use the definitions~\eqref{eq:Z2Jmu} and~\eqref{eq:Z2Q} from the $Z_2$-symmetric case to identify kink and anti-kink solutions.

\subsection{Dynamical Simulations from Random Initial Conditions}

\begin{figure}
    \centering
    \hspace*{-0.92cm}
    \includegraphics[width=0.94\textwidth]{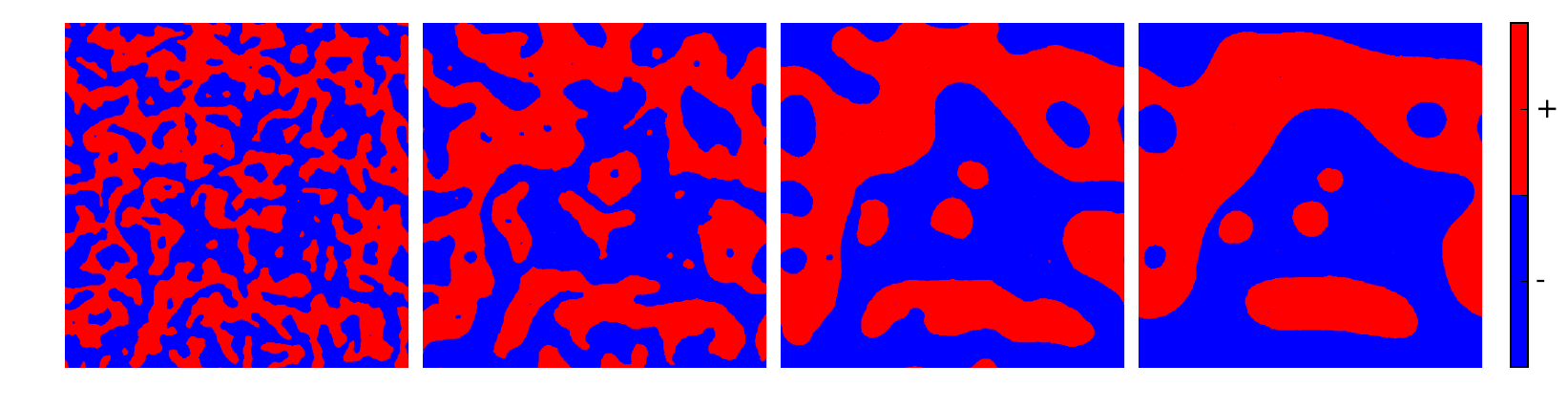} \\
    \includegraphics[width=\textwidth]{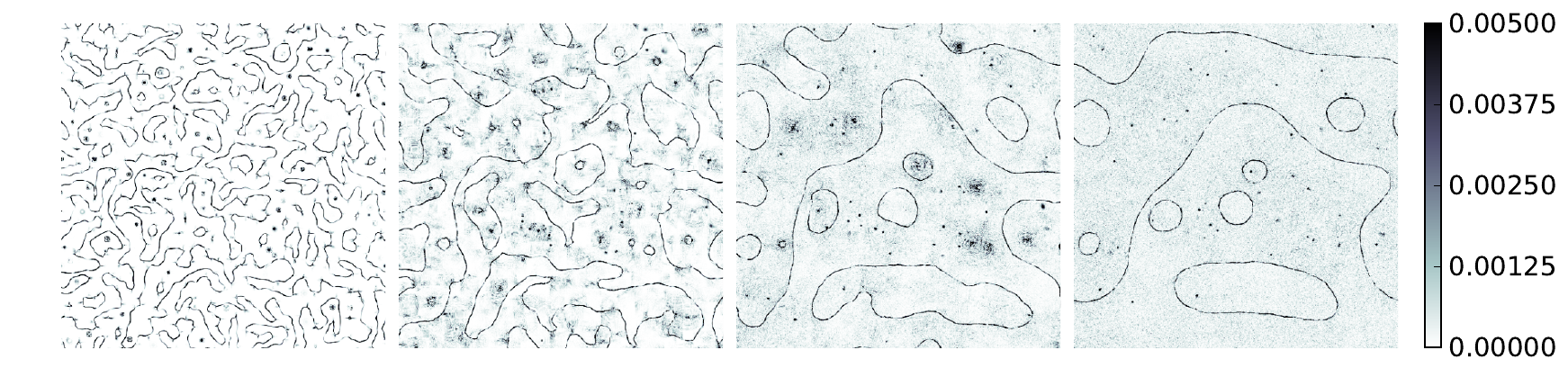}
    \caption{2D simulation of the evolution of domain walls in the diagonally reduced CP2-symmetric 2HDM with Minkowski metric. Parameters are $M_H = M_A = M_{H^\pm} = 250\;\text{GeV}$.
    Simulation was run for time, $t=480$ with temporal grid spacing, $\Delta t = 0.2$ and spatial grid size, $P = 1024$ with spacing, $\Delta x = 0.9$.
    Plots progress in time left-to-right and each plot is at double the timestep of the previous.
    The upper figures present the sign of $R^1$ and the lower ones are $R_\mu R^\mu$ clearly showing that $R_\mu R^\mu \ne 0$ on the walls where the sign of $R^1$ changes.}
    \label{fig:CP2domains}
\end{figure}

We have performed (2+1) dimensional simulations with $P = 1024$ and $P = 4096$ up to $t = \tau$ for the global field theory of the diagonally reduced CP2-symmetric 2HDM with Minkowski metric, and FRW metric in the radiation and matter dominated eras using the PRS algorithm.
The evolution of a Minkowski simulation is presented in Fig.~\ref{fig:CP2domains} along with the corresponding spatial variation of $R^\mu R_\mu$.
The domains are mapped using to the sign of $R^1$ given the solution in Fig.~\ref{fig:CP2kinkphi}.
As in the $Z_2$ and CP1 cases, we find a local violation of the neutral vacuum condition, $R_\mu R^\mu = 0$, on the domain walls in our simulations of the diagonlly reduced CP2-symmetric 2HDM as shown in Fig.~\ref{fig:CP2domains}.

\begin{figure}
    \centering
    \includegraphics[width=0.8\textwidth]{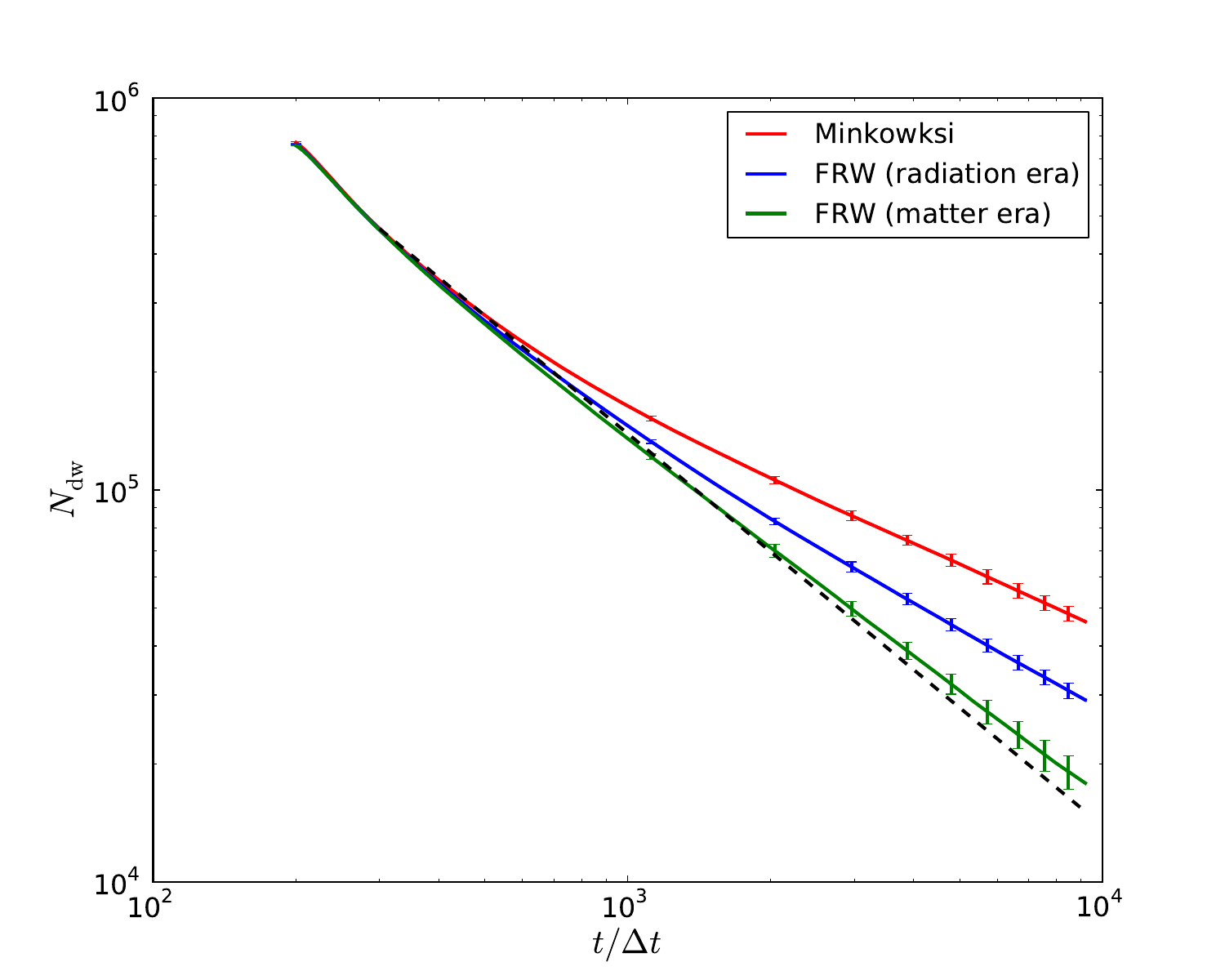}
    \caption{Evolution of the number of domain walls in the diagonally reduced CP2-symmetric 2HDM simulations averaged over 10 realizations for Minkowski, and FRW in radiation and matter dominated eras.
    Also plotted is the standard power law scaling for a domain wall network $\propto t^{-1}$.
    Parameters chosen were $M_H = M_A = M_{H^\pm} = 250\;\text{GeV}$.
    Simulations were run for time, $t=1840$ with temporal grid spacing, $\Delta t = 0.2$ and spatial grid size, $P = 4096$ with spacing, $\Delta x = 0.9$.
    Error bars, which are the standard deviation amongst the realizations, illustrate the numerical scatter.}
    \label{fig:CP2walls}
\end{figure}

The number of domain walls as a function of time in (2+1) dimensions with Minkowski and FRW metrics in radiation and matter dominated eras are presented in Fig.~\ref{fig:CP2walls}.
The time evolution of the number of domain walls is again obtained as an average over 10 realizations.
Both in Minkowski space and in FRW radiation- and matter-dominated eras, we find a similar feature that domain walls do not follow standard scaling in the diagonally reduced CP2-symmetric 2HDM.

% Discussion
\section{Conclusions}
\label{sec:discussion}

The 2HDM can in general predict a variety of topological defects when accidental symmetries are broken. Specifically, for $\text{SU}(2)_L \times \text{U}(1)_Y$ preserving symmetries, there are {\em three} domain wall solutions, {\em two} vortex solutions and {\em one} global monopole solution~\cite{Brawn2011}.
Here, our attention has been focused solely on 
2HDMs that produce domain walls after the spontaneous breakdown of the discrete $Z_2$, CP1 and CP2  symmetries.
Firstly, we have presented minimum energy kink solutions obtained via gradient flow.
We have shown that the these kinks can be visualized in the components of $R^\mu$ regardless of the specific field configuration in the linear representation.
For each case, we have been able to identify a gauge-invariant and conserved topological current, 
and the corresponding topological charge associated with these kink solutions.

For the 2HDMs with accidental $Z_2$, CP1 and CP2 symmetries, we have performed numerical simulations both in (2+1) and (3+1) dimensions.
In all three cases, simulations produce a network of domain walls which may manifest themselves in the components of the gauge-invariant $R^\mu$-vector, and are in agreement with changes of the associated topological charges.
Specifically, domain walls in the $Z_2$-symmetric 2HDM are found to flip the sign of $R^1$, domain walls in the CP1 case flip 
the sign in $R^{1,2}$ and the CP2 case has domain walls where all $R^{1,2,3}$ change sign.

In all three domain wall-forming 2HDMs under investigation for both Minkowski and FRW metrics, we have observed a deviation from the $t^{-1}$ power law scaling found in simpler models.
Specifically, in all three cases, we find more domain walls at late times in our simulations than one expects.
This feature can be attributed to the winding of the electroweak group parameters around the walls slowing their collapse.
This feature is akin to that seen in the kinky vorton model \cite{Pearson2009}.
We also observe a local violation of the neutral vacuum condition in our numerical simulations which we attribute to the relative EW rotation between domain walls.
This feature was further investigated by minimizing the energy per unit area of a field configuration with a relative EW rotation between the boundaries.
This minimization yields solutions with a local violation of the neutral vacuum conditions as seen in the random simulations.

It would be interesting to extend the present work and discuss constraints on the theoretical parameters that are obtained exclusively from the non-observation of domain walls today. Therefore, in an upcoming paper,
we plan to systematically address the particle physics phenomenological implications for all three 2HDMs with spontaneously broken discrete symmetries
that result from the absence of domain walls in the present epoch.

\subsection*{Acknowledgements}
The work of RB and AP are supported in part by the Lancaster-Manchester-Sheffield Consortium for Fundamental Physics under STFC research grant ST/L000520/1.

\vfill\eject

 \appendix
% \section{Some title}
% Please always give a title also for appendices.

\section{Physical Parametrization of $Z_2$ 2HDM}
    \label{sec:Rescaling}
    
The VEVs of the Higgs doublets can be expressed as
\begin{equation}
    v_1 = c_\beta v_\text{SM}, \quad v_2 = s_\beta v_\text{SM}
\end{equation}
where $v_\text{SM} = \sqrt{v_1^2 + v_2^2} = \text{246\;GeV}$ and $\beta$ is the mixing angle for the CP-odd scalars.
We have also introduced the short-hand notations $s_x = \sin x$ and $c_x = \cos x$.
The CP-even mass matrix \eqref{eq:CPevenmassmatrix} is diagonalized by the mixing angle, $\alpha$~\cite{Djouadi:2005gj}:
\begin{equation}
    \left(\begin{matrix}
    2 \lambda_1 v_1^2 & \Tilde{\lambda}_{345} v_1 v_2 \\
    \Tilde{\lambda}_{345} v_1 v_2 & 2 \lambda_2 v_2^2
    \end{matrix}\right) = 
    \left(\begin{matrix}
    c_\alpha & -s_\alpha \\
    s_\alpha & c_\alpha
    \end{matrix}\right)
    \left(\begin{matrix}
    M_h^2 & 0 \\
    0 & M_H^2
    \end{matrix}\right)
    \left(\begin{matrix}
    c_\alpha & s_\alpha \\
    -s_\alpha & c_\alpha
    \end{matrix}\right),
\end{equation}
from which we find
\begin{eqnarray}
\label{eq:lambda1,2_and_g}
        \lambda_1 &=& \frac{M_h^2 c_\alpha^2 + M_H^2 s_\alpha^2}{2 c_\beta^2 v_\text{SM}^2},\\
        \lambda_2 &=& \frac{M_h^2 s_\alpha^2 + M_H^2 c_\alpha^2}{2 s_\beta^2 v_\text{SM}^2},\\
\label{eq:l345}
        \Tilde{\lambda}_{345} &=& \frac{\left(M_h^2 - M_H^2\right) c_\alpha s_\alpha}{c_\beta s_\beta v_\text{SM}^2}.
\end{eqnarray}

Using the expressions for the VEVs given in \eqref{eq:Z2VEVs}, we exchange the parameters $\mu_1^2$ and $\mu_2^2$ for $v_1^2$ and $v_2^2$ as follows:
\begin{equation}
  \begin{split}
    \mu_1^2 =& \lambda_1 v_1^2 + \frac{1}{2}\tilde{\lambda}_{345}v_2^2 = \frac{1}{2}v_{\text{SM}}^2\left(2\lambda_1 c_\beta^2 + \tilde{\lambda}_{345} s_\beta^2\right),\\
    \mu_2^2 =& \lambda_2 v_2^2 + \frac{1}{2}\tilde{\lambda}_{345}v_1^2 = \frac{1}{2}v_{\text{SM}}^2\left(2\lambda_2 s_\beta^2 + \tilde{\lambda}_{345} c_\beta^2\right).
     \end{split}
\end{equation}
Hence, using \eqref{eq:lambda1,2_and_g}, we are able to write $\mu_1^2$ and $\mu_2^2$ purely in terms of physical parameters:
\begin{equation}
  \begin{split}
    \mu_1^2 =& \frac{1}{2}\left[M_h^2 c_\alpha^2 + M_H^2 s_\alpha^2 + \left(M_h^2 - M_H^2\right) c_\alpha s_\alpha \tan\beta \right],\\
    \mu_2^2 =& \frac{1}{2}\left[M_h^2 s_\alpha^2 + M_H^2 c_\alpha^2 + \left(M_h^2 - M_H^2\right) c_\alpha s_\alpha \cot\beta \right].
\end{split}
\end{equation}

The masses of the CP-odd scalar, $A$, and the charged scalars, $H^\pm$, are given by \eqref{eq:pseudomassses} and \eqref{eq:chargedmasses}, respectively.
Therefore, we can write the coefficients on the final two terms of the potential \eqref{eq:Z2Potential} as
\begin{equation}\label{eq:l4-l5}
    \left(\lambda_4 - \left|\lambda_5\right|\right) = -\frac{2 M_{H^\pm}^2}{v_\text{SM}^2}
\end{equation}
and
\begin{equation}\label{eq:l4+l5}
    \left(\lambda_4 + \left|\lambda_5\right|\right) = \frac{2\left(M_A^2 - M_{H^\pm}^2\right)}{v_\text{SM}^2}.
\end{equation}
Furthermore, we can trade $\lambda_3$ for a combination of the parameters above:
\begin{equation}\label{eq:l3}
    \lambda_3 = \tilde{\lambda}_{345} - \left(\lambda_4 - \left|\lambda_5\right|\right).
\end{equation}
Substituting \eqref{eq:l345} and \eqref{eq:l4-l5} into \eqref{eq:l3} yields
\begin{equation}
    \lambda_3 = \frac{\left(M_h^2 - M_H^2\right) c_\alpha s_\alpha + 2 M_{H^\pm}^2 c_\beta s_\beta}{c_\beta s_\beta v_\text{SM}^2}.
\end{equation}

We have now obtained expressions for all the coefficients in the potential \eqref{eq:Z2Potential} purely in terms of the Higgs masses, mixing angles and the SM VEV, $v_\text{SM}$.
Therefore, we can rewrite the $Z_2$-symmetric 2HDM potential as
\begin{align}
    V =& -\frac{1}{2}\left[M_h^2 c_\alpha^2 + M_H^2 s_\alpha^2 + \left(M_h^2 - M_H^2\right) c_\alpha s_\alpha \tan\beta \right] \Phi_1^\dagger\Phi_1\nonumber\\
    &- \frac{1}{2}\left[M_h^2 s_\alpha^2 + M_H^2 c_\alpha^2 + \left(M_h^2 - M_H^2\right) c_\alpha s_\alpha \cot\beta \right] \Phi_2^\dagger\Phi_2 + \left(\frac{M_h^2 c_\alpha^2 + M_H^2 s_\alpha^2}{2 c_\beta^2 v_\text{SM}^2}\right) (\Phi_1^\dagger\Phi_1)^2 \nonumber\\
    &+ \left(\frac{M_h^2 s_\alpha^2 + M_H^2 c_\alpha^2}{2 s_\beta^2 v_\text{SM}^2}\right) (\Phi_2^\dagger\Phi_2)^2 + \left(\frac{\left(M_h^2 - M_H^2\right) c_\alpha s_\alpha + 2 M_{H^\pm}^2 c_\beta s_\beta}{c_\beta s_\beta v_\text{SM}^2}\right) (\Phi_1^\dagger\Phi_1)(\Phi_2^\dagger\Phi_2)\nonumber\\
    &- \frac{2 M_{H^\pm}^2}{v_\text{SM}^2} \left[\text{Re}\left(\Phi_1^\dagger\Phi_2\right)\right]^2 + \frac{2\left(M_A^2 - M_{H^\pm}^2\right)}{v_\text{SM}^2} \left[\text{Im}\left(\Phi_1^\dagger\Phi_2\right)\right]^2.
\end{align}

It should be noted that two of the physical parameters are fixed by experiment; $M_h = 125 \; \text{GeV}$ and $v_\text{SM} = 246 \; \text{GeV}$ \cite{Aad2015}.
Therefore, we use these parameters to rescale for dimensionless length and energy.
We rescale to produce dimensionless length, $\Hat{x} = \rho x$, and dimensionless energy, $\Hat{\Phi}_i = \Phi_i/\eta$ \footnote{The parameters $\rho$ and $\eta$ have units of energy such that any object with a circumflex is dimensionless.}.
After making this rescaling, the energy per unit area for the $Z_2$-symmetric 2HDM is given by
\begin{align}
    &E = \int^\infty_{-\infty} \frac{d\Hat{x}}{\rho}\left\{\rho^2 \eta^2 \left|\Hat{\nabla} \Hat{\Phi}_i\right|^2 - \frac{1}{2}\left[M_h^2 c_\alpha^2 + M_H^2 s_\alpha^2 + \left(M_h^2 - M_H^2\right) c_\alpha s_\alpha \tan\beta \right] \eta^2 (\Hat{\Phi}_1^\dagger \Hat{\Phi}_1) \right.\nonumber\\
    &- \frac{1}{2}\left[M_h^2 s_\alpha^2 + M_H^2 c_\alpha^2 + \left(M_h^2 - M_H^2\right) c_\alpha s_\alpha \cot\beta \right] \eta^2 (\Hat{\Phi}_2^\dagger \Hat{\Phi}_2) + \left(\frac{M_h^2 c_\alpha^2 + M_H^2 s_\alpha^2}{2 c_\beta^2 v_\text{SM}^2}\right) \eta^4 (\Hat{\Phi}_1^\dagger \Hat{\Phi}_1)^2 \nonumber\\
    &\left. + \left(\frac{M_h^2 s_\alpha^2 + M_H^2 c_\alpha^2}{2 s_\beta^2 v_\text{SM}^2}\right) \eta^4 (\Hat{\Phi}_2^\dagger \Hat{\Phi}_2)^2 + \left(\frac{\left(M_h^2 - M_H^2\right) c_\alpha s_\alpha + 2 M_{H^\pm}^2 c_\beta s_\beta}{c_\beta s_\beta v_\text{SM}^2}\right) \eta^4 (\Hat{\Phi}_1^\dagger \Hat{\Phi}_1)(\Hat{\Phi}_2^\dagger \Hat{\Phi}_2) \right.\nonumber\\
    &\left. +\left(\frac{M_A^2 - 2M_{H^\pm}^2}{v_\text{SM}^2}\right) \eta^4 (\Hat{\Phi}_1^\dagger \Hat{\Phi}_2)(\Hat{\Phi}_2^\dagger \Hat{\Phi}_1) - \frac{M_A^2}{2 v_\text{SM}^2} \eta^4 \left[(\Hat{\Phi}_1^\dagger \Hat{\Phi}_2)^2 + \text{h.c.}\right] \right\}.
\end{align}
The most convenient choice of length and energy scales are $\rho = M_h$ and $\eta = v_\text{SM}$.
This allows an overall factor of $M_h v_\text{SM}^2$ to be taken out of the integrand.
This rescaling has now rendered the energy per unit area dimensionless:
\begin{align}
    \Hat{E} \equiv&\, \frac{E}{M_h v_\text{SM}^2} = \int^\infty_{-\infty} d\Hat{x}\left\{\left|\Hat{\nabla} \Hat{\Phi}_i\right|^2 - \frac{1}{2}\left[c_\alpha^2 + \Hat{M}_H^2 s_\alpha^2 + (1 - \Hat{M}_H^2) c_\alpha s_\alpha \tan\beta \right] (\Hat{\Phi}_1^\dagger\Hat{\Phi}_1) \right.
    \nonumber\\
    &- \frac{1}{2}\left[s_\alpha^2 + \Hat{M}_H^2 c_\alpha^2 + (1 - \Hat{M}_H^2) c_\alpha s_\alpha \cot\beta \right] (\Hat{\Phi}_2^\dagger\Hat{\Phi}_2) + \left(\frac{c_\alpha^2 + \Hat{M}_H^2 s_\alpha^2}{2 c_\beta^2}\right) (\Hat{\Phi}_1^\dagger\Hat{\Phi}_1)^2 \nonumber\\
    &\left. + \left(\frac{s_\alpha^2 + \Hat{M}_H^2 c_\alpha^2}{2 s_\beta^2}\right) (\Hat{\Phi}_2^\dagger\Hat{\Phi}_2)^2 + \left(\frac{(1 - \Hat{M}_H^2) c_\alpha s_\alpha + 2 \Hat{M}_{H^\pm}^2 c_\beta s_\beta}{c_\beta s_\beta}\right) (\Hat{\Phi}_1^\dagger\Hat{\Phi}_1)(\Hat{\Phi}_2^\dagger\Hat{\Phi}_2) \right. \nonumber\\
    &\left. +\left(\Hat{M}_A^2 - 2\Hat{M}_{H^\pm}^2\right) (\Hat{\Phi}_1^\dagger\Hat{\Phi}_2)(\Hat{\Phi}_2^\dagger\Hat{\Phi}_1) - \frac{1}{2} \Hat{M}_A^2 \left[(\Hat{\Phi}_1^\dagger \Hat{\Phi}_2)^2 + (\Hat{\Phi}_2^\dagger \Hat{\Phi}_1)^2\right] \right\}.
\end{align}
Note that we have defined the dimensionless Higgs masses $\Hat{M}_H = M_H/M_h$, $\Hat{M}_{H^\pm} = M_{H^\pm}/M_h$ and $\Hat{M}_A = M_A/M_h$ yielding the dimensionless potential
\begin{align}
    \Hat{V} =& - \frac{1}{2}\left[c_\alpha^2 + \Hat{M}_H^2 s_\alpha^2 + (1 - \Hat{M}_H^2) c_\alpha s_\alpha \tan\beta \right] (\Hat{\Phi}_1^\dagger\Hat{\Phi}_1)\nonumber\\
    &- \frac{1}{2}\left[s_\alpha^2 + \Hat{M}_H^2 c_\alpha^2 + (1 - \Hat{M}_H^2) c_\alpha s_\alpha \cot\beta \right] (\Hat{\Phi}_2^\dagger\Hat{\Phi}_2) + \frac{c_\alpha^2 + \Hat{M}_H^2 s_\alpha^2}{2 c_\beta^2} (\Hat{\Phi}_1^\dagger\Hat{\Phi}_1)^2 \nonumber\\
    &+ \frac{s_\alpha^2 + \Hat{M}_H^2 c_\alpha^2}{2 s_\beta^2} (\Hat{\Phi}_2^\dagger\Hat{\Phi}_2)^2 + \frac{(1 - \Hat{M}_H^2) c_\alpha s_\alpha + 2 \Hat{M}_{H^\pm}^2 c_\beta s_\beta}{c_\beta s_\beta} (\Hat{\Phi}_1^\dagger\Hat{\Phi}_1)(\Hat{\Phi}_2^\dagger\Hat{\Phi}_2) \nonumber\\
    &+ (\Hat{M}_A^2 - 2\Hat{M}_{H^\pm}^2) (\Hat{\Phi}_1^\dagger \Hat{\Phi}_2)(\Hat{\Phi}_2^\dagger \Hat{\Phi}_1) - \frac{1}{2} \Hat{M}_A^2 \left[(\Hat{\Phi}_1^\dagger \Hat{\Phi}_2)^2 + (\Hat{\Phi}_2^\dagger \Hat{\Phi}_1)^2\right].
\end{align}
\vfill\eject

%\acknowledgments

%This is the most common positions for acknowledgments. A macro is
%available to maintain the same layout and spelling of the heading.

%\paragraph{Note added.} This is also a good position for notes added
%after the paper has been written.


\bibliographystyle{unsrt}
\bibliography{bibliography}

% Please avoid comments such as "For a review'', "For some examples",
% "and references therein" or move them in the text. In general,
% please leave only references in the bibliography and move all
% accessory text in footnotes.

% Also, please have only one work for each \bibitem.

\end{document}